\documentclass[epj]{svjour}
\usepackage{graphics,epsfig}
\newcommand{\nn}{{\nonumber}\\}
\def\ket #1{|#1\rangle}
\def\tens{\otimes}

\def\Id{{\rm 1\kern-.3em I}}
\newcommand{\R}{{\rm I\kern-.2emR}}
\newcommand{\C}{{\ifmmode\mathchoice{{\rm C\kern-.4em\raisebox{.0ex}{\rule{.05ex}{.67em}}\kern.4em}}%
{{\rm C\kern-.4em\raisebox{.03ex}{\rule{.05ex}{.67em}}\kern.4em}}%
{{\rm C\kern-.3em\raisebox{.06ex}{\rule{.05ex}{.45em}}\kern.3em}}%
{{\rm C\kern-.3em\raisebox{.06ex}{\rule{.05ex}{.3em}}\kern.3em}}\else
{{\rm C\kern-.4em\raisebox{.03ex}{\rule{.05ex}{.67em}}\kern.4em}}\fi}}

\def\ket #1{|#1\rangle}
\def\SP #1 #2{\langle #1|#2\rangle}

\def\Expect #1{\langle #1\rangle}
\def\SpinorComp(#1,#2,#3){\Psi^{#1}_{#3}(#2_{#1})}
\def\AdSpinorComp(#1,#2,#3){\overline\Psi^{#1}_{#3}(#2_{#1})}
\def\spinorComp(#1,#2,#3){\Psi_{#3}(#2_{#1})}
\def\AdspinorComp(#1,#2,#3){\overline\Psi_{#3}(#2_{#1})}
\def\Spinor(#1,#2){\Psi^{#1}(#2_{#1})}
\def\AdSpinor(#1,#2){\bar\Psi^{#1}(#2_{#1})}
\def\SpinorI(#1,#2){\Psi_{Ip}^{#1}(#2_{#1})}
\def\AdSpinorI(#1,#2){\overline\Psi_{Ip}^{#1}(#2_{#1})}

\def\FeyProp(#1,#2,#3){S^{#1}_F(#2_{#1},#3_{#1})}
\def\FeyPropComp(#1,#2,#3,#4,#5){S^{#1}_{F\;#4 #5}(#2_{#1},#3_{#1})}
\def\metric(#1,#2){\langle #1, #2\rangle}
\def\bfgrk #1{\mbox{\boldmath$#1$}}
\def\MSD (#1,#2,#3,#4){[\Delta\frac{#1}{2}^#2]_{#3}(#4)}
\def\MSN (#1,#2,#3,#4){[N\frac{#1}{2}^#2]_{#3}(#4)}
\def\MSL (#1,#2,#3,#4){[\Lambda\frac{#1}{2}^#2]_{#3}(#4)}
\def\MSS (#1,#2,#3,#4){[\Sigma\frac{#1}{2}^#2]_{#3}(#4)}
\def\MSX (#1,#2,#3,#4){[\Xi\frac{#1}{2}^#2]_{#3}(#4)}
\def\MSO (#1,#2,#3,#4){[\Omega\frac{#1}{2}^#2]_{#3}(#4)}
\begin{document}
\onecolumn
\title{The light baryon spectrum in a relativistic quark model with
  instanton-induced quark forces}
\subtitle{The strange baryon spectrum}
\author{Ulrich L\"oring\thanks{e-mail: {\tt loering@itkp.uni-bonn.de}}, Bernard Ch.~Metsch \and  Herbert R.~Petry
}                     % Do not remove
\institute{Institut f\"ur Theoretische Kernphysik, Universit\"at Bonn, Nu{\ss}allee 14--16, D--53115 Bonn, Germany}
\date{}
\authorrunning{U.~L\"oring {\it et al.}}  

\abstract{This is the third of a series of papers treating light baryon
  resonances up to 3 GeV within a relativistically covariant quark model based
  on the Bethe-Salpeter equation with instantaneous two-
  and three-body forces. In this last paper we extend our previous work
  \cite{Loe01b} on non-strange baryons to a prediction of the complete strange
  baryon spectrum and a detailed comparison with experiment. We apply the
  covariant Salpeter framework, which we developed in the first
  paper \cite{Loe01a}, to the specific quark models introduced in ref. \cite{Loe01b}.
  Quark confinement is realized by linearly rising three-body string
  potentials with appropriate Dirac structures; to describe the hyperfine
  structure of the baryon spectrum we adopt 't~Hooft's two-quark residual
  interaction based on QCD instanton effects. The
  investigation of instanton-induced effects in the baryon mass spectrum plays
  a central role in this work. We demonstrate that several prominent features
  of the excited strange mass spectrum, {\it e.g.} the low positions of the
  strange partners of the Roper resonance or the appearance of approximate
  ''parity doublets'' in the $\Lambda$-spectrum, find a natural, uniform
  explanation in our relativistic quark model with instanton-induced forces.
 % end of PACS codes
} %end of abstract
\maketitle
%
%\tableofcontents
\section{Introduction}
\label{intro}
In this paper we want to present a description of strange baryons in a
relativistic quark model based on the three-particle Bethe-Salpeter
equation \cite{SaBe51,Tay66} with instantaneous forces (Salpeter
equation \cite{Sa52}). The model is characterized by three- and
two-particle potentials which we have already fixed in two preceding
papers \cite{Loe01a,Loe01b}; the three-particle potentials simulate
(linear) confinement \cite{CaKoPa83a,CaKoPa83b,Bal00} and the
two-particle potentials are of the form of 't~Hooft's
instanton-induced quark interaction \cite{tHo76,SVZ80}.  In
ref. \cite{Loe01b} we were able to demonstrate that non-strange
baryons can be accounted for very well (model $\mathcal{A}$); in
particular, the known Regge trajectories are correctly reproduced by
the confinement force and the hyperfine structure of the spectrum
appears with its characteristic details up to high energies as a
consequence of 't~Hooft's interaction.

The additional model parameters for models with strangeness are the strange
constituent quark mass and the coupling of strange to non-strange quarks in
't~Hooft's interaction.  We have fixed these parameters already in ref.
\cite{Loe01b} in order to reproduce the lowest baryon octet alone. This paper
contains the complete excited $\Lambda$-, $\Sigma$-, $\Xi$- and
$\Omega$-spectrum (as far as it is experimentally known) which thus is a
genuine prediction. We find excellent agreement with experiment, despite the
fact that our treatment of baryons is still incomplete in the sense that we do
not compute the strong decays of baryon resonances. We are aware that such a
calculation \cite{An97,An98a,An98b,AnSa97,AnSa96,AnProSa96} would influence
also the resonance positions computed in our model.  Indirectly, the
calculations of this paper, which are purely predictive and agree so well with
experiment, seem to indicate that such changes can probably be parameterized
within our phenomenological quark potentials and constituent quark masses.

The most significant result of this paper is the demonstration of the role of
't~Hooft's interaction in the hyperfine structure of the mass spectrum. The
details, which we shall present below, seem to establish its superior
relevance for light quark flavors; at least the characteristic operator
structure of this interaction seems to be indispensable for a satisfactory
calculation. (One-gluon-exchange potentials can be ruled out; see ref.
\cite{Loe01b}). Since we fixed the potentials strength of this interaction by
a fit to experiment, we can, however, not be completely sure that this force
is derived from QCD in a strict and unambiguous way. We include a small
theoretical consistency check in appendix \ref{sec:QCDcheck}, but believe that
more work has to be done, probably extending the efforts in ref.
\cite{SS98,DP84,DP85,DP86,DP92,NVZ89,NRZ96}. The main emphasize of this and
the two preceding papers \cite{Loe01a,Loe01b} is, however, put on the
development of a relativistic quark model and the detailed comparison with
experimental data.

The paper is organized as follows.  Section \ref{sec:Lam} is
concerned with an extensive discussion of our predictions for the excited
$\Lambda$-spectrum in comparison to the hitherto experimentally established
$\Lambda$-resonances.  A principal objective of our investigations is to
demonstrate the role of the instanton-induced 't~Hooft interaction in
generating several prominent structures seen in the experimental
$\Lambda$-spectrum.  These are for instance the low position of the Roper
analogue or the occurrence of approximate parity doublets. This discussion of
instanton effects is extended to the predictions in the other strange sectors,
{\it i.e.} for the $\Sigma$- and $\Xi$-resonances in sections  \ref{sec:Sig} and
\ref{sec:Xi}, respectively. In section \ref{sec:Om} we briefly present our
predictions for the $\Omega$-baryons, where 't~Hooft's force does not
contribute. In appendix \ref{sec:QCDcheck} we check in how far the model
parameters, which are fixed from the experimental baryon spectrum, are in
fact consistent with QCD relations from the theory of instantons.
Finally, we give a summary and conclusion in section \ref{sec:summary}.

\section{The $\Lambda$-resonance spectrum}
\label{sec:Lam}
In this section we analyze the predictions of model $\mathcal{A}$ and
$\mathcal{B}$ of ref. \cite{Loe01b} for the excited strange $\Lambda$-baryons with strangeness
$S^*=-1$ and isospin $T=0$ and compare our results with the currently
available experimental data. For the following discussion it is convenient to
begin with some general remarks concerning the action of 't~Hooft's force and
the experimental status of this flavor sector.

\subsection{Remarks -- Implications of 't~Hooft's force and the experimental situation}
Similar to the nucleon spectrum discussed in ref. \cite{Loe01b} we
expect the instanton-induced interaction to play an essential role
also for the description of the excited $\Lambda$-spectrum.  Let us
briefly comment on the influence of 't~Hooft's force on the different
states in this flavor sector.  As in the case of the excited nucleon
states, the effect of 't~Hooft's force in the different
$\Lambda$-states depends on the content of quark pairs with trivial
spin being antisymmetric in flavor. But in contrast to the nucleon
states these quark pairs can here be non-strange (nn) or
non-strange-strange (ns) and 't~Hooft's force distinguishes between
these types due to the different couplings $g_{nn}$ and $g_{ns}$.
Moreover, the constituent quark model predicts in comparison to the
nucleon sector additional degrees of freedom for the $\Lambda$-sector.
This results from inclusion of the strange quark, which leads to a
totally antisymmetric flavor singlet state $\Lambda_{\bf 1}$ in
addition to the mixed symmetric octet representations $\Lambda_{\bf
8}$. Hence, in addition to the octet states (that posses corresponding
counterparts in the nucleon spectrum) also singlet states appear in
the $\Lambda$-spectrum.  The positive and negative energy components
of the Salpeter amplitude $\Phi_{J^\pi}^\Lambda$ describing an excited
$\Lambda$-state with spin and parity $J^\pi$ are obtained by the
embedding map (see ref. \cite{Loe01a})
\begin{equation}
\label{embedSalp_Lam}
\Phi_{J^\pi}^\Lambda \;=\; T^{+++} {\varphi_{J^\pi}^\Lambda} \;+\; T^{---} {\varphi_{J^{-\pi}}^\Lambda}
\end{equation}
of totally $S_3$-symmetric Pauli spinors $\varphi^\Lambda_{J^{\pi}}$ and
$\varphi^\Lambda_{J^{-\pi}}$ which then generally decompose into the following six
different spin-flavor $SU(6)$-configurations:
\begin{eqnarray}
\label{Lam_decompPauli}
\ket{\varphi^\Lambda_{J^{\pm}}} 
&=&
\phantom{+\;}
\ket{\Lambda\;J^\pm,\; ^2 8[56]}
\;+\;
\ket{\Lambda\;J^\pm,\;^2 8[70]}
\;+\;
\ket{\Lambda\;J^\pm,\;^4 8[70]}
\;+\;
\ket{\Lambda\;J^\pm,\;^2 8[20]}\nn
& &
+\;
\ket{\Lambda\;J^\pm,\;^2 1[70]}
\;+\;
\ket{\Lambda\;J^\pm,\;^4 1[20]},
\end{eqnarray}
with the four flavor octet states (as for the nucleon
configurations)
\begin{equation}
\label{Lam8_configPauli}
\begin{array}{rcl}
\ket{\Lambda\;J^\pm,\; ^2 8[56]} &:=& 
\sum\limits_{L} \Bigg[ 
\ket{\psi^{L\;\pm}_\mathcal{S}}\tens
\frac{1}{\sqrt 2}\bigg(
\ket {\chi^\frac{1}{2}_{\mathcal{M}_\mathcal{A}}}
\tens
\ket {\phi^\Lambda_{\mathcal{M}_\mathcal{A}}}
+
\ket {\chi^\frac{1}{2}_{\mathcal{M}_\mathcal{S}}}
\tens
\ket {\phi^\Lambda_{\mathcal{M}_\mathcal{S}}}
\bigg)\Bigg]^J,\\[3mm]
\ket{\Lambda\;J^\pm,\;^2 8[70]} &:=& 
\sum\limits_{L} \Bigg[\phantom{+\;}\frac{1}{2}\; 
\ket{\psi^{L\;\pm}_{\mathcal{M}_\mathcal{A}}}\tens 
\bigg(
\ket {\chi^\frac{1}{2}_{\mathcal{M}_\mathcal{A}}}
\tens
\ket {\phi^\Lambda_{\mathcal{M}_\mathcal{S}}}
+
\ket {\chi^\frac{1}{2}_{\mathcal{M}_\mathcal{S}}}
\tens
\ket {\phi^\Lambda_{\mathcal{M}_\mathcal{A}}}
\bigg)\\[-2mm]
&& 
\phantom{\sum\limits_{L} \Bigg[}+
\frac{1}{2}\;
\ket{\psi^{L\;\pm}_{\mathcal{M}_\mathcal{S}}}\tens
\bigg(
\ket {\chi^\frac{1}{2}_{\mathcal{M}_\mathcal{A}}}
\tens
\ket {\phi^\Lambda_{\mathcal{M}_\mathcal{A}}}
-
\ket {\chi^\frac{1}{2}_{\mathcal{M}_\mathcal{S}}}
\tens
\ket {\phi^\Lambda_{\mathcal{M}_\mathcal{S}}}\bigg)\Bigg]^J,\\[3mm]
\ket{\Lambda\;J^\pm,\;^4 8[70]} &:=& 
\sum\limits_{L} \Bigg[\frac{1}{\sqrt 2} 
\bigg(
\ket{\psi^{L\;\pm}_{\mathcal{M}_\mathcal{A}}}\tens 
\ket{\chi^\frac{3}{2}_\mathcal{S}}\tens  
\ket{\phi^\Lambda_{\mathcal{M}_\mathcal{A}}}
-
\ket{\psi^{L\;\pm}_{\mathcal{M}_\mathcal{S}}}\tens 
\ket{\chi^\frac{3}{2}_\mathcal{S}}\tens  
\ket{\phi^\Lambda_{\mathcal{M}_\mathcal{S}}}
\bigg)\Bigg]^J,\\[3mm]
\ket{\Lambda\;J^\pm,\;^2 8[20]} &:=&  
\sum\limits_{L} \Bigg[\ket{\psi^{L\;\pm}_\mathcal{A}}\tens
\frac{1}{\sqrt 2}\bigg(
\ket {\chi^\frac{1}{2}_{\mathcal{M}_\mathcal{A}}}
\tens
\ket {\phi^\Lambda_{\mathcal{M}_\mathcal{S}}}
-
\ket {\chi^\frac{1}{2}_{\mathcal{M}_\mathcal{S}}}
\tens
\ket {\phi^\Lambda_{\mathcal{M}_\mathcal{A}}}
\bigg)\Bigg]^J,
\end{array}
\end{equation}
and the two flavor singlet states
\begin{equation}
\label{Lam1_configPauli}
\begin{array}{rcl}
\ket{\Lambda\;J^\pm,\;^2 1[70]} &:=&  
\sum\limits_{L} \Bigg[\frac{1}{\sqrt 2}\bigg(
\ket{\psi^{L\;\pm}_{\mathcal{M}_\mathcal{S}}}
\tens
\ket {\chi^\frac{1}{2}_{\mathcal{M}_\mathcal{A}}}
-
\ket{\psi^{L\;\pm}_{\mathcal{M}_\mathcal{A}}}
\tens
\ket {\chi^\frac{1}{2}_{\mathcal{M}_\mathcal{S}}}
\bigg)\Bigg]^J
\tens
\ket {\phi^\Lambda_\mathcal{A}},\\[3mm]
\ket{\Lambda\;J^\pm,\;^4 1[20]} &:=&  
\sum\limits_{L} \Bigg[
\ket{\psi^{L\;\pm}_\mathcal{A}}
\tens
\ket {\chi^\frac{3}{2}_\mathcal{S}}
\Bigg]^J
\tens
\ket {\phi^\Lambda_\mathcal{A}}.
\end{array}
\end{equation}
Here $\psi^{L\;\pm}_{R_L}$, $\chi^S_{R_S}$ and $\phi^N_{R_F}$ are the spatial,
spin and flavor wave functions with definite $S_3$-symmetries $R_L, R_S,
R_F\in\{\mathcal{S},\mathcal{M}_\mathcal{S}, \mathcal{M}_\mathcal{A},
\mathcal{A}\}$. The sum runs over the possible orbital angular momenta $L$ that
can be coupled with the internal spin $S$ to the total spin $J$ as denoted by
the angular brackets $[\ldots]^J$. To explore the implications of the strong
selection rules of 't~Hooft's force for the different $\Lambda$-states let us
discuss qualitatively what one naively expects in a simplified picture
(disregarding the negative energy component and the relativistic effects from
the embedding map of the Salpeter amplitudes (non-relativistic limit)).
Recalling the selection rules of 't~Hooft's force for the flavor octet states
from our earlier discussion of the nucleon sector (see ref. \cite{Loe01b}), we
expect the dominantly $^4 8[70]$ and $^2 8[20]$ states to be hardly
influenced, whereas dominantly $^2 8[56]$ and $^2 8[70]$ states should
be shifted downward mass shift.  Moreover, 't~Hooft's force generally mixes the
configurations $^2 8[56]$ and $^2 8[70]$.  Concerning the additional flavor
singlet states we expect dominantly $^4 1[20]$ states to remain
essentially unaffected due to the internal totally symmetric spin function
$\chi^{3/2}_\mathcal{S}$ in the $^4 1[20]$ configurations.  But similar to
dominantly $^2 8[56]$ and $^2 8[70]$ states we likewise expect a lowering of
dominantly $^2 1[70]$ states.  Finally, we should mention here that the
difference $g_{nn}-g_{ns}>0$ between the 't~Hooft couplings as required by the
$\Sigma-\Lambda$ ground-state splitting (see ref. \cite{Loe01b}) implies
further flavor $SU(3)$ symmetry breaking effects in addition to those arising
already from the difference $m_s-m_n>0$ between the non-strange and strange
quark masses. This leads to a further mixing of the flavor singlet
configuration $^2 1[70]$ with the flavor octet configurations $^2 8[70]$ and
$^2 8[56]$, which depends on the difference $g_{nn}-g_{ns}$ of the two
couplings and vanishes in the case $g_{ns}=g_{nn}$.\\ Once again we should be
aware of the simplicity of these naive non-relativistic considerations: In the
same manner as observed for the excited nucleon states \cite{Loe01b}, the relativistic
effects in our fully relativistic framework, especially the interplay of
't~Hooft's force with relativistic effects from confinement should also here
be very crucial for the influence of 't~Hooft's residual force on the excited
$\Lambda$-states.  In the course of the following discussion we therefore will
again analyze how instanton-induced effects in our fully relativistic approach
do really shape the hyperfine structures in the excited $\Lambda$-spectrum.
From the discussion of the nucleon spectrum we expect again
substantial differences between the results of the confinement models
$\mathcal{A}$ and
$\mathcal{B}$.\\

\begin{figure}[!h]
  \begin{center}
    \epsfig{file={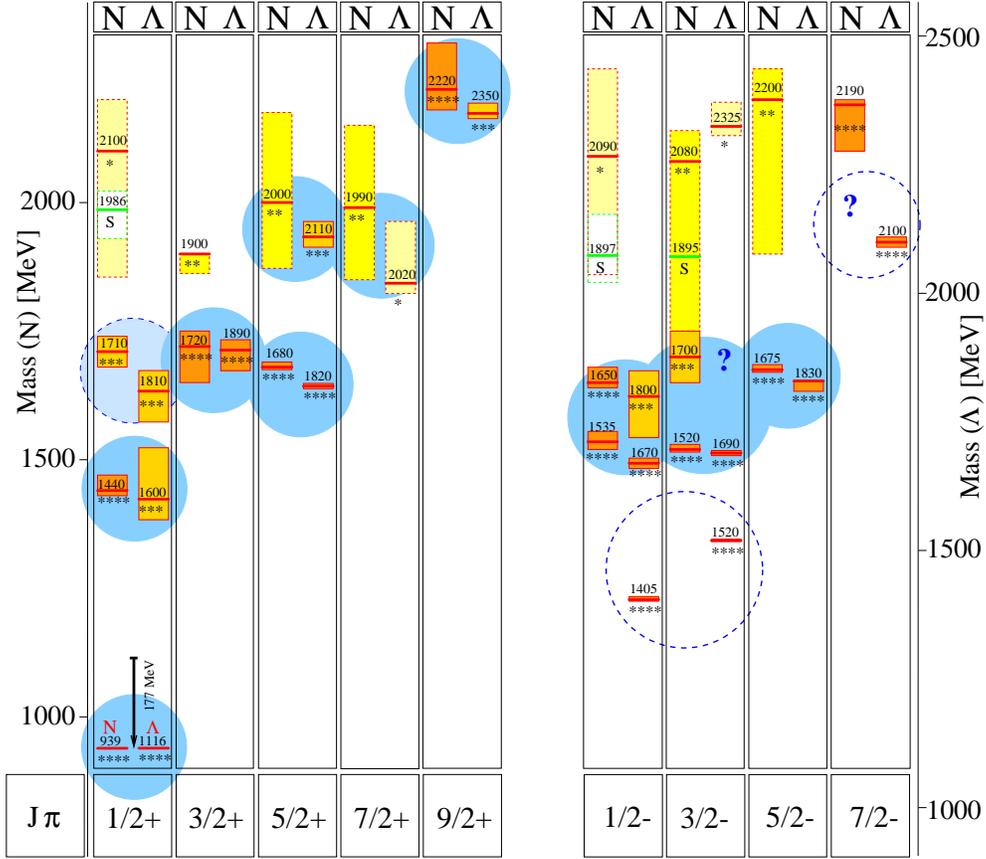},width=130mm}
  \end{center}
\caption{Comparison of the present experimental situation for nucleon- and
  $\Lambda$-resonances. The resonances are classified due to their total spin
  and parity $J^\pi$. On the left in each column, the nucleon resonances are
  shown. For comparison, the $\Lambda$-resonances are shown on the right hand
  side in each column. Note, that the mass scale for the $\Lambda$-states is
  shifted downwards with respect to the mass scale of nucleon states by $177$ MeV, so
  that the ground-states appear on the same level. In, fact there are a lot of
  $\Lambda$-states (expected to be dominantly flavor octet), that have their
  direct counterparts in the nucleon spectrum. See text for further
  explanations.}
\label{fig:NucLamComp}
\end{figure}

Before quoting our predictions let us first discuss heuristically what we do
expect from our earlier investigations of the nucleon sector (see ref.
\cite{Loe01b}) in view of the rather similar structures that can be found in the
experimental $\Lambda$- and nucleon spectra. In this respect, it is
instructive to consider the flavor $SU(3)$ symmetric limit, {\it i.e.}
$m_s=m_n$ and $g_{ns}=g_{nn}$. In this limit the flavor octet and flavor
singlet states completely decouple due to the explicit flavor-independence of
the confinement kernel and due to the flavor $SU(3)$ invariance of the
embedding map, the kinetic energy operator and 't~Hooft's force in this case.
Consequently, the flavor octet states of the $\Lambda$-spectrum and the
nucleon spectrum have exactly the same masses and configuration mixings. The
singlet states just additionally appear with the $^2 1[70]$ states lowered
with respect to the $^4 1[20]$ states.  Of course, in the realistic case (with
$m_s>m_n$ and $g_{ns}<g_{nn}$) this degeneracy is lifted and singlet and octet
states mix. Nonetheless, we expect the dominantly flavor octet states of the
$\Lambda$-spectrum forming hyperfine structures that have their direct
counterparts in the excited nucleon spectrum.  This, in fact, one really
observes in the experimental $\Lambda$-resonance spectrum. Figure
\ref{fig:NucLamComp} shows a direct comparison of the present experimental
situation for the nucleon- and $\Lambda$-resonances for each sector with spin
and parity $J^\pi$.
The nucleon states are displayed on the left hand side in each column and the
$\Lambda$-states on the right hand side. To correct approximately for the flavor
$SU(3)$ breaking effects in the $\Lambda$-states, the
mass scales for $N$- and $\Lambda$-resonances are mutually shifted by 177 MeV, such
that the ground-states appear at the same level. The figure nicely
demonstrates that the positive- and negative-parity $\Lambda$-spectra indeed
exhibit several structures showing similar hyperfine splittings as in the nucleon
spectrum. The corresponding states thus presumably are the flavor
octet counterparts of the nucleon spectrum. According to fig.
\ref{fig:NucLamComp}, let us briefly summarize the most striking features of
the experimentally observed $\Lambda$-spectrum that have (and have not)
counterparts in the experimental nucleon spectrum:
\begin{itemize}
\item The pattern of four low-lying $\Lambda$-resonances in the
  positive-parity $2\hbar\omega$ band indeed shows a very striking similarity
  to the structure of the four low-lying states in $2\hbar\omega$ shell of the
  nucleon spectrum: The lowest lying resonance
  $\Lambda\frac{1}{2}^+(1600,\mbox{***})$, which is the first isoscalar/scalar
  excitation of the $\Lambda$-ground-state, may be viewed as the strange counterpart of the
  Roper resonance $N\frac{1}{2}^+(1440,\mbox{****})$.  The remaining
  three well-established resonances $\Lambda\frac{1}{2}^+(1810,\mbox{***})$,
  $\Lambda\frac{3}{2}^+(1890,\mbox{****})$ and
  $\Lambda\frac{5}{2}^+(1820,\mbox{****})$ are approximately degenerate
  at around 1850 MeV quite similar to the three nucleon resonances
  $N\frac{1}{2}^+(1710,\mbox{***})$, $N\frac{3}{2}^+(1720,\mbox{****})$ and
  $N\frac{5}{2}^+(1680,\mbox{****})$ which are nearly degenerate at
  around 1700 MeV in the nucleon $2\hbar\omega$ band. Here the $\Lambda\frac{3}{2}^+(1890,\mbox{****})$ and the
  $\Lambda\frac{5}{2}^+(1820,\mbox{****})$  are presumably the octet partners of
  the $N\frac{3}{2}^+(1720,\mbox{****})$ and the
  $N\frac{5}{2}^+(1680,\mbox{****})$, respectively. The correspondence of
  $\Lambda\frac{1}{2}^+(1810,\mbox{***})$ with
  $N\frac{1}{2}^+(1710,\mbox{***})$, however, seems to be less clear.
\item In the upper part of the negative-parity $1\hbar\omega$ band between
  1600 and 1900 MeV the four well-established three- and four-star
  $\Lambda$-resonances stated by the Particle Data Group exhibit
  hyperfine splittings which are similar to those of the five resonances
  observed in the $1\hbar\omega$ shell of the nucleon spectrum: The two lower
  lying resonances $\Lambda\frac{1}{2}^-(1670,\mbox{****})$ and
  $\Lambda\frac{3}{2}^-(1690,\mbox{****})$, which form a nearly degenerate
  doublet at around 1680 MeV, may be interpreted as the octet partners of the
  two approximately degenerate (dominantly $^2 8[70]$) states
  $N\frac{1}{2}^-(1535,\mbox{****})$ and $N\frac{3}{2}^-(1520,\mbox{***})$ in
  the lower part of the nucleon $1\hbar\omega$ shell.  The two higher lying
  resonances $\Lambda\frac{1}{2}^-(1800,\mbox{***})$ and
  $\Lambda\frac{5}{2}^-(1830,\mbox{****})$ positioned approximately degenerate
  at roughly 1800 MeV should then be the octet counterparts to the
  $J^\pi=\frac{1}{2}^-$ and $J^\pi=\frac{5}{2}^-$ nucleon states of the
  triplet formed by the nearly degenerate (dominantly $^4 8[70]$)
  resonances $N\frac{1}{2}^-(1650,\mbox{****})$,
  $N\frac{3}{2}^-(1700,\mbox{***})$ and $N\frac{5}{2}^-(1675,\mbox{****})$ in
  the upper part of the nucleon $1\hbar\omega$ shell.  The octet partner of
  $N\frac{3}{2}^-(1700,\mbox{***})$, however, which likewise should appear at
  roughly 1800 - 1900 MeV in the $\Lambda\frac{3}{2}^-$ sector, has not been
  observed so far.
  
  The two lowest lying $\Lambda$-resonances in the $1\hbar\omega$ band, {\it
    i.e.} the two four-star states $\Lambda\frac{1}{2}^-(1405,\mbox{****})$
  and $\Lambda\frac{3}{2}^-(1520,\mbox{****})$, have no counterparts in the
  nucleon spectrum. Consequently, they are expected to be dominantly flavor
  singlet states. Apart from the significantly lower position with respect to
  the octet states, a very striking feature of these two resonances is the
  very low position of the $\Lambda\frac{1}{2}^-(1405,\mbox{****})$ relative
  to the $\Lambda\frac{3}{2}^-(1520,\mbox{****})$. The
  $\Lambda\frac{1}{2}^-(1405,\mbox{****})$ lies even below the lowest nucleon
  excitations and is the lowest negative parity excitation in the light baryon
  spectrum measured at all. The failure of several constituent quark models in
  reproducing the low position of this well-established four-star state is a
  long-standing problem and there is still controversy about the real physical
  nature of this state, see for instance \cite{Dal00}.
\item
Similar to the nucleon spectrum one observes overlapping parts of alternating 
even- and odd-parity bands which likewise lead to the appearance of
approximate ''parity doublets'' in the experimental $\Lambda$-spectrum.
The best established parity doublet structure is formed by the two 
lowest four-star excitations in the sectors with total spin $J=\frac{5}{2}$:
\begin{center}
\begin{tabular}{lcl}
$\Lambda\frac{5}{2}^+(1820,\mbox{****})$   &--& $\Lambda\frac{5}{2}^-(1830,\mbox{****})$.
\end{tabular}
\end{center}
This doublet is due to the overlap of the negative-parity $1\hbar\omega$ and the
positive-parity $2\hbar\omega$ shell.
A further parity doublet in the same energy range between 1800 and 1850
MeV is found in the sectors with total spin $J=\frac{1}{2}$ with the two
three-star states
\begin{center}
\begin{tabular}{lcl}
$\Lambda\frac{1}{2}^+(1810,\mbox{***})$   &--& $\Lambda\frac{1}{2}^-(1800,\mbox{***})$.
\end{tabular}
\end{center}
In the higher mass region of the $\Lambda$-spectrum such a parity
doublet pattern is indicated by the lowest resonances observed in
the $\Lambda\frac{7}{2}^\pm$ sectors\footnote{It is worth to note in
this respect the striking difference to the present experimental
situation in the nucleon spectrum (see fig. \ref{fig:NucLamComp})
and to refer back to our predictions for the lowest excitations in the
corresponding $N\frac{7}{2}^\pm$ sectors (see ref. \cite{Loe01b}): 
In contrast to our model calculation the
currently available experimental data do not exhibit a clear parity
doublet structure of the lowest excitations in $N\frac{7}{2}^\pm$ due
to the rather high position of the $N\frac{7}{2}^-(2190,\mbox{****})$
with respect to the $N\frac{7}{2}^+(1990,\mbox{**})$.  In view of the
otherwise striking similarities between the experimental nucleon and
$\Lambda$-spectra the comparatively low position of
$\Lambda\frac{7}{2}^-(2100,\mbox{****})$ might be a further strong
{\it experimental} indication for a $N\frac{7}{2}^-$ resonance
even below the $N\frac{7}{2}^-(2190,\mbox{****})$.}:
\begin{center}
\begin{tabular}{lcl}
$\Lambda\frac{7}{2}^+(2020,\mbox{*})$   &--& $\Lambda\frac{7}{2}^-(2100,\mbox{****})$.
\end{tabular}
\end{center}
\end{itemize}
Since all corresponding structures of the excited nucleon spectrum
could be nicely reproduced by means of 't~Hooft's residual
interaction, our approach (model $\mathcal{A}$) looks very promising to work as well for the
counterparts in the $\Lambda$-spectrum. So let us see now how far the
predictions of both models can really account for these observed
structures using the parameter sets of ref. \cite{Loe01b} as
they were fixed on the phenomenology of the $\Delta$- and the
ground-state spectrum alone. In this respect, we should stress once more
that also in the $\Lambda$-sector all calculated positions of excited
states are {\it real parameter-free predictions}.  As already in the
nucleon spectrum, {\it no} structure in the $\Lambda$-spectrum has
been explicitly adjusted.

\subsection{Discussion of the complete $\Lambda$-spectrum}
\begin{figure}[!h]
  \begin{center}
    \epsfig{file={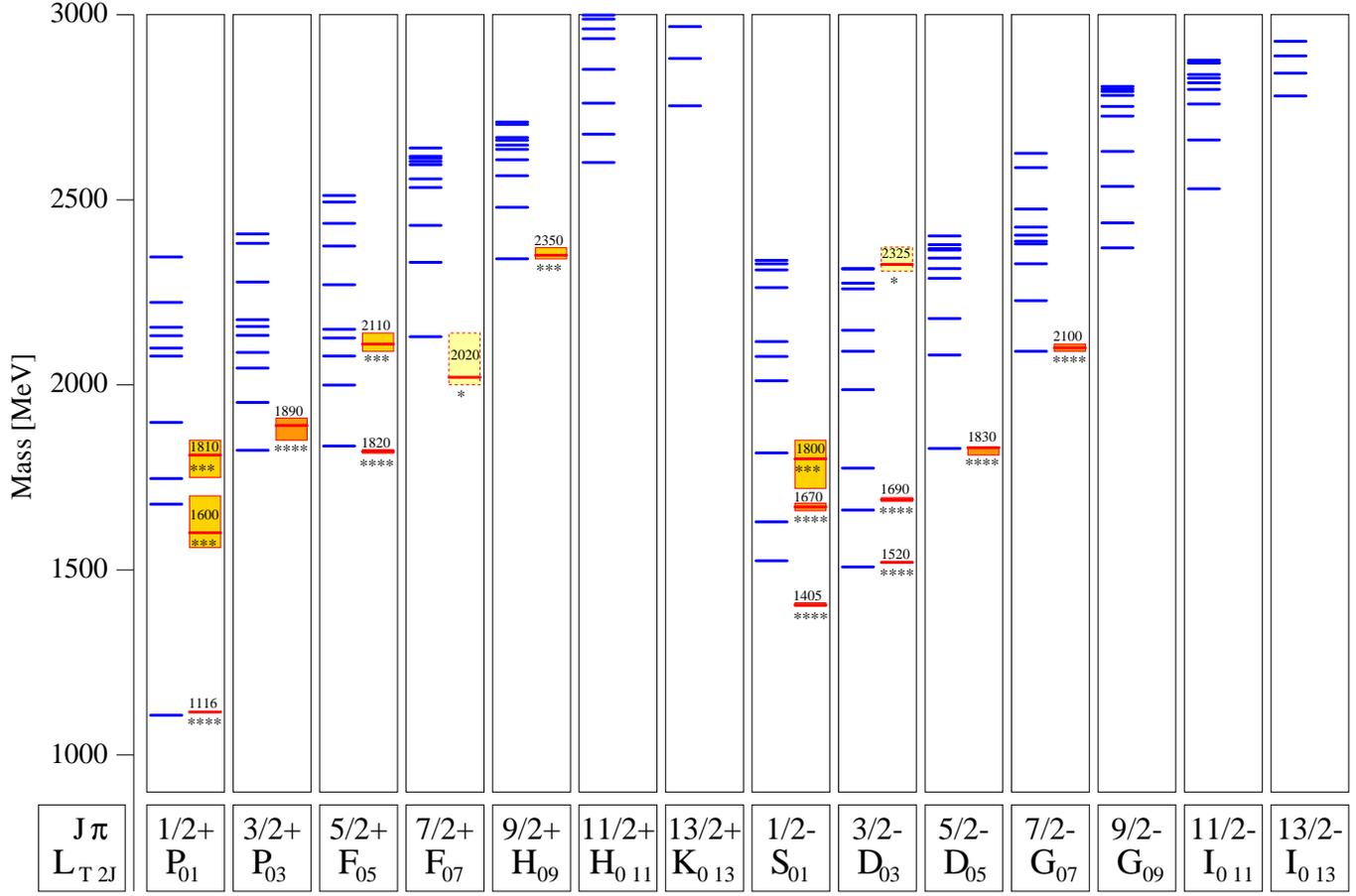},width=180mm}
  \end{center}
\caption{The calculated positive and negative parity
\textbf{$\Lambda$-resonance spectrum} with isospin $T=0$ and
strangeness $S^* = -1$ in \textbf{model $\mathcal{A}$} (left part of each column)
in comparison to the experimental spectrum taken from Particle Data Group \cite{PDG00} (right part of each
column).  The resonances are classified by the total spin
$J$ and parity $\pi$. The experimental resonance position is indicated
by a bar, the corresponding uncertainty by the shaded box, which is
darker for better established resonances; the status of each
resonance is additionally indicated by stars.}
\label{fig:LamM2}
\end{figure}

\begin{figure}[!h]
  \begin{center}
    \epsfig{file={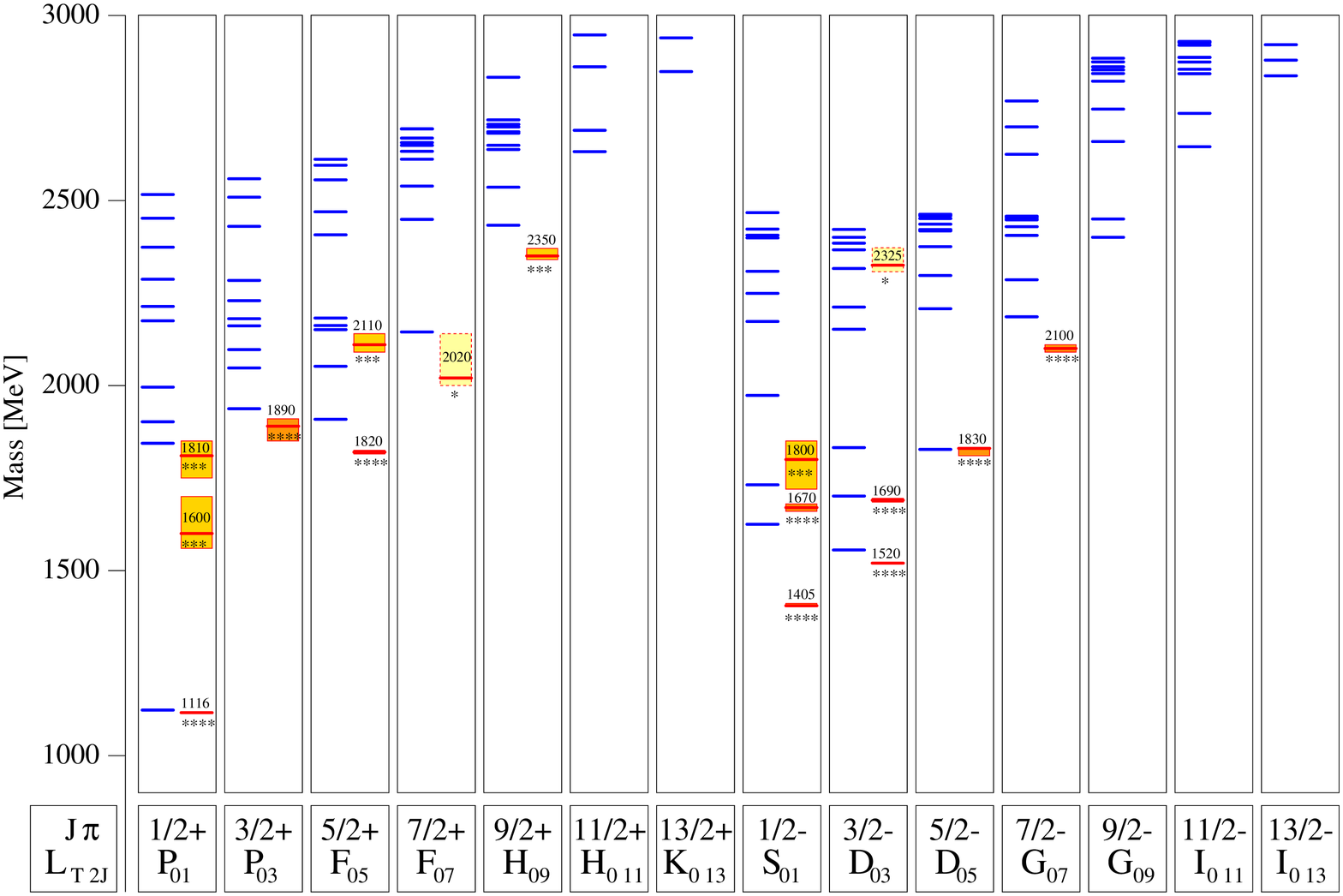},width=180mm}
  \end{center}
\caption{The calculated positive and negative parity
\textbf{$\Lambda$-resonance spectrum} with isospin $T=0$ and
strangeness $S^* = -1$ in \textbf{model $\mathcal{B}$} (left part of each column)
in comparison to the experimental spectrum taken from Particle Data Group \cite{PDG00} (right part of each
column).  The resonances are classified by the total spin
$J$ and parity $\pi$. See also caption to fig. \ref{fig:LamM2}.}
\label{fig:LamM1}
\end{figure}

Our predictions for the $\Lambda$-spectrum in both model variants
$\mathcal{A}$ and $\mathcal{B}$ are depicted in figs. \ref{fig:LamM2}
and \ref{fig:LamM1}, respectively. The resonances in each column are
classified by their total spin $J$ and parity $\pi$. For both parities
the predictions are shown up to $J=\frac{13}{2}$. For each sector
$[\Lambda\;J^\pi]$ at most ten radially excited states are displayed
on the left hand side of the column.  In comparison, the currently
experimentally known $\Lambda$-resonances \cite{PDG00} are shown on
the right hand side in each column. The corresponding uncertainties in
the measured resonance positions are indicated by the shaded
areas. The status of each resonance is denoted by the corresponding
number of stars following the notation of the Particle Data Group
\cite{PDG00} and moreover, by the shading of the error box which is
darker for better established resonances.  The following
discussion is organized according to a separate investigation of each
shell. The predicted masses for the states of each shell in comparison
with corresponding resonance positions measured experimentally are
explicitly given in the tables \ref{tab:L2hwband}, \ref{tab:L1hwband},
\ref{tab:L3hwband} and \ref{tab:L4hwband}, respectively.

\subsubsection{A first glimpse of the resulting spectra--comparing model $\mathcal{A}$ and model $\mathcal{B}$}
Comparing globally the predictions of model $\mathcal{A}$ and model
$\mathcal{B}$ in figs. \ref{fig:LamM2} and \ref{fig:LamM1}, respectively, we
again observe that model $\mathcal{A}$ leads to consistently better agreement
with experiment than model $\mathcal{B}$: As in the non-strange nucleon
sector, model $\mathcal{A}$ can excellently explain the positions of several
well-established four- and three-star resonances of the complete
$\Lambda$-spectrum presently known (most of them belonging to the $1\hbar\omega$
and $2\hbar\omega$ shell): Since this is in fact a {\it parameter-free
  prediction}, the excellent quantitative agreement is very remarkable and
strongly supports the credibility and predictive power of our model
$\mathcal{A}$. 

The only feature which model $\mathcal{A}$ unfortunately
cannot account for is the puzzling low position of
$\Lambda\frac{1}{2}^-(1405,\mbox{****})$.  Hence, the notorious difficulty of
{\it all} previous constituent quark models in explaining the position of
$\Lambda\frac{1}{2}^-(1405,\mbox{****})$ remains unsolved also in our fully
relativistic approach which uses instanton-induced, flavor-dependent forces.
In view of the otherwise excellent results, this shortcoming strongly
indicates that something in the present dynamics is missing which must be very
specific to that single state. We will come back to this question
during the following more detailed discussion.

As one would already anticipate from our discussion of the nucleon
spectrum \cite{Loe01b}, the most distinct deviations between model $\mathcal{A}$ and
$\mathcal{B}$ again show up in the sectors with total spin $J=\frac{1}{2}$.
In particular, model $\mathcal{B}$ once again strongly fails in describing the
striking low position of the first scalar/isoscalar excitation in the
$\frac{1}{2}^+$--sector, which in this flavor sector is the counterpart
$\Lambda\frac{1}{2}^+(1600,\mbox{***})$ of the Roper resonance.
Furthermore, several positions of higher mass states (in the $2\hbar\omega$
shell and beyond) are generally predicted
too high in model $\mathcal{B}$.
This result once more confirms that model $\mathcal{A}$ is more realistic
and thus the favored model for describing light baryons. For this reason, the
following detailed comparison of our predictions with experiment will
henceforth mainly focus to the results of the more successful model
$\mathcal{A}$.  Now let us discuss and investigate in detail the hyperfine
structures in each shell. We start with the predictions in the positive parity
$2\hbar\omega$ shell.
\subsubsection{States of the positive-parity 2${\bfgrk\hbar}{\bfgrk\omega}$ band}
The $2\hbar\omega$ band includes states with spin $J^\pi =
\frac{1}{2}^+$, $\frac{3}{2}^+$, $\frac{5}{2}^+$ and
$\frac{7}{2}^+$. The  positions predicted for these states (in both
models) are summarized in table \ref{tab:L2hwband} together with the
assignment to the observed states according to a comparison of the
predicted and measured masses.  
\begin{table}[!h]
\center
\begin{tabular}{ccccccc}
\hline
Exp. state   &PW&${J^\pi}$        & Rating     & Mass range [MeV]& Model state &  Model state\\
\cite{PDG00} &  &                      &            & \cite{PDG00}    & in model $\mathcal{A}$& in model $\mathcal{B}$  \\
\hline
$\Lambda(1600)$&$P_{01}$&${\frac{1}{2}^+}$& ***     &1560-1700  &$\MSL(1,+,2,1677)$   &$\MSL(1,+,2,1844)$ \\
$\Lambda(1810)$&$P_{01}$&${\frac{1}{2}^+}$& ***     &1750-1850  &$\MSL(1,+,3,1747)$   &$\MSL(1,+,3,1902)$ \\[3mm]
               &        &                 &         &           &$\begin{array}{c}
                                                                  \MSL(1,+,4,1898)\\
                                                                  \MSL(1,+,5,2077)\\
                                                                  \MSL(1,+,6,2099)\\
                                                                  \MSL(1,+,7,2132)\\
                                                                  \end{array}$&
                                                                  $\begin{array}{c}
                                                                  \MSL(1,+,4,1996)\\
                                                                  \MSL(1,+,5,2175)\\
                                                                  \MSL(1,+,6,2214)\\
                                                                  \MSL(1,+,7,2287)\\
                                                                  \end{array}$\\
\hline
$\Lambda(1890)$&$P_{03}$&${\frac{3}{2}^+}$& ****     &1850-1910  &$\MSL(3,+,1,1823)$   &$\MSL(3,+,1,1937)$ \\[3mm]
               &        &                 &         &           &$\begin{array}{c}
                                                                  \MSL(3,+,2,1952)\\
                                                                  \MSL(3,+,3,2045)\\
                                                                  \MSL(3,+,4,2087)\\
                                                                  \MSL(3,+,5,2133)\\
                                                                  \MSL(3,+,6,2157)\\
                                                                  \MSL(3,+,7,2167)\\
                                                                  \end{array}$&
                                                                   $\begin{array}{c}
                                                                  \MSL(3,+,2,2047)\\
                                                                  \MSL(3,+,3,2097)\\
                                                                  \MSL(3,+,4,2161)\\
                                                                  \MSL(3,+,5,2180)\\
                                                                  \MSL(3,+,6,2229)\\
                                                                  \MSL(3,+,7,2285)\\
                                                                  \end{array}$\\
\hline
$\Lambda(1820)$&$F_{05}$&${\frac{5}{2}^+}$& ****     &1815-1825  &$\MSL(5,+,1,1834)$   &$\MSL(5,+,1,1909)$ \\[3mm]
               &        &                 &          &           &$\MSL(5,+,2,1999)$   &$\MSL(5,+,2,2052)$ \\
$\Lambda(2110)$&$F_{05}$&${\frac{5}{2}^+}$& ***     &2090-2140  &$\begin{array}{c}
                                                                  \MSL(5,+,3,2078)\\
                                                                  \MSL(5,+,4,2127)\\
                                                                  \MSL(5,+,5,2150)\\
                                                                  \end{array}$&
                                                                  $\begin{array}{c}
                                                                  \MSL(5,+,3,2151)\\
                                                                  \MSL(5,+,4,2162)\\
                                                                  \MSL(5,+,5,2182)\\
                                                                  \end{array}$\\
\hline
$\Lambda(2020)$&$F_{07}$&${\frac{7}{2}^+}$& *       &2000-2140  &$\MSL(7,+,1,2130)$   &$\MSL(7,+,1,2145)$ \\
\hline
\end{tabular}
\caption{Calculated positions of all $\Lambda$-states assigned to the positive parity $2\hbar\omega$ shell
  in comparison to the corresponding experimental mass values taken from
  \cite{PDG00}. PW denotes the partial wave and the rating is
         given according to the PDG classification \cite{PDG00}. Here and throughout this work we use the notation
         $[B\;J^\pi]_n({M})$ for the predicted model states in model $\mathcal{A}$ and
         $\mathcal{B}$, respectively, where $B$ denotes the baryon ({\it i.e.} the
         flavor), $J^\pi$ are spin and parity, and $M$ is the predicted mass
         given in MeV. 
         $n = 1, 2, 3, \ldots$ is the principal quantum
         number counting the states in each sector $J^\pi$ beginning with the
         lowest state.}
\label{tab:L2hwband}
\end{table}
The number of predicted $2\hbar\omega$ states is even larger than in the
nucleon spectrum owing to the additional flavor singlet states.
On the other hand there are considerably fewer states observed experimentally,
especially in the upper part of this band (beyond 2 GeV), where so far only
two resonances have been seen in the $\Lambda\frac{5}{2}^+$ and
$\Lambda\frac{7}{2}^+$ sectors.  Hence a quite large number of
$\Lambda$-resonances in this mass region is expected to be ``missing'', {\it
  i.e.} hitherto these states have not been seen in multichannel phase-shift
analysis of $\bar K N$ scattering data. In this respect, the study of strong two-body
decay amplitudes of baryons within our covariant Bethe-Salpeter
framework would be very favorable again.  This should offer the possibility to
explain that model states, assigned to observed resonances (according to their
position), strongly couple to the $\bar K N$ channel, whereas the others do
not (or only weakly couple to $\bar K N$) and thus escape from observation
\cite{CaRo00}.  Since these investigations are still in progress, here the
assignment of model states to experimental states has again to be made on the
basis of the masses alone.  To distinguish between dominantly flavor octet and
flavor singlet states we additionally tabulated for each state of model
$\mathcal{A}$ the corresponding spin-flavor $SU(6)$ contributions in table
\ref{tab:conf_mix_Lam_2hw}. This information is useful for identifying those
structures, that have their direct counterparts in the nucleon spectrum.

\begin{figure}[!h]
  \begin{center}
    \epsfig{file={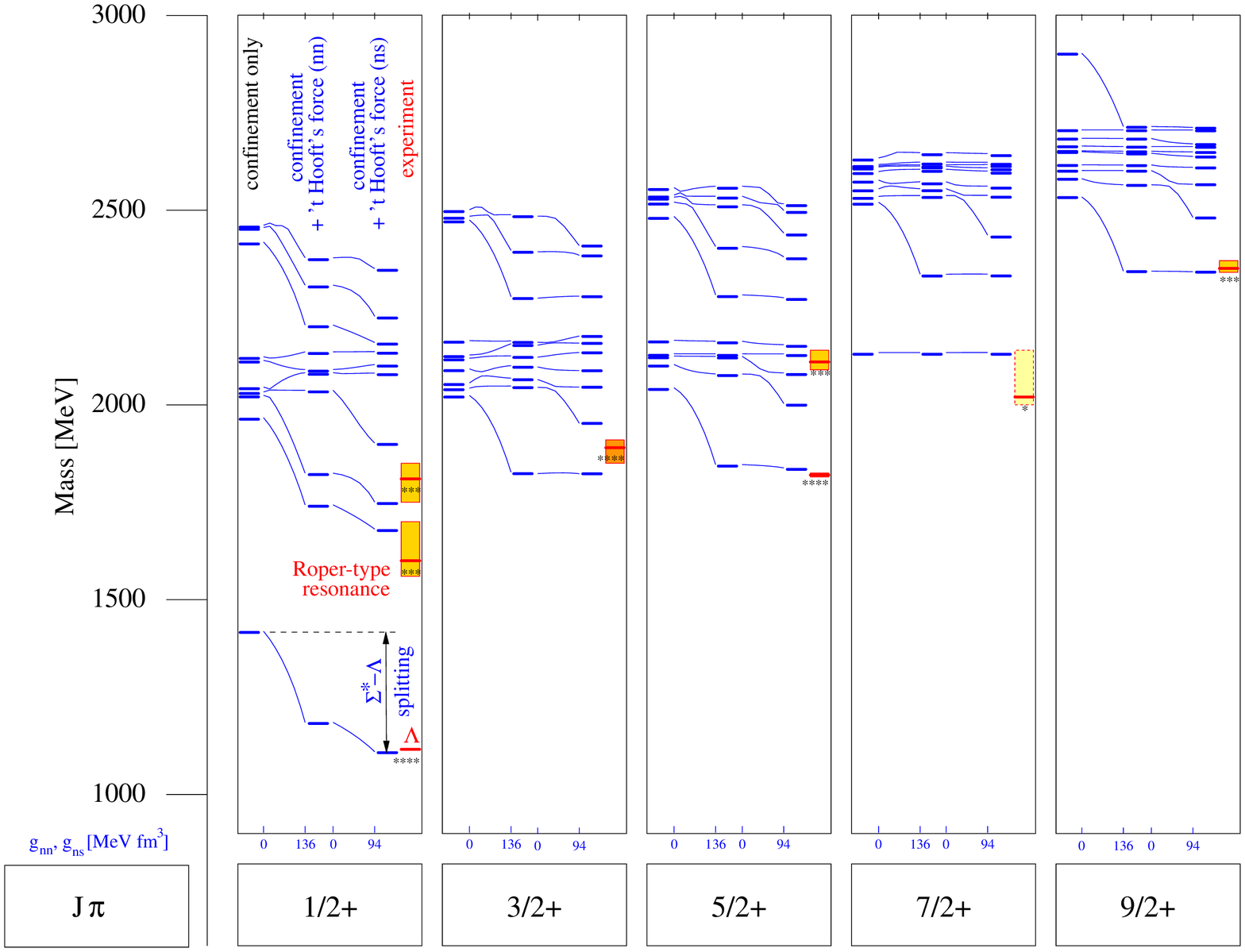},width=160mm}
  \end{center}
\caption{Influence of the instanton-induced interaction on the energy levels
  of the positive-parity $\Lambda$-states in model $\mathcal{A}$. The effects
  of non-strange (nn) and non-strange-strange (ns) diquark correlations are
  shown separately: In each column the leftmost spectrum shows the result with
  confinement only. The curves illustrate the variation of the spectrum with
  increasing 't~Hooft couplings $g_{nn}$ and $g_{ns}$, respectively. The
  middle spectrum shows first the result with 't~Hooft's force acting only for
  non-strange quark pairs ($g_{nn}=136$ MeV fm$^3$, $g_{ns}=0$ MeV fm$^3$). Finally, the
  right spectrum shows the prediction with 't~Hooft's force acting for both,
  non-strange and non-strange-strange quark pairs ($g_{nn}=136$ MeV fm$^3$,
  $g_{ns}=94$ MeV fm$^3$). For comparison the rightmost spectrum shows the
  experimental data with the corresponding uncertainties.}
\label{fig:Model2L+gvar}
\end{figure}

\begin{table}[!h]
\begin{center}
\begin{tabular}[h]{|c||c||c||cccc||cc|}
\hline
$J^\pi$&
Model state & 
pos. & 
$\!\!{}^2 8[56]\!\!$&
$\!\!{}^2 8[70]\!\!$&
$\!\!{}^4 8[70]\!\!$&
$\!\!{}^2 8[20]\!\!$&
$\!\!{}^2 1[70]\!\!$&
$\!\!{}^4 1[20]\!\!$\\[1mm]
&
in model $\mathcal{A}$& 
neg. & 
$\!\!{}^2 8[56]\!\!$&
$\!\!{}^2 8[70]\!\!$&
$\!\!{}^4 8[70]\!\!$&
$\!\!{}^2 8[20]\!\!$&
$\!\!{}^2 1[70]\!\!$&
$\!\!{}^4 1[20]\!\!$\\
\hline
\hline
$\frac{1}{2}^+$    &$\MSL(1,+,1,1108)$ &   98.6 & {\bf \underline{    94.3}} &      3.9 &      0.0 &      0.0 &      0.3 &      0.0\\
    & $\Lambda$ ground-state      &    1.4 &      0.4 &      0.6 &      0.3 &      0.0 &      0.0 &      0.0\\
\hline
\hline
    & $\MSL(1,+,2,1677)$ &   98.6 & {\bf \underline{    88.4}} &      6.2 &      0.1 &      0.2 &      3.7 &      0.1\\
    &       &    1.4 &      0.3 &      0.6 &      0.4 &      0.0 &      0.1 &      0.0\\
\cline{2-9}
    & $\MSL(1,+,3,1747)$ &   98.9 &      5.1 &      2.1 &      0.0 &      0.1 & {\bf \underline{    90.6}} &      0.9\\
    &       &    1.1 &      0.0 &      0.0 &      0.0 &      0.0 &      1.0 &      0.1\\
\cline{2-9}
    & $\MSL(1,+,4,1898)$ &   99.1 &      9.1 & {\bf \underline{    84.2}} &      1.0 &      0.8 &      3.8 &      0.2\\
$\frac{1}{2}^+$    &       &    0.9 &      0.2 &      0.3 &      0.3 &      0.1 &      0.1 &      0.0\\
\cline{2-9}
    & $\MSL(1,+,5,2077)$ &   99.0 &      0.5 &      1.2 & {\bf \underline{    85.8}} &     11.2 &      0.1 &      0.3\\
    &       &    1.0 &      0.0 &      0.1 &      0.8 &      0.0 &      0.0 &      0.0\\
\cline{2-9}
    & $\MSL(1,+,6,2099)$ &   98.9 &      1.1 &      0.6 &      1.5 &     11.7 &      0.9 & {\bf \underline{    83.0}}\\
    &       &    1.1 &      0.0 &      0.1 &      0.1 &      0.0 &      0.8 &      0.1\\
\cline{2-9}
    & $\MSL(1,+,7,2132)$ &   98.9 &      2.2 &      1.6 &     11.2 & {\bf \underline{    69.8}} &      0.6 &     13.5\\
    &       &    1.1 &      0.0 &      0.3 &      0.4 &      0.2 &      0.1 &      0.0\\
\hline
\hline
    & $\MSL(3,+,1,1823)$ &   98.5 & {\bf \underline{    60.0}} &     28.2 &      0.3 &      0.1 &      9.9 &      0.1\\
    &       &    1.5 &      0.4 &      0.3 &      0.6 &      0.0 &      0.2 &      0.0\\
\cline{2-9}
    & $\MSL(3,+,2,1952)$ &   98.4 &      3.8 &      7.6 &      0.8 &      0.1 & {\bf \underline{    84.0}} &      2.2\\
    &       &    1.6 &      0.0 &      0.0 &      0.1 &      0.0 &      1.4 &      0.1\\
\cline{2-9}
    & $\MSL(3,+,3,2045)$ &   99.2 &      0.5 &      0.2 & {\bf \underline{    96.9}} &      1.1 &      0.3 &      0.2\\
    &       &    0.8 &      0.1 &      0.2 &      0.5 &      0.0 &      0.0 &      0.0\\
\cline{2-9}
$\frac{3}{2}^+$ & $\MSL(3,+,4,2087)$ &   99.0 &      1.3 &      1.6 & {\bf \underline{    84.0}} &     11.2 &      0.6 &      0.3\\
    &       &    1.0 &      0.0 &      0.4 &      0.6 &      0.0 &      0.0 &      0.0\\
\cline{2-9}
    & $\MSL(3,+,5,2133)$ &   99.1 &     25.2 & {\bf \underline{    56.2}} &      7.2 &      9.1 &      1.2 &      0.2\\
    &       &    0.9 &      0.1 &      0.4 &      0.4 &      0.0 &      0.0 &      0.0\\
\cline{2-9}
    & $\MSL(3,+,6,2157)$ &   99.0 &      5.1 &      8.0 &      8.5 & {\bf \underline{    70.8}} &      1.1 &      5.5\\
    &       &    1.0 &      0.0 &      0.3 &      0.4 &      0.1 &      0.0 &      0.1\\
\cline{2-9}
    & $\MSL(3,+,7,2176)$ &   99.0 &      0.3 &      0.4 &      0.7 &      5.8 &      4.2 & {\bf \underline{    87.5}}\\
    &       &    1.0 &      0.0 &      0.0 &      0.0 &      0.0 &      0.1 &      0.8\\
\hline
\hline
    & $\MSL(5,+,1,1834)$ &   98.5 & {\bf \underline{    57.8}} &     28.3 &      0.2 &      0.1 &     12.1 &      0.0\\
    &       &    1.5 &      0.4 &      0.4 &      0.4 &      0.1 &      0.1 &      0.1\\
\cline{2-9}
    & $\MSL(5,+,2,1999)$ &   98.7 &      4.5 &      8.9 &      1.0 &      0.1 & {\bf \underline{    84.1}} &      0.2\\
    &       &    1.3 &      0.1 &      0.1 &      0.1 &      0.0 &      0.6 &      0.5\\
\cline{2-9}
$\frac{5}{2}^+$    & $\MSL(5,+,3,2078)$ &   99.0 &      9.0 &      9.9 & {\bf \underline{    77.1}} &      0.0 &      2.0 &      0.9\\
    &       &    1.0 &      0.4 &      0.3 &      0.3 &      0.0 &      0.0 &      0.0\\
\cline{2-9}
    & $\MSL(5,+,4,2127)$ &   99.0 &     20.9 & {\bf \underline{    45.9}} &     12.9 &      0.0 &      0.6 &     18.7\\
    &       &    1.0 &      0.1 &      0.2 &      0.1 &      0.3 &      0.1 &      0.1\\
\cline{2-9}
    & $\MSL(5,+,5,2150)$ &   99.0 &      4.6 &      7.6 &      7.8 &      0.0 &      0.7 & {\bf \underline{    78.3}}\\
    &       &    1.0 &      0.0 &      0.1 &      0.0 &      0.1 &      0.5 &      0.4\\
\hline
\hline
$\frac{7}{2}^+$    & $\MSL(7,+,1,2130)$ &   99.2 &      0.0 &      0.0 & {\bf \underline{    99.1}} &      0.0 &      0.0 &      0.1\\
    &       &    0.8 &      0.1 &      0.1 &      0.6 &      0.1 &      0.0 &      0.0\\
\hline
\end{tabular}
\end{center}
\caption{Configuration mixing of positive-parity $\Lambda$ states in model $\mathcal{A}$
assigned to the  $2\hbar\omega$ band.}
\label{tab:conf_mix_Lam_2hw}
\end{table}
Figure \ref{fig:LamM2} shows that
the agreement between the predictions of model $\mathcal{A}$ and the few
presently known resonances of the $2\hbar\omega$ shell is of similar
good quality as that of the corresponding nucleon states. Due to
't~Hooft's force, we again observe in the sectors
$J^\pi=\frac{1}{2}^+$, $\frac{3}{2}^+$ and $\frac{5}{2}^+$ a selective
lowering of particular states with respect to the majority of
states grouped between 2000 and 2200 MeV, which essentially remain
unaffected by 't~Hooft's interaction.  
The centroid of the bulk of unaffected states 
fairly agrees with the positions of the $\Lambda\frac{5}{2}^+(2110,\mbox{***})$ and the
$\Lambda\frac{7}{2}^+(2020,\mbox{*})$ observed so far in the upper energy range.
With the couplings $g_{nn}$
and $g_{ns}$ fixed from the ground-state hyperfine pattern, four of
the separated states are lowered even deeply enough to fit rather
well into the conspicuous low-lying pattern formed by the four
experimentally observed resonances
$\Lambda\frac{1}{2}^+(1600,\mbox{***})$,
$\Lambda\frac{1}{2}^+(1810,\mbox{***})$,
$\Lambda\frac{3}{2}^+(1890,\mbox{****})$ and
$\Lambda\frac{5}{2}^+(1820,\mbox{****})$. Their structure is very
similar to that found in the nucleon spectrum (see fig. \ref{fig:NucLamComp}).
Figure \ref{fig:Model2L+gvar} demonstrates how in model $\mathcal{A}$ the
instanton-induced 't~Hooft interaction shapes this pattern along with the
right position of the ground-state $\Lambda\frac{1}{2}^+(1116,\mbox{****})$.

The figure illustrates the influence of 't~Hooft's force on the energy levels
of all positive-parity $\Lambda$-states, where the effects of non-strange (nn)
and non-strange-strange (ns) diquark correlations are shown separately\footnote{
The different behavior of the states under influence of the
two distinct (nn) and (ns) parts of 't~Hooft's force then reflects
the (scalar) diquark content of both flavor types within the
states.}. Starting from the case with confinement only
($g_{nn}=g_{ns}=0$) which is shown in the leftmost spectrum of each column,
first the non-strange coupling $g_{nn}$ is gradually increased up to
its value $g_{nn}=136$ MeV fm$^3$ fixed from the $\Delta-N$-splitting
while the non-strange-strange coupling still is kept at
$g_{ns}=0$ MeV fm$^3$. Then, the non-strange-strange coupling
$g_{ns}$ is increased until the $\Lambda$-ground-state together with
the other octet hyperons $\Sigma$ and $\Xi$ (here compare to ref. \cite{Loe01b}) is correctly reproduced ($g_{nn}=136$ MeV
fm$^3$, $g_{ns}=94$ MeV fm$^3$). The spectrum predicted finally in
model $\mathcal{A}$ is shown on the right hand side of each column in
comparison to the experimental resonance positions.  Indeed, it turns
out that, once the couplings are fixed to reproduce the hyperfine
pattern of ground-states, the positions of the four comparatively
low-lying states $\Lambda\frac{1}{2}^+(1600,\mbox{***})$,
$\Lambda\frac{1}{2}^+(1810,\mbox{***})$,
$\Lambda\frac{3}{2}^+(1890,\mbox{****})$ and
$\Lambda\frac{5}{2}^+(1820,\mbox{****})$ are  {\it simultaneously} well described.  
Apart from the distinct action of 't~Hooft's force
in the non-strange and non-strange-strange diquark channels the
systematics observed here is quite similar to that found for the
corresponding $2\hbar\omega$ nucleon states. For a more detailed
discussion let us investigate the situation for each spin sector from
$\frac{1}{2}^+$ to $\frac{7}{2}^+$ in turn:\\

In the ${\bf\Lambda\frac{7}{2}^+}$ \textbf{sector} with maximal possible spin
$J_{\rm max}(N) = N +\frac{3}{2}$ in the $N=2$ oscillator shell, our model
predicts a single well isolated state in the upper part of the shell, which
coincides with the only resonance $\Lambda\frac{7}{2}^+(2020,\mbox{*})$ 'seen'
in this mass range of the $F_{07}$ partial wave. The mass predicted at 2130
MeV lies within the quite large range of possible values of this poorly
determined one-star resonance.  To achieve this maximal total spin, this state
contains a symmetric spin-quartet wave function.  Consequently, this state,
{\it i.e.}  its mass as well as its Salpeter amplitude remains totally
unaffected by 't~Hooft's force as confirmed in fig. \ref{fig:Model2L+gvar}.
In fact, this state shows an almost pure $^4 8[70]$ configuration ($> 97\%$,
see table \ref{tab:conf_mix_Lam_2hw}) and thus is the octet partner of the
resonance $N\frac{7}{2}^+(1990,\mbox{**})$ in the corresponding
$N\frac{7}{2}^+$ sector of the nucleon spectrum. 

In the ${\bf\Lambda\frac{5}{2}^+}$ {\bf sector} model $\mathcal{A}$ predicts
altogether five states. As in the corresponding nucleon sector, the dominantly
$^2 8[56]$ state shows the largest downward mass shift of roughly 200 MeV
relative to its position in the pure confinement spectrum. Note that this
lowering originates mainly from an attractive correlation in the scalar {\it
  non-strange} (nn) diquark channel, whereas the contribution from the
non-strange-strange (ns) diquark correlation is almost negligible.  With the
couplings $g_{nn}$ and $g_{ns}$ fixed from the ground-state hyperfine pattern
the predicted mass of this lowest $\Lambda\frac{5}{2}^+$ excitation at 1834
MeV is in nice agreement with the lowest observed four-star resonance
$\Lambda\frac{5}{2}^+(1820,\mbox{****})$.  The admixture of flavor singlet
contributions amounts to only $\sim 12\%$.  Hence, this low-lying, dominantly
$^2 8[56]$ state may be considered as the octet counterpart of the
$N\frac{5}{2}^+(1720,\mbox{****})$. The second excited state is predicted at
1999 MeV. It exhibits a dominant $^2 1[70]$ configuration ($~85\%$) and
't~Hooft's force induces a moderate downward mass shift of roughly 120 MeV,
where now this lowering originates mainly due to an attractive
non-strange-strange (ns) scalar diquark correlation.  This dominantly flavor
singlet state has no counterpart in the $N^*$ spectrum and a corresponding
resonance observed experimentally in this region is still ''missing'' although
its position lies rather isolated between the
$\Lambda\frac{5}{2}^+(1820,\mbox{****})$ and the three remaining states
predicted in this sector.  The three other states are hardly influenced by
't~Hooft's force and are all predicted to lie in the narrow range between
$\sim 2080$ and $\sim 2150$ MeV around the three-star resonance
$\Lambda\frac{5}{2}^+(2110,\mbox{***})$.

In the ${\bf\Lambda\frac{3}{2}^+}$ {\bf sector} our model predicts seven
$2\hbar\omega$ states. Here the situation is quite similar to the
$\Lambda\frac{5}{2}^+$ sector.  Again, the dominantly $^2 8[56]$ state is
lowered by an equally big downward shift of roughly 200 MeV. As in the
$\Lambda\frac{5}{2}^+$ sector, this state apparently has no scalar
non-strange-strange (ns) diquark contributions and the whole mass shift turns
out to be due to an attractive non-strange (nn) diquark correlation alone.
With the 't~Hooft couplings fixed on the ground-state spectrum, the lowest
excited state is then predicted at 1823 MeV and thus can be uniquely
identified with the well-established low-lying resonance
$\Lambda\frac{3}{2}^+(1890,\mbox{****})$ observed experimentally in the
$P_{03}$ partial wave. Once again, the flavor singlet admixtures of roughly
$10\%$ are moderate, and thus the $\Lambda\frac{3}{2}^+(1890,\mbox{****})$ is
the octet partner of the $N\frac{3}{2}^+(1720,\mbox{****})$. Unfortunately,
apart from the low-lying $\Lambda\frac{3}{2}^+(1890,\mbox{****})$, no further
resonance has been seen in the $P_{03}$ partial wave so far.  Similar to the
$\Lambda\frac{5}{2}^+$ sector the second state predicted at 1952 MeV again
reveals a dominantly flavor singlet $^2 1[70]$ configuration ($\sim85\%$).
Compared to the dominantly $^2 8[56]$ state, it exhibits again a rather
moderate downward mass shift of roughly 130 MeV due to a dominantly
non-strange-strange (ns) scalar diquark correlation. Hence this state, which
has no counterpart in the nucleon spectrum, comes to lie fairly in between the
lowest excitation and the five remaining states in this sector, whose
positions are virtually not affected by 't~Hooft's force and are predicted to
lie in the range between $\sim 2045$ and $\sim 2180$ MeV.

Finally let us focus on the ${\bf\Lambda\frac{1}{2}^+}$ {\bf sector} with the
scalar/isoscalar excitations of the $\Lambda$-ground-state. Here we expect six
excited $2\hbar\omega$ states. Again, the lowest state predicted at 1677 MeV
turns out to be dominantly $^2 8[56]$ ($\sim 89\%$). Due to 't~Hooft's force
this state is strongly lowered by almost the same amount (roughly 300 MeV) as
the ground-state. In fact, this downward mass shift is even large enough to
account for the striking low position of
$\Lambda\frac{1}{2}^+(1600,\mbox{***})$. This state behaves quite similar to
the corresponding lowest state in the nucleon sector which is assigned to the
Roper resonance $N\frac{1}{2}^+(1440,\mbox{****})$ and hence it may readily be
associated with the octet partner of the Roper resonance.  In the nucleon
sector we found \cite{Loe01b} the Roper state in model $\mathcal{A}$ showing a very similar
behavior under the influence of 't~Hooft's instanton-induced interaction like
the ground-state. Here we make a similar observation: Both, the
$\Lambda$-ground-state and the Roper-type state, show almost the same
contributions of non-strange and non-strange-strange scalar diquarks.
Consequently, in both cases, the equally large energy shift originates mainly
from the non-strange and partly from the non-strange-strange diquark
correlations (here compare to the discussion of the ground-state spectrum in
ref. \cite{Loe01b}).  It is interesting to note that exactly the
same behavior is also observed for the second radially excited state predicted
at 1747 MeV, which lies at the lower end of the uncertainty range quoted for
the resonance $\Lambda\frac{1}{2}^+(1810,\mbox{***})$. As the Roper-type
state, this resonance likewise reveals the almost same strong downward mass
shift of about 300 MeV. Hence, the calculated mass splitting of about 70 MeV
between the first two radial excitations in $\Lambda\frac{1}{2}^+$ is
primarily a relativistic spin-orbit effect of the confinement force.  Unlike
the nucleon $N\frac{1}{2}^+$ sector, this second excitation predicted is {\it
  not} the dominantly flavor octet $^2 8[70]$ state, but a dominantly flavor
singlet $^2 1[70]$ state ($\sim 92\%$).  The dominantly flavor octet $^2
8[70]$ state, however, appears here as the third radial excitation predicted
at 1898 MeV.  Consequently, we would identify the
$\Lambda\frac{1}{2}^+(1810,\mbox{***})$ as being a dominantly flavor singlet
state, hence {\it not} being the octet counterpart of the
$N\frac{1}{2}^+(1710,\mbox{***})$ in the corresponding $N\frac{1}{2}^+$
sector.  However, our assignment is less clear, since the reported central
value of 1710 MeV of the second observed excitation lies fairly in between our
two model predictions at 1747 and 1898 MeV. Under circumstances the
$\Lambda\frac{1}{2}^+(1810,\mbox{***})$ might be resolved into two separate
states by future experiments.

\subsubsection{States of the negative-parity 1${\bfgrk\hbar}{\bfgrk\omega}$ band}
\begin{figure}[!h]
  \begin{center}
    \epsfig{file={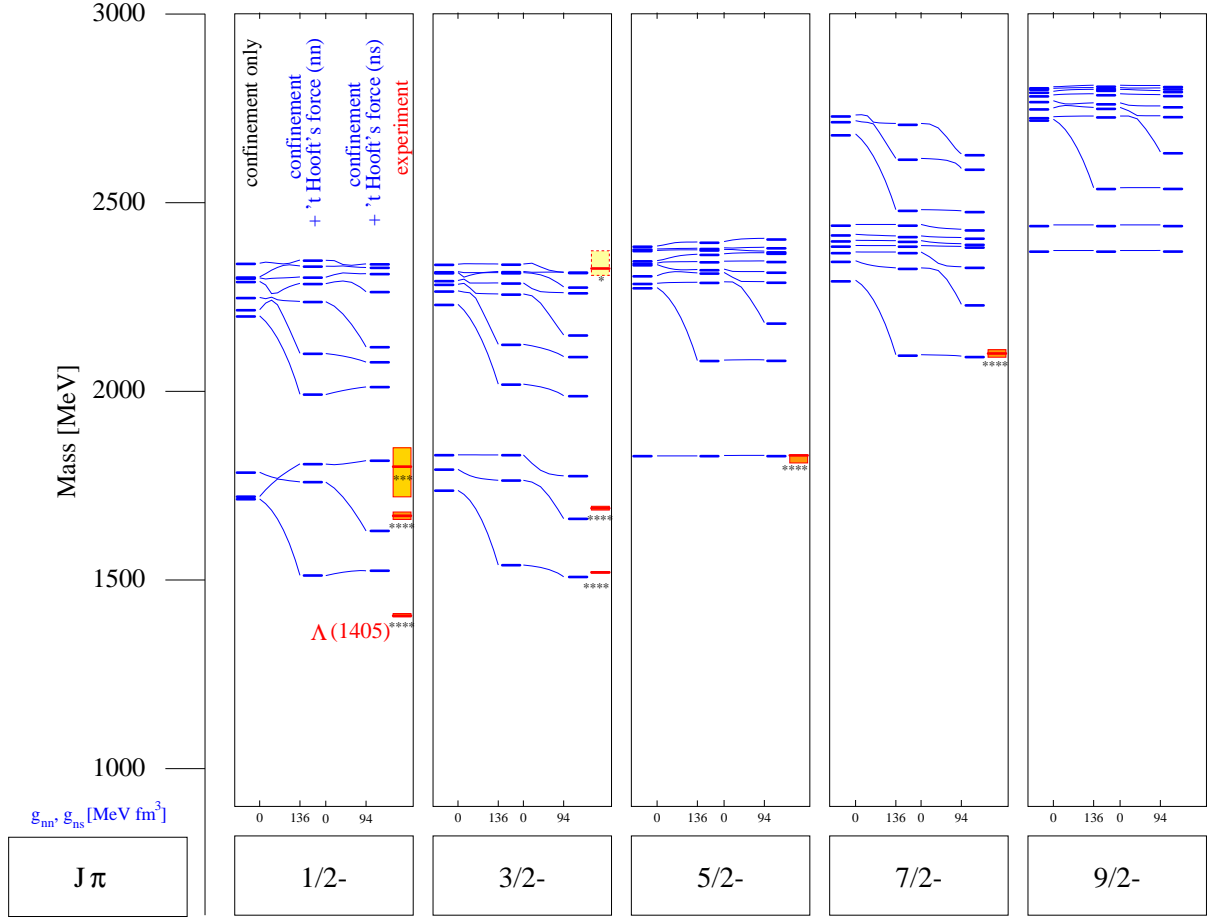},width=160mm}
  \end{center}
\caption{Influence of the instanton-induced interaction on the energy levels
  of the negative-parity $\Lambda$-states in model $\mathcal{A}$. The effects
  of non-strange (nn) and non-strange-strange (ns) diquark correlations are
  shown separately. See also caption of fig. \ref{fig:Model2L+gvar} and text for further explanations.}
\label{fig:Model2L-gvar}
\end{figure}

We now turn to the discussion of  negative-parity $\Lambda$-states in the $1\hbar\omega$ shell. 
As usual in constituent quark models for baryons, our models predict
altogether seven states with spins $J^\pi=\frac{1}{2}^-$, $\frac{3}{2}^-$ and
$\frac{5}{2}^-$ in this shell. In the baryon summary table of the Particle
Data Group \cite{PDG00} six well-established four- and three-star
resonances with these spins below 2 GeV are listed. In model $\mathcal{A}$ the assignment
to these observed resonances due to a comparison of predicted and measured
masses is unambiguous and a clear identification of the states is thus
readily possible. Only one resonance of the $1\hbar\omega$ shell has not been observed
hitherto: the third radial excitation in the $\Lambda\frac{3}{2}^-$ is still
missing. 
\begin{table}[!h]
\center
\begin{tabular}{ccccccc}
\hline
Exp. state   &PW&${J^\pi}$        & Rating     & Mass range [MeV]& Model state &  Model state \\
\cite{PDG00} &  &                      &            & \cite{PDG00}    & in model $\mathcal{A}$& in model $\mathcal{B}$  \\
\hline
$\Lambda(1405)$&$S_{01}$&${\frac{1}{2}^-}$& ****     &1402-1411  &$\MSL(1,-,1,1524)$   & \\
$\Lambda(1670)$&$S_{01}$&${\frac{1}{2}^-}$& ****     &1660-1680  &$\MSL(1,-,2,1630)$   &$\MSL(1,-,1,1625)$\\
$\Lambda(1800)$&$S_{01}$&${\frac{1}{2}^-}$& ***      &1720-1850  &$\MSL(1,-,3,1816)$   &$\MSL(1,-,2,1732)$\\
              &        &                 &          &            &                     &$\MSL(1,-,3,1973)$\\
\hline
$\Lambda(1520)$&$D_{03}$&${\frac{3}{2}^-}$& ****     &1518-1522  &$\MSL(3,-,1,1508)$   &$\MSL(3,-,1,1555)$ \\
$\Lambda(1690)$&$D_{03}$&${\frac{3}{2}^-}$& ****     &1685-1695  &$\MSL(3,-,2,1662)$   &$\MSL(3,-,2,1701)$ \\
               &        &                 &          &           &$\MSL(3,-,3,1775)$   &$\MSL(3,-,3,1832)$ \\
\hline
$\Lambda(1830)$&$D_{05}$&${\frac{5}{2}^-}$& ****     &1810-1830  &$\MSL(5,-,1,1828)$   &$\MSL(5,-,1,1827)$ \\
\hline
\end{tabular}
\caption{Calculated positions of all $\Lambda$-states assigned to the negative parity $1\hbar\omega$ shell
  in comparison to the corresponding experimental mass values taken from \cite{PDG00}. Notation as in table \ref{tab:L2hwband}.}
\label{tab:L1hwband}
\end{table}
\begin{table}[!h]
\begin{center}
\begin{tabular}[h]{|c||c||c||cccc||cc|}
\hline
$J^\pi$&
Model state & 
pos. & 
$\!\!{}^2 8[56]\!\!$&
$\!\!{}^2 8[70]\!\!$&
$\!\!{}^4 8[70]\!\!$&
$\!\!{}^2 8[20]\!\!$&
$\!\!{}^2 1[70]\!\!$&
$\!\!{}^4 1[20]\!\!$\\[1mm]
&
in model $\mathcal{A}$& 
neg. & 
$\!\!{}^2 8[56]\!\!$&
$\!\!{}^2 8[70]\!\!$&
$\!\!{}^4 8[70]\!\!$&
$\!\!{}^2 8[20]\!\!$&
$\!\!{}^2 1[70]\!\!$&
$\!\!{}^4 1[20]\!\!$\\
\hline
\hline
     &$\MSL(1,-,1,1524)$ &   98.8 &      2.9 &     26.0 &      0.3 &      0.0 & {\bf \underline{    69.4}} &      0.2\\
     &      &    1.2 &      0.3 &      0.1 &      0.0 &      0.1 &      0.3 &      0.5\\
\cline{2-9}
$\frac{1}{2}^-$ &$\MSL(1,-,2,1630)$ &   98.7 &      5.5 & {\bf \underline{    61.6}} &      2.1 &      0.3 &     29.2 &      0.1\\
     &      &    1.3 &      0.6 &      0.2 &      0.0 &      0.2 &      0.1 &      0.2\\
\cline{2-9}
     &$\MSL(1,-,3,1816)$ &   99.4 &      0.2 &      3.1 & {\bf \underline{    94.9}} &      0.6 &      0.1 &      0.6\\
     &      &    0.6 &      0.0 &      0.0 &      0.5 &      0.1 &      0.0 &      0.0\\
\hline
\hline
     &$\MSL(3,-,1,1508)$ &   98.6 &      2.0 &     18.7 &      0.1 &      0.0 & {\bf \underline{    77.7}} &      0.1\\
     &      &    1.4 &      0.1 &      0.1 &      0.1 &      0.0 &      0.9 &      0.3\\
\cline{2-9}
$\frac{3}{2}^-$  &$\MSL(3,-,2,1662)$ &   98.9 &      4.4 & {\bf \underline{    72.0}} &      2.2 &      0.2 &     20.1 &      0.0\\
     &      &    1.1 &      0.1 &      0.2 &      0.3 &      0.1 &      0.2 &      0.0\\
\cline{2-9}
     &$\MSL(3,-,3,1775)$ &   99.3 &      0.8 &      1.5 & {\bf \underline{    96.1}} &      0.0 &      0.4 &      0.4\\
     &      &    0.7 &      0.1 &      0.2 &      0.3 &      0.1 &      0.0 &      0.0\\
\hline
\hline
$\frac{5}{2}^-$ &$\MSL(5,-,1,1828)$ &   99.4 &      0.0 &      0.0 & {\bf \underline{    99.0}} &      0.0 &      0.0 &      0.4\\
     &      &    0.6 &      0.0 &      0.1 &      0.4 &      0.1 &      0.0 &      0.0\\
\hline
\end{tabular}
\end{center}
\caption{Configuration mixing of negative-parity $\Lambda$ states in model $\mathcal{A}$
assigned to the $1\hbar\omega$ band.}
\label{tab:conf_mix_Lam_1hw}
\end{table}\\
Table \ref{tab:L1hwband} gives the predicted masses of both models in
comparison to the six corresponding experimental resonance positions. In
addition, table \ref{tab:conf_mix_Lam_1hw} shows the contributions of the
different spin-flavor $SU(6)$ configurations to each state of model
$\mathcal{A}$. Again let us focus in this discussion to the (better) results of
model $\mathcal{A}$.  As shown in fig. \ref{fig:LamM2} the  masses predicted in model
$\mathcal{A}$ can considerably well account for all resonance positions quoted
by the Particle Data Group, however, as already mentioned, with one
striking exception: The notorious difficulty in explaining the low position of
the $\Lambda\frac{1}{2}^-(1405,\mbox{****})$ with respect to
$\Lambda\frac{3}{2}^-(1520,\mbox{****})$ cannot be resolved by using instanton-induced 
forces also within our fully relativistic framework based on the Salpeter equation. Otherwise, the hyperfine structure of the other
$1\hbar\omega$ $\Lambda$-states can be nicely explained by 't~Hooft's residual
force. Figure \ref{fig:Model2L-gvar} shows how 't~Hooft's residual interaction
shapes this hyperfine structure of the negative-parity $1\hbar\omega$ shell
due to the correlation of non-strange and non-strange-strange diquarks within
these states. 
\\
Analogous to fig. \ref{fig:Model2L+gvar} which shows instanton-induced
effects for the positive-parity $\Lambda$-states, fig.
\ref{fig:Model2L-gvar} likewise  shows the influence of 't~Hooft's force on the
energy levels of all negative-parity $\Lambda$-states. As before, we start
with the pure confinement spectrum and in succession the couplings $g_{nn}$ and
$g_{ns}$ are gradually increased.  Once again we observe the remarkable
feature of the instanton-induced interaction that once the 't~Hooft couplings
are fixed to account for the octet-decuplet ground-state splittings, the
excited states, in this case the well-established resonances
$\Lambda\frac{1}{2}^-(1800,\mbox{***})$,
$\Lambda\frac{5}{2}^-(1830,\mbox{****})$,
$\Lambda\frac{1}{2}^-(1670,\mbox{****})$,
$\Lambda\frac{3}{2}^-(1690,\mbox{****})$ and
$\Lambda\frac{3}{2}^-(1520,\mbox{****})$ of the $1\hbar\omega$ shell, are {\it
  at the same time} excellently described. Again let us discuss the situation
for each spin sector $\frac{1}{2}^-$, $\frac{3}{2}^-$ and $\frac{5}{2}^-$
in turn:\\

In the ${\bf\Lambda\frac{5}{2}^-}$ {\bf sector} our model predicts a single
$1\hbar\omega$ state at 1828 MeV. Its mass exactly agrees with the resonance
position of the single four-star resonance
$\Lambda\frac{5}{2}^-(1830,\mbox{****})$ observed in the $D_{05}$ partial
wave. Note that $J=\frac{5}{2}$ is the highest possible spin in the
$1\hbar\omega$ shell and thus requires an internal spin-quartet function for
this state.  Table \ref{tab:conf_mix_Lam_1hw} shows the $\Lambda\frac{5}{2}^-$
resonance being a pure flavor octet $^4 8[70]$ state, which is by no means
influenced by 't~Hooft's force as illustrated in fig.
\ref{fig:Model2L-gvar}.  Similar to its counterpart
$N\frac{5}{2}^-(1675,\mbox{****})$ in the nucleon spectrum, this state is thus
determined by the confinement force alone. This selection rule of 't~Hooft's
force once again nicely conforms with the experimental findings, since already
in the pure confinement spectrum the $\Lambda\frac{5}{2}^-(1830,\mbox{****})$
is well described.  The excellent agreement of our prediction with the
resonance position of $\Lambda\frac{5}{2}^-(1830,\mbox{****})$ thus provides
good support for the flavor-independent confinement force, whose
parameters have been fixed on the phenomenology of the (non-strange)
$\Delta$-spectrum alone: concerning the positioning of the shells the
confinement force works obviously equally well in this strange ($S^*=-1$)
sector.

In the ${\bf\Lambda\frac{3}{2}^-}$ {\bf sector} the hyperfine splitting
between the first resonance $\Lambda\frac{3}{2}^-(1520,\mbox{****})$ and the
second resonance $\Lambda\frac{3}{2}^-(1690,\mbox{****})$ can be fairly well
reproduced by 't~Hooft's force: The first excitation is predicted at 1508 MeV
and the second one at 1662 MeV, thus both states are predicted close to the
resonance positions quoted by the Particle Data Group \cite{PDG00}.  Indeed,
the first state associated with the $\Lambda\frac{3}{2}^-(1520,\mbox{****})$
turns out be dominantly $^2 1[70]$ ($\sim 79\%$) with an additional moderate
admixture of a flavor octet $^2 8[70]$ contribution ($\sim 19\%$). This
dominantly flavor singlet state, which has no counterpart in the $N^*$
spectrum, shows the largest downward mass shift of almost the same amount as
the Roper-type state and the $\Lambda$-ground-state in the positive parity
sector. The biggest part of this shift originates from the attractive
non-strange and only a small fraction comes from the non-strange-strange
diquark correlation.  The second state predicted reveals a dominant $^2 8[70]$
configuration ($\sim 72\%$) with an additional moderate contribution of a
flavor singlet $^2 1[70]$ configuration ($\sim 20\%$).  Hence, the associated
resonance $\Lambda\frac{3}{2}^-(1690,\mbox{****})$ indeed may be viewed as the
octet partner of $N\frac{3}{2}^-(1520,\mbox{****})$. Compared to the lowest
state predicted, here the lowering by 't~Hooft's force is rather moderate and
is primarily due to an attractive non-strange-strange diquark correlation.
The third $\Lambda\frac{3}{2}^-$ state, which so far has not been observed
experimentally, is predicted at 1816 MeV nearly degenerate to the single state
in $\Lambda\frac{5}{2}^-$. This state turns out to be an almost pure $^4
8[56]$ state ($\sim 97\%$) and hence is the expected flavor octet partner of
the nucleon resonance $N\frac{3}{2}^-(1700,\mbox{***})$.

Finally we turn to the
${\bf\Lambda\frac{1}{2}^-}$ {\bf sector}, where the situation is quite
similar to the $\Lambda\frac{3}{2}^-$ sector. The lowest state predicted at 1524 MeV
again turns out to be a dominantly ($\sim 70\%$) flavor singlet $^2 1[70]$
state with a rather strong ($27\%$) contribution of a flavor octet $^2
8[70]$ configurations. As in the
$\Lambda\frac{3}{2}^-$ sector, this state exhibits the strongest
lowering due to 't~Hooft's force, but unfortunately {\it not} deeply
enough to account for the puzzling low position of the
well-established four-star resonance
$\Lambda\frac{1}{2}^-(1405,\mbox{****})$ below the
$\Lambda\frac{3}{2}^-(1520,\mbox{****})$. Actually, the behavior of
this state under influence of 't~Hooft's force is rather similar to
the lowest state in the $\Lambda\frac{3}{2}^-$ sector. Both states
show an equally big lowering which primarily originates from a scalar
non-strange diquark correlation, thus both becoming degenerate close
to the position of the $\Lambda\frac{3}{2}^-(1520,\mbox{****})$.
Nonetheless, there is no doubt to associate the lowest predicted state
to the $\Lambda\frac{1}{2}^-(1405,\mbox{****})$, since 't~Hooft's
force otherwise provides a quite good explanation of the remaining
states in this sector. The second state is predicted at 1630 MeV and
may readily be associated with the $\Lambda\frac{1}{2}^-(1670,
\mbox{****})$. This state exhibits a dominantly flavor octet $^2
8[70]$ configuration ($\sim 62\%$) with an admixed  $^2 1[70]$ contribution ($\sim 29 \%$) 
and it is moderately lowered by 't~Hooft's
interaction mainly due to a scalar non-strange-strange correlation.
The third state predicted at 1816 MeV nicely agrees with the
$\Lambda\frac{1}{2}^-(1800, \mbox{***})$ and turns out to be an almost
pure flavor octet $^4 8[70]$ state.  Hence,
$\Lambda\frac{1}{2}^-(1670, \mbox{****})$ and
$\Lambda\frac{1}{2}^-(1800, \mbox{***})$ indeed may be viewed as the
octet partners of $N\frac{1}{2}^-(1535,\mbox{****})$ and
$N\frac{1}{2}^-(1650,\mbox{****})$ in the corresponding
$N\frac{1}{2}^-$ sector.  Note that similar to the $N\frac{1}{2}^-$
sector the second state reveals an {\it upwards} mass shift due to the
repulsive part in the relativistic version of 't~Hooft's force acting
in the {\it pseudo-scalar} diquark sector.  This upwards mass shift in
fact is necessary here to match the $\Lambda\frac{1}{2}^-(1800,
\mbox{***})$ and finally to correctly reproduce the hyperfine
splitting between $\Lambda\frac{1}{2}^-(1670, \mbox{****})$ and
$\Lambda\frac{1}{2}^-(1800, \mbox{***})$.\\

Altogether we thus find a rather good explanation for the hyperfine
structure of the negative-parity $\Lambda$-states in the
$1\hbar\omega$ shell due to instanton effects. The only striking
exception is the $\Lambda\frac{1}{2}^-(1405,\mbox{****})$: although
't~Hooft's force provides an explanation of the relatively low energy
of the dominantly flavor singlet states below the dominantly flavor
octet states the instanton-induced interaction\footnote{We should add
a remark here concerning the genuine three-body force of the
instanton-induced interaction: In the literature one sometimes finds
the statement that this force, which only acts in the $uds$ singlet
states, should additionally affect the $\Lambda\frac{1}{2}^-(1405)$.
However, as discussed in ref. \cite{Loe01b}, this statement is not
correct: The three-body force can be shown to act only on
flavor-antisymmetric, spatial-symmetric three-quark states.  There
are, however, {\it no physical} three-quark states (baryons) with this
property, since these are color-antisymmetric and finally the Pauli
principle can only be satisfied by spin-antisymmetric three-quark wave
functions, which do not exist. Hence this three-body force does not
contribute to physical color singlet, but only to color octet or
decuplet three-quark states.}, even if treated fully relativistically,
cannot explain the exceptionally large splitting between the
'spin-orbit partners' $\Lambda\frac{1}{2}^-(1405,\mbox{****})$ and
$\Lambda\frac{3}{2}^-(1520,\mbox{****})$. Both states are roughly
degenerate with a mass close to the
$\Lambda\frac{3}{2}^-(1520,\mbox{****})$.  This shortcoming is
observed in {\it all} other constituent quark models for baryons as
well and remains one of the outstanding problems.  It is worthwhile to
comment here on this somewhat puzzling state.  In view of the
otherwise very consistent description of hyperfine structures in the
$1\hbar\omega$ and $2\hbar\omega$ bands we believe that the too high
predicted mass does not really reflect a fundamental flaw in our model
but rather the fact that other dynamical effects that are specific to
that state are presently not taken into account in our model. There is
still controversy about the physical nature of the
$\Lambda\frac{1}{2}^-(1405,\mbox{****})$ and its role in the $\bar K
N$ interaction at low energies. A review of this discussion is given
by Dalitz \cite{Dal00} in the latest edition of the \textit{'Review of
Particle Physics'}.  The mass splitting of
$\Lambda\frac{1}{2}^-(1405)$ and $\Lambda\frac{3}{2}^-(1520)$ is often
interpreted as a possible hint for the relevance of spin-orbit forces
in baryon spectroscopy. In our opinion such an interpretation is
rather questionable. The splitting between these states is
exceptionally large and otherwise the light baryon spectra show hardly
any evidence for strong spin-orbit effects. This was just the reason
why we have chosen those confinement Dirac structures which induce
only moderate spin-orbit effects. Larger spin-orbit interactions would
generally spoil the agreement for the spectrum in other sectors.
Another extreme possibility discussed in literature is that the
$\Lambda\frac{1}{2}^-(1405)$, which is lying about 30 MeV below the
$N\bar K$ threshold, might be an unstable $N\bar K$ bound state. But,
such an interpretation requires the observation of another state lying
close to the $\Lambda\frac{3}{2}^-(1520)$ MeV which then completes the
$\Lambda\frac{1}{2}^-$ states of the $1\hbar\omega$ shell.  However,
this energy region is already well explored by $N\bar K$ scattering
experiments with no sign of such a resonance \cite{Dal00}. The most
appropriate dynamical explanation for the low position
$\Lambda\frac{1}{2}^-(1405)$ seems to be a strong coupling of the bare
three-quark state to virtual meson-baryon ($\bar K N$) decay channels
due to its proximity to $\bar K N$ threshold. A confirmation of this
conjecture on the basis of our covariant model requires the
calculation of the $\bar K N$ decay amplitude of the model state
$\MSL(1,-,1,1524)$, which then should be exceptionally large. But a
really quantitative statement would require a fully dynamical
treatment of various external meson-baryon decay channels in general.
Such investigations have been made in non-relativistic quark models
\cite{AMS94}, showing indeed a naturally shift of the lowest
$\Lambda\frac{1}{2}^-$ three-quark state towards the $\bar K N$
threshold. An incorporation of these effects in our fully relativistic
Bethe-Salpeter framework, however, seems presently to be technically
too much involved.
\subsubsection{Beyond the 2${\bfgrk\hbar}{\bfgrk\omega}$ band}
The high energy part of the positive- and negative-parity $\Lambda$-spectrum
beyond the mass region of the $2\hbar\omega$ and $1\hbar\omega$ shell is
experimentally hardly explored. Here the Particle Data Group states only
three further resonances with established quantum numbers. In the negative-parity
sector these are the well-established four-star resonance
$\Lambda\frac{7}{2}^-(2100,\mbox{****})$ and the one-star resonance
$\Lambda\frac{3}{2}^-(2325,\mbox{*})$ for which evidence, however, is very poor.
According to its spin $J^\pi=\frac{7}{2}^-$ the
$\Lambda\frac{7}{2}^-(2100,\mbox{****})$ has to be assigned to the
$3\hbar\omega$ band, but its position is comparatively low and nearly
degenerate with the first $\Lambda\frac{7}{2}^+$ resonance
$\Lambda\frac{7}{2}^+(2020,\mbox{*})$ in the upper part of the positive-parity
$2\hbar\omega$ shell.  In the positive parity sector the well-established
three-star resonance $\Lambda\frac{9}{2}^+(2350,\mbox{***})$ is seen
experimentally in the $H_{09}$ partial wave and due to its spin and position this state is assigned to the
$4\hbar\omega$ band. Apart from a bump $\Lambda ?^?(2585,\mbox{**})$ with
undefined spin and parity no resonances have been observed so far which could
be assigned to the higher $5\hbar\omega$ and $6\hbar\omega$ shells.  Our
predictions of model $\mathcal{A}$ and $\mathcal{B}$ for the lightest few
states in the $3\hbar\omega$ and $4\hbar\omega$ bands are summarized in tables
\ref{tab:L3hwband} and \ref{tab:L4hwband} respectively. 
\begin{table}[!h]
\center
\begin{tabular}{ccccccc}
\hline
State&  &${J^\pi}$              & Rating     & Experiment [MeV]& Mass [MeV]   &  Mass [MeV] \\
     &  &                       &            & \cite{PDG00}    & in model $\mathcal{A}$& in model $\mathcal{B}$  \\
\hline
               &$S_{01}$&${\frac{1}{2}^-}$&          &           &$\MSL(1,-,4,2011)$   &$\MSL(1,-,4,2173)$ \\
               &        &                 &          &           &$\MSL(1,-,5,2076)$   &$\MSL(1,-,5,2249)$\\
               &        &                 &          &           &$\MSL(1,-,6,2117)$   &$\MSL(1,-,6,2308)$\\
               &        &                 &          &           &$\MSL(1,-,7,2263)$   &$\MSL(1,-,7,2399)$\\
               &        &                 &          &           &$\MSL(1,-,8,2310)$   &$\MSL(1,-,8,2406)$\\
\hline
               &        &                 &          &           &$\MSL(3,-,4,1987)$   &$\MSL(3,-,4,2152)$ \\
               &        &                 &          &           &$\MSL(3,-,5,2090)$   &$\MSL(3,-,5,2212)$\\
               &        &                 &          &           &$\MSL(3,-,6,2147)$   &$\MSL(3,-,6,2316)$\\
               &        &                 &          &           &$\MSL(3,-,7,2259)$   &$\MSL(3,-,7,2366)$\\
               &        &                 &          &           &$\MSL(3,-,8,2275)$   &$\MSL(3,-,8,2384)$\\
$\Lambda(2325)$&$D_{03}$&${\frac{3}{2}^-}$& *        &2307-2372  &$\MSL(3,-,9,2313)$   &$\MSL(3,-,9,2400)$\\
               &        &                 &          &           &$\MSL(3,-,10,2314)$   &$\MSL(3,-,10,2422)$\\
\hline
               &$D_{05}$&${\frac{5}{2}^-}$&          &           &$\MSL(5,-,2,2080)$   &$\MSL(5,-,2,2207)$ \\
               &        &                 &          &           &$\MSL(5,-,3,2179)$   &$\MSL(5,-,3,2297)$\\
               &        &                 &          &           &$\MSL(5,-,4,2287)$   &$\MSL(5,-,4,2375)$\\
               &        &                 &          &           &$\MSL(5,-,5,2314)$   &$\MSL(5,-,5,2417)$\\
\hline
$\Lambda(2100)$&$G_{07}$&${\frac{7}{2}^-}$& ****     &2090-2110  &$\MSL(7,-,1,2090)$   &$\MSL(7,-,1,2185)$\\[3mm]
               &        &                 &          &           &$\MSL(7,-,2,2227)$   &$\MSL(7,-,2,2286)$\\
               &        &                 &          &           &$\MSL(7,-,3,2327)$   &$\MSL(7,-,3,2405)$\\
               &        &                 &          &           &$\MSL(7,-,4,2380)$   &$\MSL(7,-,4,2429)$\\
\hline
               &$G_{09}$&${\frac{9}{2}^-}$&          &           &$\MSL(9,-,1,2370)$   &$\MSL(9,-,1,2400)$ \\
               &        &                 &          &           &$\MSL(9,-,2,2437)$   &$\MSL(9,-,2,2450)$\\
\hline
\end{tabular}
\caption{Calculated positions of the lightest few $\Lambda$ states assigned to
  the negative parity $3\hbar\omega$ shell. Notation as in table \ref{tab:L2hwband}.}
\label{tab:L3hwband}
\end{table}
\begin{table}[!h]
\center
\begin{tabular}{ccccccc}
\hline
Exp. state   &PW&${J^\pi}$        & Rating     & Mass range [MeV]& Model state &  Model state \\
\cite{PDG00} &  &                      &            & \cite{PDG00}    & in model $\mathcal{A}$& in model $\mathcal{B}$  \\
\hline
               &$P_{01}$&${\frac{1}{2}^+}$&          &           &$\MSL(1,+,8,2156)$   &$\MSL(1,+,8,2374)$ \\
               &        &                 &          &           &$\MSL(1,+,9,2223)$   &$\MSL(1,+,9,2452)$\\
               &        &                 &          &           &$\MSL(1,+,10,2345)$  &$\MSL(1,+,10,2516)$\\
\hline
               &$P_{03}$&${\frac{3}{2}^+}$&          &           &$\MSL(3,+,8,2278)$   &$\MSL(3,+,8,2430)$ \\
               &        &                 &          &           &$\MSL(3,+,9,2382)$   &$\MSL(3,+,9,2509)$\\
               &        &                 &          &           &$\MSL(3,+,10,2408)$  &$\MSL(3,+,10,2558)$\\
\hline
               &$F_{05}$&${\frac{5}{2}^+}$&          &           &$\MSL(5,+,6,2270)$   &$\MSL(5,+,6,2407)$ \\
               &        &                 &          &           &$\MSL(5,+,7,2375)$   &$\MSL(5,+,7,2469)$\\
               &        &                 &          &           &$\MSL(5,+,8,2436)$   &$\MSL(5,+,8,2556)$\\
               &        &                 &          &           &$\MSL(5,+,9,2494)$   &$\MSL(5,+,9,2595)$\\
\hline
               &$F_{07}$&${\frac{7}{2}^+}$&          &           &$\MSL(7,+,2,2331)$   &$\MSL(7,+,2,2449)$ \\
               &        &                 &          &           &$\MSL(7,+,3,2431)$   &$\MSL(7,+,3,2539)$\\
               &        &                 &          &           &$\MSL(7,+,4,2533)$   &$\MSL(7,+,4,2611)$\\
               &        &                 &          &           &$\MSL(7,+,5,2556)$   &$\MSL(7,+,5,2632)$\\
\hline
$\Lambda(2350)$&$H_{09}$&${\frac{9}{2}^+}$&  ***     & 2340-2370 &$\MSL(9,+,1,2340)$   &$\MSL(9,+,1,2433)$ \\[3mm]
               &        &                 &          &           &$\MSL(9,+,2,2479)$   &$\MSL(9,+,2,2536)$\\
               &        &                 &          &           &$\MSL(9,+,3,2565)$   &$\MSL(9,+,3,2637)$\\
               &        &                 &          &           &$\MSL(9,+,4,2608)$   &$\MSL(9,+,4,2649)$\\
\hline
               &$H_{0\;11}$&${\frac{11}{2}^+}$&          &           &$\MSL(11,+,1,2601)$  & $\MSL(11,+,1,2632)$  \\
               &        &                 &          &           &$\MSL(11,+,2,2677)$  & $\MSL(11,+,2,2689)$ \\
\hline
\end{tabular}
\caption{Calculated positions of the lightest few $\Lambda$ states assigned to
  the positive parity $4\hbar\omega$ shell.
  Notation as in table \ref{tab:L2hwband}.}
\label{tab:L4hwband}
\end{table}
Again let us restrict our detailed discussion to model $\mathcal{A}$. As can
be seen in fig. \ref{fig:LamM2} and tables \ref{tab:L3hwband} and
\ref{tab:L4hwband} model $\mathcal{A}$ predicts the first excitations in the
$\Lambda\frac{7}{2}^-$ and $\Lambda\frac{9}{2}^+$ sectors in excellent
agreement with the measured positions of the well-established resonances
observed in these sectors. The predicted masses for the first excited state in
$\Lambda\frac{7}{2}^-$ at 2090 MeV and for the first excitation in
$\Lambda\frac{9}{2}^+$ at 2340 MeV are in both cases very close to the
reported resonance positions of $\Lambda\frac{7}{2}^-(2100,\mbox{****})$ and
$\Lambda\frac{9}{2}^+(2350,\mbox{***})$, respectively.  According to fig.
\ref{fig:NucLamComp} we would expect that the
$\Lambda\frac{9}{2}^+(2350,\mbox{***})$ is the octet partner of
$N\frac{9}{2}^+(2220,\mbox{****})$.  In fact the model state associated to the
$\Lambda\frac{9}{2}^+(2350,\mbox{***})$ turns out being dominantly flavor
octet ($~45\%\;^2 8[56]$ and $32\%\;^2 8[70]$) with an additional moderate
admixture ($22\%$) of a flavor singlet $^2 1[70]$ configuration. The model
state associated to the $\Lambda\frac{7}{2}^-(2100,\mbox{****})$ shows almost
equally big flavor octet ($\sim 16\%\;^2 8[56]$ and $\sim 31\%\;^2 8[70]$) and
flavor singlet ($\sim 51\%\; ^2 1[70]$) contributions.  Unfortunately, the
lack of data allows no detailed investigation of the instanton-induced
hyperfine structures in $3\hbar\omega$ and $4\hbar\omega$ shells in comparison
with experiment as was possible in the lower-lying $1\hbar\omega$ and
$2\hbar\omega$ shells.  Nevertheless, the remarkably good predictions of the two single
resonances $\Lambda\frac{7}{2}^-(2100,\mbox{****})$ and
$\Lambda\frac{9}{2}^+(2350,\mbox{***})$ already clearly demonstrate the
importance of instanton-induced effects also in this higher mass region of the
$\Lambda$-spectrum, since
't~Hooft's force nicely explains the comparatively low positions of these
states. This is convincingly illustrated in figs. \ref{fig:Model2L+gvar} and \ref{fig:Model2L-gvar}. 
't~Hooft's interaction shifts both associated model
states relatively strongly downward by roughly 200 MeV with respect to the
other states and hence these become well isolated.  As can be seen in figs.
\ref{fig:Model2L+gvar} and \ref{fig:Model2L-gvar} this mass shift originates
for both states merely from an attractive non-strange scalar diquark
correlation, while the non-strange-strange diquark contributions is
negligible.  Once again we observe the remarkable feature that the shift of
the states is exactly the right size to match the experimentally measured
positions just when the 't~Hooft couplings $g_{nn}$ and $g_{ns}$ are fixed to
reproduce the ground-state hyperfine-splittings.  Otherwise, 't~Hooft's force
causes in the $3\hbar\omega$ and $4\hbar\omega$ shells in general  quite
similar effects as in the lower lying $1\hbar\omega$ and $2\hbar\omega$ shells
as can be seen in figs. \ref{fig:Model2L+gvar} and \ref{fig:Model2L-gvar}.
Therefore it is worthwhile to comment at this stage on the instanton-induced
hyperfine structures of the shells in general.
\subsubsection{Instanton-induced hyperfine structures and approximate parity doublets}
In the foregoing discussion we convincingly demonstrated that 't~Hooft's force
with its couplings $g_{nn}$ and $g_{ns}$ fixed from the ground-state baryons
excellently explains the most prominent features in the single even- and
odd-parity bands of the experimental $\Lambda$-spectrum. Here let us summarize
how 't~Hooft's force globally reorganizes the pure confinement
$\Lambda$-spectrum.  The systematics is essentially the same as already
exposed in the detailed discussion of the nucleon spectrum in ref. \cite{Loe01b}. Without
residual instanton force the three-body confinement arranges the
$\Lambda$-spectrum into alternating even- and odd-parity bands, where in each
band the states are clustered in rather narrow energy
ranges.  Due to the strong selection rules of 't~Hooft's force a particular
set of states in each $N\hbar\omega$ shell, namely those states which are
dominantly $^2 8 [56]$, $^2 8 [70]$ or $^2 1[70]$, are selectively lowered
relative to the bulk of dominantly $^4 8 [70]$, $^2 8 [20]$ and $^4 1[70]$
states, which essentially remain unaffected. In particular, the states of the
shell with maximally possible total spin $J_{\rm max}(N)=N+\frac{3}{2}$ which
posses an internal spin-quartet ($S=3/2$) function in order to achieve this
maximal total spin, remain unshifted.  Thus each $N\hbar\omega$ shell
systematically splits into an upper part with total spins up to $J_{\rm
  max}(N)$ and a lower part with total spins up to $J_{\rm max}(N)-1$.  With
the couplings fixed from the ground-states the lower part comes to lie fairly
in between the upper parts of the two adjacent $N\hbar\omega$ and
$(N-2)\hbar\omega$ shells with the same parity and thus nearly degenerate with
the unshifted part of the $(N-1)\hbar\omega$ shell with opposite parity.
Concerning the lower part of the shells, the main difference to the nucleon
spectrum is firstly, that in general more states are shifted downwards due to
the additional flavor singlet $^2 1[70]$ configuration, and secondly, that both
types of diquarks, non-strange and non-strange-strange, emerge. In the
$\Lambda$-spectrum we thus observe in figs. \ref{fig:Model2L+gvar} and
\ref{fig:Model2L-gvar} the lowered part of the shells actually separating into
two further parts. The excited states with dominant non-strange diquark
contribution are generally more strongly lowered than those with dominant
non-strange-strange diquark contribution. This level ordering is qualitatively
understandable in the following manner. Firstly, the scalar non-strange
diquark correlation is stronger than the non-strange-strange correlation due
to $g_{nn}>g_{ns}$. Apart from that, the second, still more important effect
is due to the different kinematics of the two distinct diquark plus quark
systems. In a strongly simplified, non-relativistic picture with harmonic
three-body confinement forces ($\rho$- plus $\lambda$-oscillator) one can
approximately consider these states as two-particle bound states of a scalar
diquark with mass $M_D$ and a quark with mass $m_q$, {\it i.e.} as non-strange
diquark plus strange quark ($M_{nn}$, $m_s$) or as non-strange-strange diquark
plus non-strange quark ($M_{ns}$, $m_n$). Diquark and quark move in a two-body
confining potential which then is given by the $\lambda$-oscillator alone. The two distinct
diquark-quark systems form oscillator shells with {\it different} mass gaps
$\hbar\omega_{(nn)s}$ and $\hbar\omega_{(ns)n}$, where the frequencies
$\omega_{Dq}\sim (\frac{M_D+m_q}{M_D m_q})^\frac{1}{2}$ are proportional to
the square root of the inverse reduced diquark-quark masses.  Assuming weakly bound
diqarks, {\it i.e.} $M_{nn}\approx 2 m_{n}$ and $M_{ns}\approx m_n+m_s$, one
finds $\omega_{(nn)s}\approx\omega_{(ns)n}(\frac{m_n+m_s}{2
  m_n})^\frac{1}{2}$, thus in fact $\hbar\omega_{(nn)s}<\hbar\omega_{(ns)n}$.
Hence this level ordering is predominantly a flavor $SU(3)$ symmetry breaking
effect from the quark mass difference $m_s>m_n$.\\

\begin{figure}[!h]
  \begin{center}
    \epsfig{file={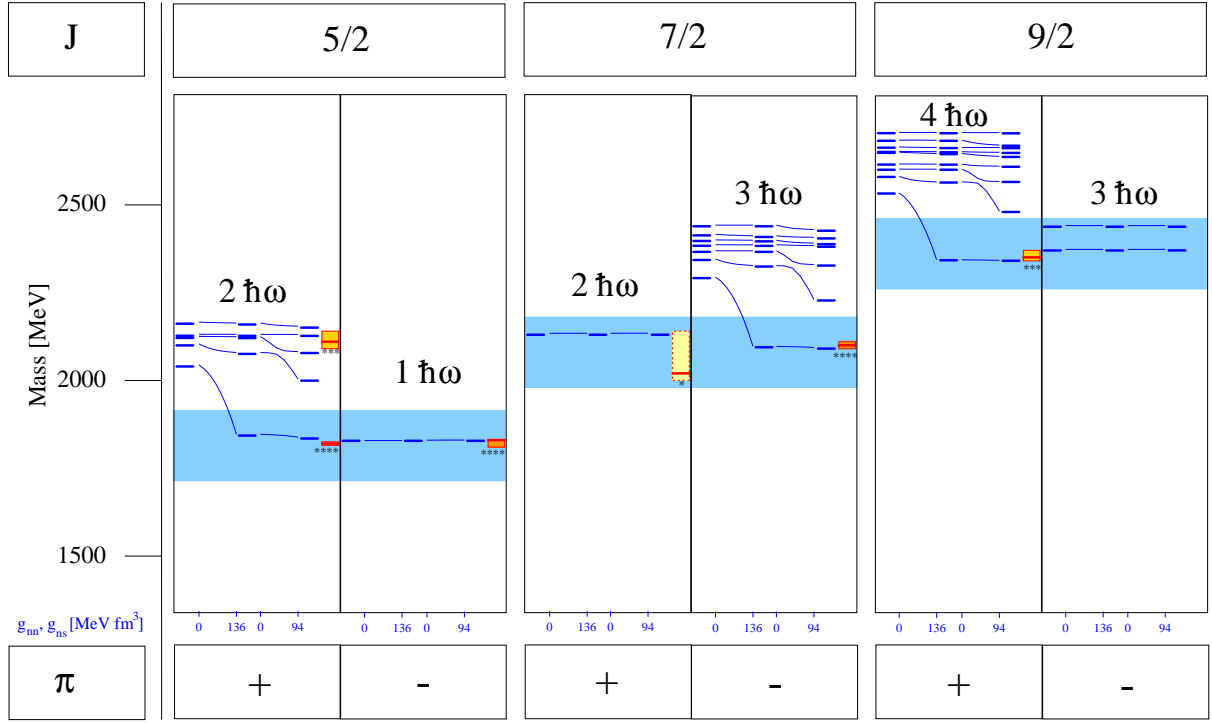},width=160mm}
  \end{center}
\caption{Instanton-induced generation of approximate parity doublets of lowest
  lying $\Lambda$-states in the sectors with
    $J=\frac{5}{2}$, $\frac{7}{2}$ and $\frac{7}{2}$ (in model $\mathcal{A}$).}
\label{fig:PDreggeLam}
\end{figure}

Similar to the nucleon spectrum the overlapping positive- and
negative-parity bands lead to approximate parity doublets. Figure
\ref{fig:PDreggeLam} shows how 't~Hooft's force induces such
approximates doublets for the lowest exited states in the sectors with
total spin $J=\frac{5}{2}$, $\frac{7}{2}$ and $\frac{9}{2}$.
The lowest states in $\frac{5}{2}^-$, $\frac{7}{2}^+$ $\frac{9}{2}^-$ are
those with the maximum total spin in the $1\hbar\omega$, $2\hbar\omega$ and
$3\hbar\omega$ bands, respectively, and thus are determined by confinement
alone. In the corresponding sectors $\frac{5}{2}^+$, $\frac{7}{2}^-$
$\frac{9}{2}^+$ with the same spin but opposite parity there are always
exactly two\footnote{one state with a non-strange diquark contribution and
  another with a non-strange-strange diquark contribution} states of the
higher lying $2\hbar\omega$, $3\hbar\omega$ and $4\hbar\omega$ shells which
are selectively lowered by 't~Hooft's force and thus become well isolated from
the other states. It is quite remarkable that the shift of the lowest
excitations (with the non-strange diquark contribution) once again is such
that these in fact become degenerate with the unshifted first excited states of opposite
parity. The experimentally best established parity doublet in the
$\Lambda$-spectrum, which nicely confirms this scenario, is that of the two
four-star resonances $\Lambda\frac{5}{2}^+(1820,\mbox{****})$ and
$\Lambda\frac{5}{2}^-(1830,\mbox{****})$ in the $2\hbar\omega$ and
$1\hbar\omega$ shell, respectively. It is the counterpart to the
well-established parity doublet structure $N\frac{5}{2}^+(1680,\mbox{****})$
-- $N\frac{5}{2}^-(1675,\mbox{****})$ in the nucleon spectrum (see ref. \cite{Loe01b}). The two single
resonances $\Lambda\frac{9}{2}^+(2350,\mbox{***})$ and
$\Lambda\frac{7}{2}^-(2100,\mbox{****})$ observed in the high energy region of
the $\Lambda$-spectrum are just the states with maximally possible total spin
$J=J_{\rm max}(N)-1$ in the lowered substructures of the $N=4$ and $N=3$
shells, respectively.  In the $\Lambda\frac{9}{2}^+$ sector the situation is
similar to the corresponding $N\frac{9}{2}^+$ nucleon sector, where this mass
shift of the lowest state turned out to be important for generating the
striking parity doublet structure
$N\frac{9}{2}^+(2220,\mbox{****})$--$N\frac{9}{2}^-(2250,\mbox{****})$ in the
upper part of the nucleon spectrum \cite{Loe01b}.  The counterpart to the doublet partner
$N\frac{9}{2}^-(2250,\mbox{****})$ of $N\frac{9}{2}^+(2220,\mbox{****})$ has
not yet been observed in the $\Lambda\frac{9}{2}^-$ sector. But it is
interesting that our model predicts the lowest state in this sector at 2370
MeV, indeed approximately degenerate to the
$\Lambda\frac{9}{2}^+(2350,\mbox{***})$. In the sectors with spin
$J=\frac{7}{2}$ the same mechanism can nicely explain the parity doublet
$\Lambda\frac{7}{2}^+(2020,\mbox{*})$--$\Lambda\frac{7}{2}^-(2100,\mbox{****})$
observed experimentally.  Here 't~Hooft's force induces a strong attractive
scalar non-strange diquark correlation in the $\Lambda\frac{7}{2}^-$ state
shifting this $3\hbar\omega$ state down to 2090 MeV, deeply enough to match
the well-established $\Lambda\frac{7}{2}^-(2100,\mbox{****})$ and thus to
become nearly degenerate to the unaffected $\Lambda\frac{7}{2}^+$ state in the
$2\hbar\omega$ shell predicted at 2130 MeV. It is interesting to compare this
nice result to that of the corresponding nucleon sectors $N\frac{7}{2}^\pm$.
In ref. \cite{Loe01b} we showed that 't~Hooft's force induced in the same manner the degeneracy of
the lowest $N\frac{7}{2}^\pm$ nucleon states.
In particular we likewise found a similar strong downward shift of the
$N\frac{7}{2}^-$ state. However, the currently available experimental data in
$N\frac{7}{2}^\pm$ do not show a clear parity doublet structure according to
the comparatively high position of the $N\frac{7}{2}^-(2190,\mbox{****})$ with
respect to the $N\frac{7}{2}^+(1990,\mbox{**})$. While in the
$\Lambda\frac{7}{2}^-$ sector the instanton-induced effect is able to
explain the low-lying $\Lambda\frac{7}{2}^-(2100,\mbox{****})$, in the
$N\frac{7}{2}^-$ sector the same effect shifts the first state predicted far
below the $N\frac{7}{2}^-(2190,\mbox{****})$. In view of the otherwise
striking similarities between the nucleon and $\Lambda$-states, that {\it all}
can be simultaneously explained by 't~Hooft's force in both flavor sectors,
this discrepancy in the $N\frac{7}{2}^-$ sector is rather surprising.  We
therefore consider the remarkably good prediction for the low-lying
$\Lambda\frac{7}{2}^-(2100,\mbox{****})$ as a further strong support for our
conjecture in ref. \cite{Loe01b} that there might be a $N\frac{7}{2}^-$ state below the
$N\frac{7}{2}^-(2190,\mbox{****})$ at roughly 2015 MeV which is the expected
parity doublet partner to
$N\frac{7}{2}^+(1990,\mbox{**})$.\\

Note that the lowest states in the sectors $J^\pi=\frac{5}{2}^+$ and
$\frac{9}{2}^+$ belong to the positive-parity $\Lambda$-Regge
trajectory. Since in our model both states are quite strongly lowered by the instanton
interaction as the ground-state, it is again interesting to investigate how far the
't~Hooft's force influences the linear Regge characteristics $M^2 \sim J$
observed empirically for the Regge states
$\Lambda\frac{1}{2}^+(1116,\mbox{****})$,
$\Lambda\frac{5}{2}^+(1820,\mbox{****})$ and
$\Lambda\frac{9}{2}^+(2350,\mbox{***})$.

\subsubsection{The positive-parity $\Lambda$-Regge trajectory}
\begin{figure}[!h]
  \begin{center}
    \input{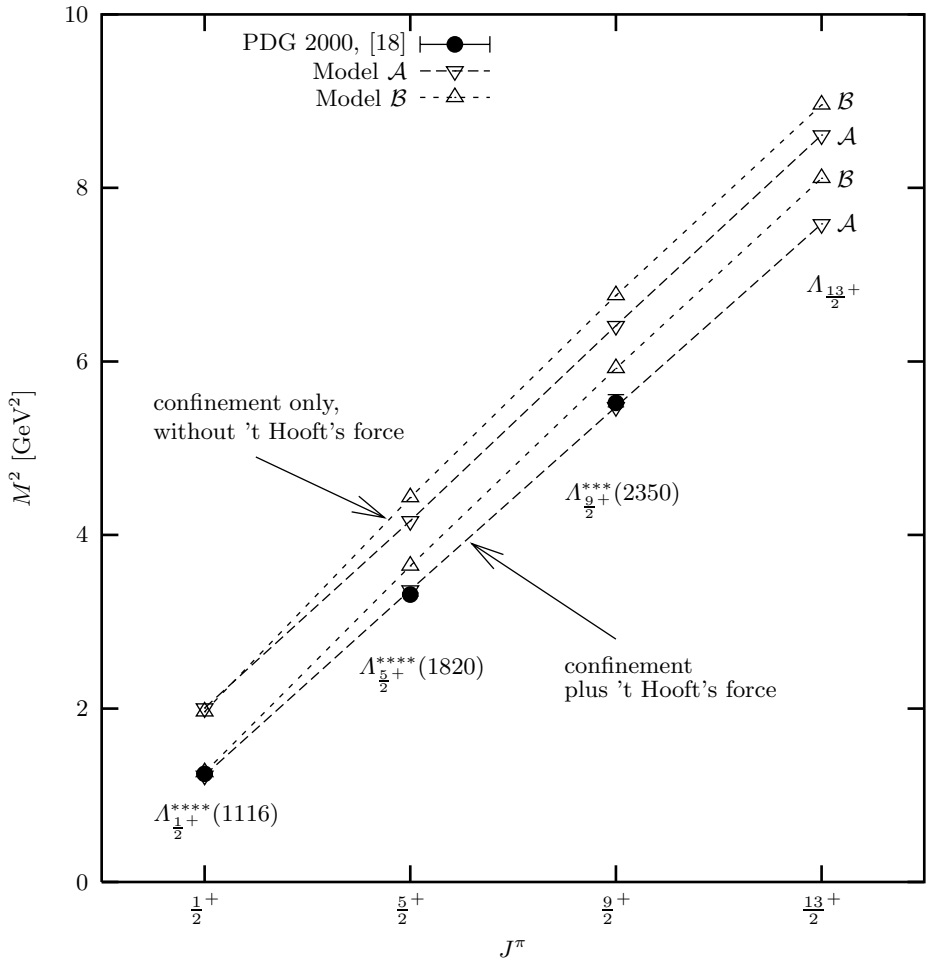}
  \end{center}
\caption{Chew-Frautschi plot ($M^2$ {\it vs.} $J$) of the
  positive-parity $\Lambda$-Regge trajectory $\Lambda\frac{1}{2}^+$,
  $\Lambda\frac{5}{2}^+$, $\Lambda\frac{9}{2}^+$, $\Lambda\frac{13}{2}^+$, $\ldots$, in the
  models $\mathcal{A}$ and $\mathcal{B}$ (lower curves) compared to
  experimental masses from the Particle Data Group (see \cite{PDG00}).
  The upper curves show the trajectories without influence of 't~Hooft's
  instanton-induced interaction.}
\label{fig:lam_regge}
\end{figure}

Figure \ref{fig:lam_regge} depicts the Chew-Frautschi plot ($M^2$
{\it vs.} $J$) of the positive-parity $\Lambda$-Regge trajectory in both
confinement models $\mathcal{A}$ and $\mathcal{B}$ in comparison to the states
$\Lambda\frac{1}{2}^+(1116,\mbox{****})$,
$\Lambda\frac{5}{2}^+(1820,\mbox{****})$ and
$\Lambda\frac{9}{2}^+(2350,\mbox{***})$ observed experimentally.
To illustrate the influence of 't~Hooft's force we displayed
in addition the trajectory as determined by the confinement force alone.
\\
In consistence with the well-described $\Delta$-trajectory which is determined
by the string-like confinement alone \cite{Loe01b}, the flavor-independent confinement force
produces likewise a linear characteristics for the pure confinement
$\Lambda$-trajectories. In model $\mathcal{A}$ this trajectory already shows
the quantitatively correct slope, whereas the slope in model $\mathcal{B}$ turns
out too large right from start. Similar to the nucleon Regge trajectory
we observe once more the remarkable and nontrivial feature of 't~Hooft's force
to be compatible with the linear Regge characteristics. Again the
downward shift of the Regge states, which is largest for the
ground-states and moderate for the higher states, is such that it preserves
the linear behavior $M^2\sim J$ with almost the same slope as in the pure
confinement case.  In model $\mathcal{A}$ the equally large shift of all
trajectory members in the mass square $M^2$ leads simultaneously to an excellent
agreement with the observed states $\Lambda\frac{1}{2}^+(1116,\mbox{****})$,
$\Lambda\frac{5}{2}^+(1820,\mbox{****})$ and
$\Lambda\frac{9}{2}^+(2350,\mbox{***})$ and also the next (still ''missing'')
member predicted in $\Lambda\frac{13}{2}^+$ at 2754 MeV lies fairly well on
this linear trajectory. In model $\mathcal{B}$, however, the slope remains too
big to account for the observed states. In table \ref{tab:lam_regge} the
predicted positions for the $\Lambda$-Regge states are once more
explicitly summarized in comparison to the experimental findings.

\begin{table}[!h]
\center
\begin{tabular}{cccccc}
\hline
Regge-          & Rating & $J^\pi$           & exp. Mass [MeV] & Mass [MeV]         & Mass [MeV]\\
state               &        &                   &                 &  Model $\mathcal{A}$& Model $\mathcal{B}$\\
\hline
$\Lambda(1116)$      & ****   & $\frac{1}{2}^+$   & 1116            & 1108             & 1123 \\ 
$\Lambda(1820)$      & ****   & $\frac{5}{2}^+$   & 1815-1825       & 1834             & 1909\\
$\Lambda(2350)$      & ***   & $\frac{9}{2}^+$    &  2340-2370      & 2340             & 2433\\
    ''missing''          &       & $\frac{13}{2}^+$   &                 & 2754             & 2848\\
\hline
\end{tabular}
\caption{Position of states belonging to the positive-parity $\Lambda$-Regge trajectory calculated in the models $\mathcal{A}$ and
$\mathcal{B}$ in comparison to the experimental resonance positions \cite{PDG00}.
For a graphical presentation see fig. \ref{fig:lam_regge}.}
\label{tab:lam_regge}
\end{table}
This result once again illustrates the importance of instanton effects in the
higher mass regions of baryon spectra. In particular, it demonstrates that our
string-like, flavor-dependent confinement ansatz $\mathcal{A}$ with its
parameters $a$ (offset) and $b$ (slope) fixed entirely from the phenomenology of
the {\it non-strange} $\Delta$-spectrum works equally well also in the strange
flavor sector. In this respect model $\mathcal{B}$, however, strongly fails:
although the confinement ansatz $\mathcal{B}$ could equally well as model
$\mathcal{A}$ account for the non-strange nucleon and $\Delta$-Regge
trajectories, it produces here, in the strange $\Lambda$-sector, a wrong slope
for the trajectory. At the end of this section we shall now briefly comment on the shortcomings
of model $\mathcal{B}$ in general.
\subsubsection{Shortcomings of model $\mathcal{B}$} 
In accordance with our results for the nucleon spectrum in
ref. \cite{Loe01b}, also the $\Lambda$-spectrum is consistently better
described in model $\mathcal{A}$ than in model $\mathcal{B}$. For this
reason our preceding detailed discussion of the $\Lambda$-spectrum has
been restricted mainly to the results of model $\mathcal{A}$.  But for
the sake of completeness we should conclude our investigations of the
$\Lambda$-spectrum by briefly summarizing the main shortcomings of
model $\mathcal{B}$.  In the $\Lambda$-sector the discrepancies are
even more distinctive than in the nucleon sector.  As mentioned, model
$\mathcal{B}$ already fails in describing the $\Lambda$-Regge
trajectory. The slope of the trajectory turns out too large. This is
quite in contrast to the non-strange sectors, where $\Delta$- and
nucleon trajectories still could be quantitatively explained also by
model $\mathcal{B}$ \cite{Loe01b}.
\begin{figure}[!h]
  \begin{center}
    \epsfig{file={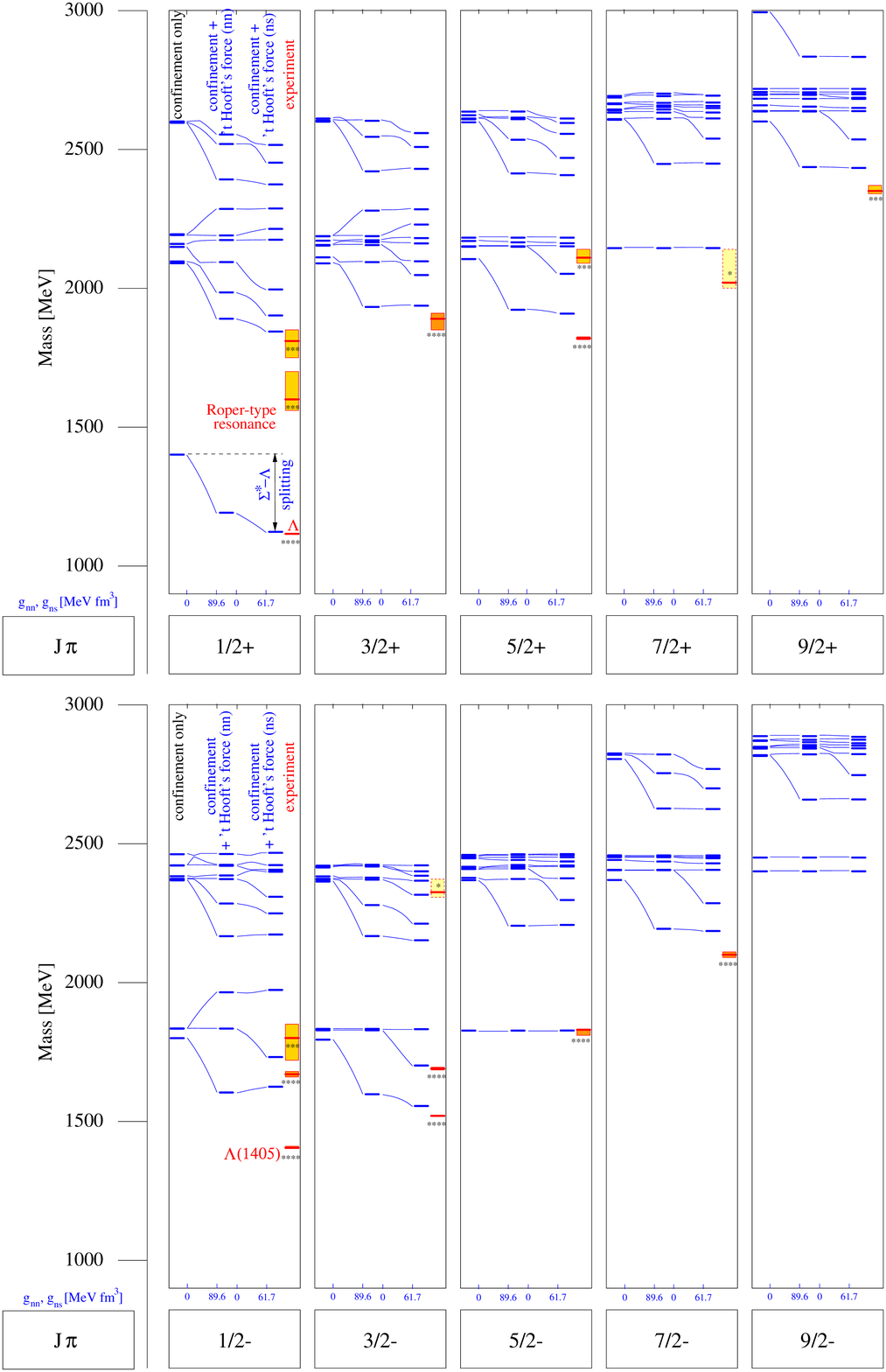},width=135mm}
  \end{center}
\caption{Influence of the instanton-induced interaction on the energy levels
  of the positive-parity (ahead) and negative-parity (below) $\Lambda$-states
  in model $\mathcal{B}$. The curves illustrate the variation of energy levels
  with increasing 't~Hooft couplings $g_{nn}$ and $g_{ns}$ which are finally
  fixed to $g_{nn}=89.6$ MeV fm$^3$ and $g_{ns}=61.7$ MeV fm$^3$.}
\label{fig:Mod1aLgvar}
\end{figure}

Figure \ref{fig:Mod1aLgvar} shows for model $\mathcal{B}$ the influence of
't~Hooft's force on the positive- and negative-parity $\Lambda$-states in
analogy to the corresponding effects of model $\mathcal{A}$ in figs.
\ref{fig:Model2L+gvar} and \ref{fig:Model2L-gvar}.  Looking at the pure
confinement spectra on the left in each column, we find the centroids of the
$2\hbar\omega$, $3\hbar\omega$ and $4\hbar\omega$ band structures generally
positioned higher than in model $\mathcal{A}$. Moreover, we again observe the
confinement Dirac structure of model $\mathcal{B}$ inducing different
spin-orbit effects than that of model $\mathcal{A}$.  Similar to the nucleon
sector this leads to different intra-band splittings, level orderings and
configuration mixings in each shell what finally implies a different effect of
't~Hooft's force on the states. All in all, the 't~Hooft couplings $g_{nn}$
and $g_{ns}$ as chosen to reproduce the positions of flavor octet
ground-states (here the $\Lambda(1116,\mbox{****})$) are not sufficiently
large to account for the masses of several comparatively low-lying states in
each shell.  In this respect, the largest discrepancies to experiment show up
in the $\Lambda\frac{1}{2}^+$ sector, where the first scalar/isoscalar
excitation predicted appears far above the low-lying Roper-type resonance
$\Lambda\frac{1}{2}^+(1600,\mbox{***})$. In the $\Lambda\frac{1}{2}^-$ the
interplay of 't~Hooft's residual interaction with the relativistic effects of
the confinement force even produces a completely different hyperfine structure
of the three $1\hbar\omega$ states. Here the two first states predicted agree
with the second and third resonances observed, while there is no state that
could be associated with the low-lying
$\Lambda\frac{1}{2}^-(1405,\mbox{****})$. Instead, the third $1\hbar\omega$
state is predicted at roughly 2 GeV due to a strong upward shift of this state
by the attractive part of 't~Hooft's force that acts in the pseudo-scalar
diquark channel. It is interesting that this result would indeed allow for the
alternative interpretation of $\Lambda\frac{1}{2}^-(1405,\mbox{****})$ as an
additional $\bar K N$ bound state below the $\bar K N$ threshold, since all
other states of the $1\hbar\omega$ shell are reasonably well reproduced. But
in view of the otherwise rather poor results of model $\mathcal{B}$ we do not
take this alternative seriously.

\subsection{Summary for the $\Lambda$-spectrum}
To summarize our discussion of the $\Lambda$-sector, we presented our results
of models $\mathcal{A}$ and $\mathcal{B}$ for the complete $\Lambda$-resonance
spectrum. With the parameters being fixed all calculated states were true
parameter-free predictions which we compared with the presently available
experimental data. In fact we found excellent agreement between our
predictions of model $\mathcal{A}$ and the positions of the hitherto
experimentally observed resonances in the $\Lambda$-sector.  Moreover, we
again analyzed in detail the role of 't~Hooft's residual force for the
hyperfine structures of the excited $\Lambda$-spectrum and  similar to the
nucleon spectrum, we could convincingly demonstrate that also in the
strange $\Lambda$-sector instanton-induced effects in fact provide a
consistent and uniform explanation for almost all prominent features in the
lower as well as in the higher mass region of the spectrum.  Once the 't~Hooft
couplings are fixed to account for the correct position of the octet-ground
states (here the $\Lambda$-ground-state), several excited states can be simultaneously
well described, most of these even in completely quantitative agreement:
\begin{itemize}
\item We found an excellent description of the states
  $\Lambda\frac{1}{2}^+(1116,\mbox{****})$,
  $\Lambda\frac{5}{2}^+(1820,\mbox{****})$ and $\Lambda\frac{9}{2}^+(2350,
  \mbox{***})$ belonging to the positive parity $\Lambda$-Regge trajectory .
  The model yields the correct empirical Regge characteristics $M^2\sim J$
  with the quantitatively right slope of the trajectory. Once more we could
  demonstrate the non-trivial property of 't~Hooft's force to be compatible
  with the observed linear Regge characteristics.
\item The hyperfine intra-band structure of the positive-parity $2\hbar\omega$
  shell could be nicely reproduced. As in the $2\hbar\omega$ shell of the nucleon
  spectrum the analogous pattern of four comparatively low lying states is
  explained due to a selective lowering by 't~Hooft's force: Similar to the
  Roper state model $\mathcal{A}$ likewise accounts for the strikingly
  low position of the strange partner of the Roper resonance, the
  $\Lambda\frac{1}{2}^+(1600, \mbox{***})$. Moreover, also the model states
  associated to the three other low-lying resonances
  $\Lambda\frac{1}{2}^+(1810, \mbox{***})$,
  $\Lambda\frac{3}{2}^+(1890,\mbox{****})$ and
  $\Lambda\frac{5}{2}^+(1820,\mbox{****})$ are predicted close to the
  experimental resonance positions.
\item The intra-band structure of the negative-parity $1\hbar\omega$ shell
  could be likewise explained by 't~Hooft's force. In the upper part of this
  shell the model states fit quantitatively the structure of the four
  well-established states $\Lambda\frac{1}{2}^-(1670,\mbox{****})$,
  $\Lambda\frac{1}{2}^-(1800,\mbox{***})$,
  $\Lambda\frac{3}{2}^-(1690,\mbox{****})$ and $\Lambda\frac{5}{2}^-(1830,\mbox{****})$.
  These states in fact turned out to be dominantly flavor octet and hence
  could be identified as the octet counterparts of the $1\hbar\omega$ nucleon
  states. Moreover 't~Hooft's force provides an explanation for the position
  of the dominantly flavor singlet states below the dominantly flavor octet
  states. The position of the $\Lambda\frac{3}{2}^-(1520,\mbox{****})$ could
  be nicely explained, however, the first $\Lambda\frac{1}{2}^-$ state turned
  out to be degenerate with the $\Lambda\frac{3}{2}^-(1520,\mbox{****})$ and
  thus the notorious problem in quark models to explain the exceptionally large
  splitting between the $\Lambda\frac{1}{2}^-(1405,\mbox{****})$ and
  $\Lambda\frac{3}{2}^-(1520,\mbox{****})$ unfortunately remains unresolved also
  in our fully relativistic approach.
\item Due to 't~Hooft's force we again found overlapping substructures of
  shells with opposite parity leading to the occurrence of approximately
  degenerate states with the same spin but opposite parity in the same manner as
  in the nucleon spectrum. In this way our model is able to account for
  approximate parity doublets observed in the $\Lambda$-spectrum, {\it e.g.} 
  $\Lambda\frac{5}{2}^+(1820,\mbox{****})$--$\Lambda\frac{5}{2}^-(1830,\mbox{****})$
  and $\Lambda\frac{7}{2}^+(2020,\mbox{*})$--$\Lambda\frac{7}{2}^-(2100,\mbox{****})$. 
\end{itemize} 
Concerning the results of model $\mathcal{B}$ we found similar shortcomings as
discussed in detail for the nucleon spectrum in ref. \cite{Loe01b}. But, in addition, the centroids
of the band-structures are generally predicted too high showing that the
confinement ansatz $\mathcal{B}$ works even less well in this strange flavor
sector.\\

At the end of this section we should finally mention that the fully
relativistic treatment of the quark dynamics within our covariant Salpeter 
framework of model $\mathcal{A}$ leads again to large improvements
of the results as compared to the corresponding non-relativistic quark model
of \cite{BBHMP90,Met92} which employed instanton-induced forces as well. Although
similar effects of 't~Hooft's force likewise emerged in this non-relativistic
version, the lowering of particular $\Lambda$-states due to the scalar
diquark correlations were in general too small to account quantitatively for
prominent features as {\it e.g.} the low position of the Roper partner
$\Lambda\frac{3}{2}^+(1600,\mbox{***})$. Positive-parity excited states tended
to be too massive by roughly 200 MeV.  All in all the results of
ref. \cite{BBHMP90,Met92} for the $\Lambda$-spectrum are rather similar to those of our
inferior model $\mathcal{B}$.\\

Let us now continue
our investigations with the discussion of the strange
$\Sigma$ baryons ($S^*=-1$, $T=1$), where in contrast to the $\Lambda$-states
't~Hooft's force acts only in the non-strange-strange diquark channel but is
absent for the flavor-symmetric ($T=1$) non-strange diquarks.

\section{The $\Sigma$-resonance spectrum}
\label{sec:Sig}
In this section we will discuss our predictions for the excited
$\Sigma$-baryons with strangeness $S^*=-1$ and isospin $T=1$ and compare our
results with the corresponding experimental data quoted by the Particle Data Group
\cite{PDG00}. Again let us start with some general remarks concerning the
effects of 't~Hooft's force expected in this flavor sector.
\subsection{Remarks -- Implications of 't~Hooft's force and the experimental situation}
The $\Sigma$ baryons have the same flavor content of two non-strange quarks and one
strange quark as the $\Lambda$ baryons. But in contrast to the $\Lambda$ states
the non-strange quarks form symmetric isovector ($T=1$) quark pairs and
therefore 't~Hooft's force does not act in the non-strange diquark channel but 
exclusively in the flavor-antisymmetric non-strange-strange diquark channel. 
For the $\Sigma$-baryons the flavor wave functions can be combined to the mixed symmetric octet
representations $\Sigma_{\bf 8}$ corresponding to the non-strange $N$ states and to the
totally symmetric flavor decuplet  representations $\Sigma_{\bf 10}$ which correspond
to the non-strange $\Delta$ states.      
The positive and negative energy components
of the Salpeter amplitude $\Phi_{J^\pi}^\Sigma$ describing an excited
$\Sigma$-state with spin and parity $J^\pi$ are obtained by the embedding
map (see ref. \cite{Loe01a})
\begin{equation}
\label{embedSalp_Sig}
\Phi_{J^\pi}^\Sigma \;=\; T^{+++} {\varphi_{J^\pi}^\Sigma} \;+\; T^{---} {\varphi_{J^{-\pi}}^\Sigma}
\end{equation}
of totally $S_3$-symmetric Pauli spinors $\varphi^\Sigma_{J^{\pi}}$ and
$\varphi^\Sigma_{J^{-\pi}}$, which then in general can be decomposed into the following six
different spin-flavor $SU(6)$-configurations:
\begin{eqnarray}
\label{Sig_decompPauli}
\ket{\varphi^\Sigma_{J^{\pm}}} 
&=&
\phantom{+\;}
\ket{\Sigma\;J^\pm,\; ^2 8[56]}
\;+\;
\ket{\Sigma\;J^\pm,\;^2 8[70]}
\;+\;
\ket{\Sigma\;J^\pm,\;^4 8[70]}
\;+\;
\ket{\Sigma\;J^\pm,\;^2 8[20]}\nn
& &
+\;
\ket{\Sigma\;J^\pm,\;^4 10[56]}
\;+\;
\ket{\Sigma\;J^\pm,\;^2 10[70]},
\end{eqnarray}
with the four flavor octet contributions
\begin{equation}
\label{Sig8_configPauli}
\begin{array}{rcl}
\ket{\Sigma\;J^\pm,\; ^2 8[56]} &:=& 
\sum\limits_{L} \Bigg[ 
\ket{\psi^{L\;\pm}_\mathcal{S}}\tens
\frac{1}{\sqrt 2}\bigg(
\ket {\chi^\frac{1}{2}_{\mathcal{M}_\mathcal{A}}}
\tens
\ket {\phi^\Sigma_{\mathcal{M}_\mathcal{A}}}
+
\ket {\chi^\frac{1}{2}_{\mathcal{M}_\mathcal{S}}}
\tens
\ket {\phi^\Sigma_{\mathcal{M}_\mathcal{S}}}
\bigg)\Bigg]^J,\\[3mm]
\ket{\Sigma\;J^\pm,\;^2 8[70]} &:=& 
\sum\limits_{L} \Bigg[\phantom{+\;}\frac{1}{2}\; 
\ket{\psi^{L\;\pm}_{\mathcal{M}_\mathcal{A}}}\tens 
\bigg(
\ket {\chi^\frac{1}{2}_{\mathcal{M}_\mathcal{A}}}
\tens
\ket {\phi^\Sigma_{\mathcal{M}_\mathcal{S}}}
+
\ket {\chi^\frac{1}{2}_{\mathcal{M}_\mathcal{S}}}
\tens
\ket {\phi^\Sigma_{\mathcal{M}_\mathcal{A}}}
\bigg)\\[-2mm]
&& 
\phantom{\sum\limits_{L} \Bigg[}+
\frac{1}{2}\;
\ket{\psi^{L\;\pm}_{\mathcal{M}_\mathcal{S}}}\tens
\bigg(
\ket {\chi^\frac{1}{2}_{\mathcal{M}_\mathcal{A}}}
\tens
\ket {\phi^\Sigma_{\mathcal{M}_\mathcal{A}}}
-
\ket {\chi^\frac{1}{2}_{\mathcal{M}_\mathcal{S}}}
\tens
\ket {\phi^\Sigma_{\mathcal{M}_\mathcal{S}}}\bigg)\Bigg]^J,\\[3mm]
\ket{\Sigma\;J^\pm,\;^4 8[70]} &:=& 
\sum\limits_{L} \Bigg[\frac{1}{\sqrt 2} 
\bigg(
\ket{\psi^{L\;\pm}_{\mathcal{M}_\mathcal{A}}}\tens 
\ket{\chi^\frac{3}{2}_\mathcal{S}}\tens  
\ket{\phi^\Sigma_{\mathcal{M}_\mathcal{A}}}
-
\ket{\psi^{L\;\pm}_{\mathcal{M}_\mathcal{S}}}\tens 
\ket{\chi^\frac{3}{2}_\mathcal{S}}\tens  
\ket{\phi^\Sigma_{\mathcal{M}_\mathcal{S}}}
\bigg)\Bigg]^J,\\[3mm]
\ket{\Sigma\;J^\pm,\;^2 8[20]} &:=&  
\sum\limits_{L} \Bigg[\ket{\psi^{L\;\pm}_\mathcal{A}}\tens
\frac{1}{\sqrt 2}\bigg(
\ket {\chi^\frac{1}{2}_{\mathcal{M}_\mathcal{A}}}
\tens
\ket {\phi^\Sigma_{\mathcal{M}_\mathcal{S}}}
-
\ket {\chi^\frac{1}{2}_{\mathcal{M}_\mathcal{S}}}
\tens
\ket {\phi^\Sigma_{\mathcal{M}_\mathcal{A}}}
\bigg)\Bigg]^J,
\end{array}
\end{equation}
and the two flavor decuplet contributions
\begin{equation}
\label{Sig10_configPauli}
\begin{array}{rcl}
\ket{\Sigma\;J^\pm,\;^4 10[56]} &:=&  
\sum\limits_{L} \Bigg[
\ket{\psi^{L\;\pm}_\mathcal{S}}
\tens
\ket {\chi^\frac{3}{2}_\mathcal{S}}\Bigg]^J
\tens
\ket {\phi^\Sigma_\mathcal{S}},\\[3mm]
\ket{\Sigma\;J^\pm,\;^2 10[70]} &:=&  
\sum\limits_{L} \Bigg[
\frac{1}{\sqrt 2}\bigg(
\ket{\psi^{L\;\pm}_{\mathcal{M}_\mathcal{S}}}
\tens
\ket {\chi^\frac{1}{2}_{\mathcal{M}_\mathcal{S}}}
+
\ket{\psi^{L\;\pm}_{\mathcal{M}_\mathcal{A}}}
\tens
\ket {\chi^\frac{1}{2}_{\mathcal{M}_\mathcal{A}}}
\bigg)
\Bigg]^J
\tens
\ket {\phi^\Sigma_\mathcal{S}}.
\end{array}
\end{equation}
Here we used the same notation for the spatial, spin and flavor wave functions
as in the preceding section. For each shell the constituent quark model thus
predicts the same number of states as in the combined spectrum of $N$- and
$\Delta$-states. Due to the flavor-$SU(3)$ symmetry breaking
quark mass difference $m_s-m_n>0$, the decuplet and octet states
mix. Since the instanton induced force acts on flavor-antisymmetric 
quark pairs only, it does not affect the totally symmetric flavor
decuplet contributions $^4 10[56]$ and $^2 10[70]$ (similar to the non-strange
$\Delta$ states). But 't~Hooft's force again affects the flavor octet
contributions in the same manner as for the nucleon states: From
the strong selection rules of 't~Hooft's force the states with dominant $^4
8[70]$ and $^2 8[20]$ spin-flavor $SU(6)$ contributions are expected to be
hardly influenced, whereas the dominantly $^2 8[56]$ and $^2 8[70]$ states
shift downward due to the attractive scalar
non-strange-strange diquark correlation within these states. Apart from the
mixing of flavor-octet and flavor-decuplet configurations owing to the
flavor-$SU(3)$ symmetry breaking effects from the mass difference of the
non-strange and strange quark masses, we nonetheless anticipate a spectrum of
overlapping dominantly flavor-decuplet and dominantly flavor-octet states,
which by themselves form similar intra-band structures as their corresponding
counterparts in the non-strange $\Delta$- and $N$- spectra, respectively.
In particular, we expect 't~Hooft's force generating hyperfine splittings of
the dominantly octet states with the same systematics as observed for the
nucleon states (and the dominantly flavor-octet $\Lambda$ states).  But note
in this respect the following substantial differences to the $N$ and
$\Lambda$ spectra: The instanton induced hyperfine splittings, which here
arise solely from the non-strange-strange diquark correlation, are expected to
be considerably smaller and therefore, these structures generally might be
hidden according to the overlapping
flavor-decuplet states.\\
This might be the reason why also the experimental situation concerning the
intra-band splittings is rather unclear and inconclusive. In contrast to the
$N$- and $\Lambda$- sectors the $\Sigma$ sector lacks well-established
experimental data. Many resonances in the lower energy regions of the
$1\hbar\omega$ and $2\hbar\omega$ bands are only poorly established. The
clearest evidence for a hyperfine structure with an equivalent in the $N$- and
$\Lambda$-spectrum is the low-lying three-star resonance
$\Sigma\frac{1}{2}^+(1660,\mbox{***})$ as the counterpart to the Roper 
resonance $N\frac{1}{2}^+(1440,\mbox{****})$ and its partner $\Lambda\frac{1}{2}^+(1600,\mbox{***})$ .  Altogether there are
(apart from the $\Sigma\frac{1}{2}^+$ and $\Sigma^*\frac{3}{2}^+$
ground-states) only four well-established resonances with four-star rating,
{\it i.e.} the $\Sigma\frac{5}{2}^+(1915,\mbox{****})$ and the
$\Sigma\frac{7}{2}^+(2030,\mbox{****})$ in the positive-parity sector, as well
as the $\Sigma\frac{3}{2}^-(1670,\mbox{****})$ and the
$\Sigma\frac{5}{2}^-(1775,\mbox{****})$ in the negative-parity sector.  It
should be noted here that, unlike the $N$- and $\Lambda$-sector, the two lowest 
states in the $J^\pi=\frac{5}{2}^\pm$ sectors do {\it not} form a parity doublet
structure and moreover, there are no strong evidences for parity
doublets in general.
\subsection{Discussion of the complete $\Sigma$-spectrum}
\begin{figure}[!h]
  \begin{center}
    \epsfig{file={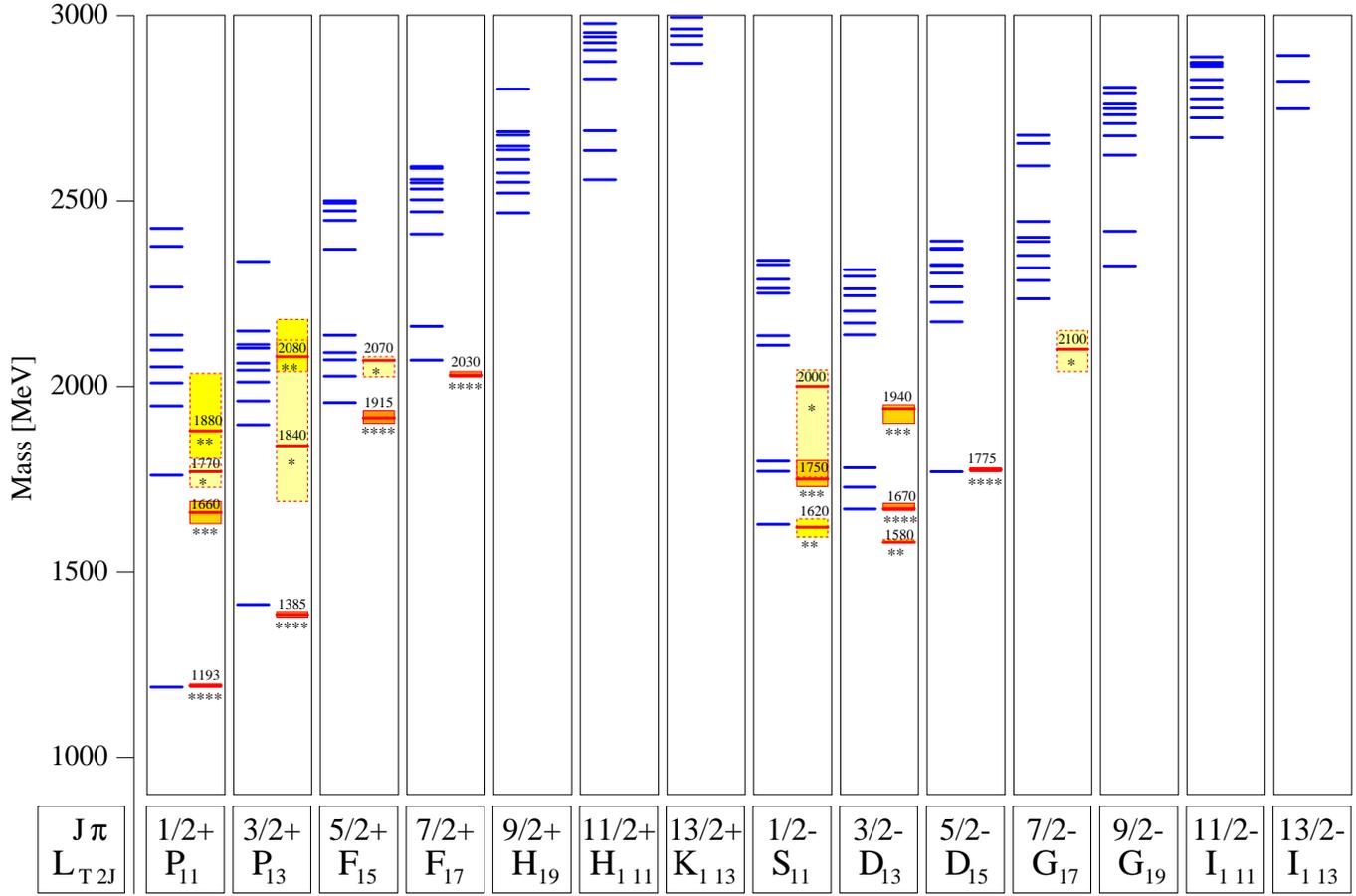},width=180mm}
  \end{center}
\caption{The predicted positive- and negative-parity
\textbf{$\Sigma$-resonance spectrum} with isospin $T=1$ and
strangeness $S^* = -1$ in \textbf{model $\mathcal{A}$} (left part of each column)
in comparison to the experimental spectrum taken from Particle Data Group 
\cite{PDG00} (right part of each
column).  The resonances are classified by the total spin
$J$ and parity $\pi$. The experimental resonance position is indicated
by a bar, the corresponding uncertainty by the shaded box, which is
darker for better established resonances; the status of each
resonance is additionally indicated by stars.}
\label{fig:SigM2}
\end{figure}

\begin{figure}[!h]
  \begin{center}
    \epsfig{file={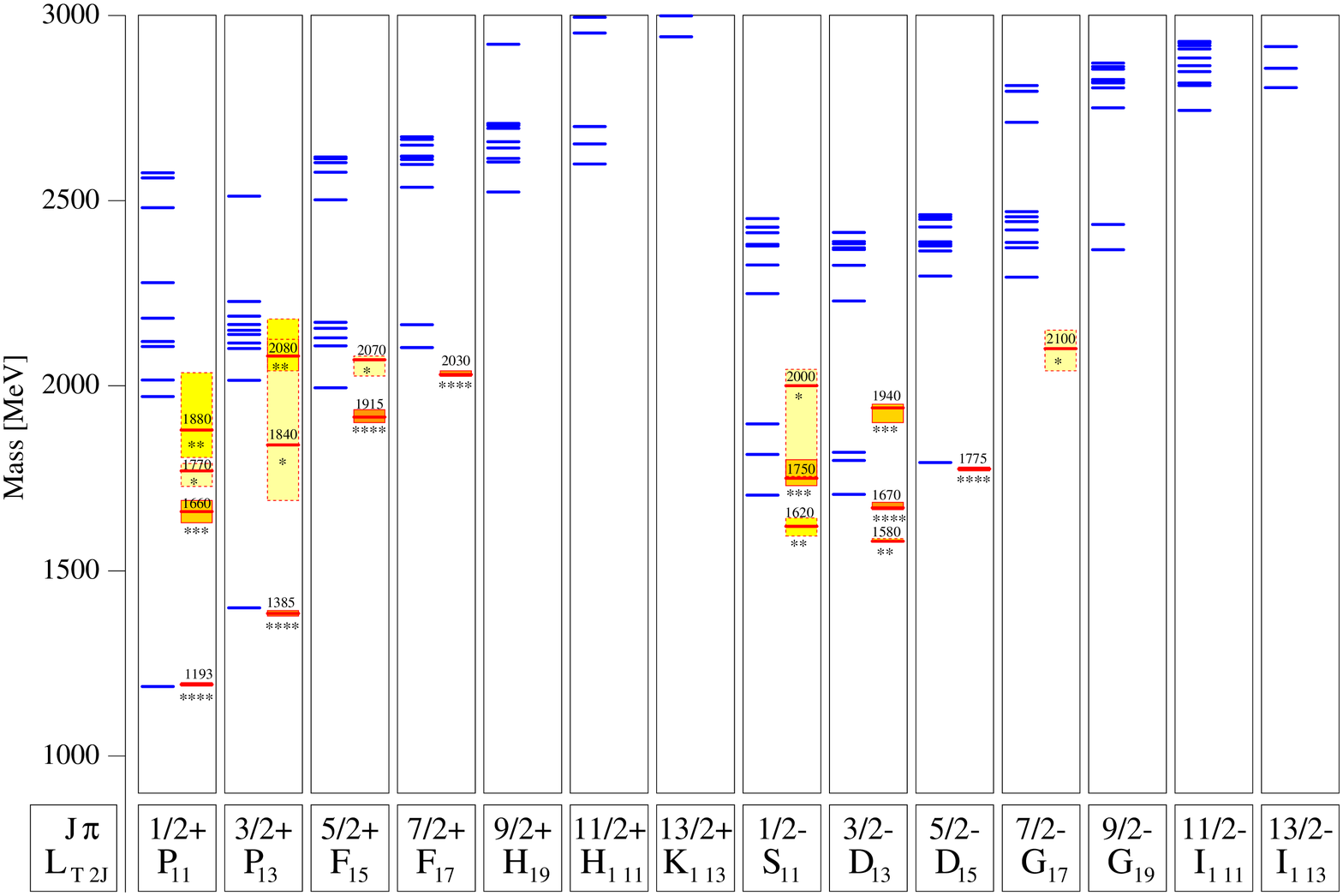},width=180mm}
  \end{center}
\caption{The predicted positive- and negative-parity
\textbf{$\Sigma$-resonance spectrum} with isospin $T=1$ and
strangeness $S^* = -1$ in \textbf{model $\mathcal{B}$} (left part of each column)
in comparison to the experimental spectrum taken from Particle Data Group \cite{PDG00} (right part of each
column).  The resonances are classified by the total spin
$J$ and parity $\pi$. See also caption to fig. \ref{fig:SigM2}.}
\label{fig:SigM1}
\end{figure}

Figures \ref{fig:SigM2} and \ref{fig:SigM1} show our predictions for the
$\Sigma$ spectrum in models $\mathcal{A}$ and $\mathcal{B}$, respectively.
These are compared with currently available experimental data as quoted by the
Particle Data Group \cite{PDG00}. As before, the states depicted in each
column are classified by their total spin and parity $J^\pi$. For each sector
the predictions for at most ten radial excitations are shown on the left hand
side of each column. The experimental $\Sigma$-resonance positions are
displayed on the right, where we use the same notation concerning the status
and the uncertainty of each resonance as before. For both parities figs. \ref{fig:SigM2} and
\ref{fig:SigM1} show our predictions for spins up to $J=\frac{13}{2}$. In
addition, the calculated masses of positive- and negative-parity states 
are given explicitly in tables \ref{tab:S2hwband},
\ref{tab:S1hwband} and \ref{tab:S3hwband}.

\subsubsection{Positive-parity excited $\Sigma$ states}
All observed positive-parity excited $\Sigma$ baryons quoted by the
Particle Data Group \cite{PDG00} lie in the energy region between 1600
and 2100 MeV and posses total spins from $J^\pi=\frac{1}{2}^+$ to
$\frac{7}{2}^+$. Hence, they all should belong to the positive-parity
$2\hbar\omega$ band.  There are no candidates with established quantum
numbers for the higher lying $4\hbar\omega$ or even $6\hbar\omega$ bands.  Our
predictions in models $\mathcal{A}$ and $\mathcal{B}$ for $\Sigma$ states of
the $2\hbar\omega$ shell are summarized in table \ref{tab:S2hwband}, where the
assignment to observed states again is made according to a comparison of the
predicted and experimentally determined resonance positions. 
\begin{table}[!h]
\center
\begin{tabular}{ccccccc}
\hline
Exp. state   &PW&${J^\pi}$        & Rating     & Mass range [MeV]& Model state &  Model state \\
\cite{PDG00} &  &                      &       & \cite{PDG00}    & in model $\mathcal{A}$& in model $\mathcal{B}$  \\
\hline
$\Sigma(1660)$&$P_{11}$&${\frac{1}{2}^+}$& ***    &1630-1690  &$\MSS(1,+,2,1760)$   &$\MSS(1,+,2,1971)$\\
$\Sigma(1760)$&$P_{11}$&${\frac{1}{2}^+}$& *      &1728-1790  &                     &              \\
$\Sigma(1880)$&$P_{11}$&${\frac{1}{2}^+}$& **     &1806-2035  &$\MSS(1,+,3,1947)$   &$\MSS(1,+,3,2015)$\\
              &        &                 &        &           &$\MSS(1,+,4,2009)$   &$\MSS(1,+,4,2106)$\\
              &        &                 &        &           &$\MSS(1,+,5,2052)$   &$\MSS(1,+,5,2119)$\\ 
              &        &                 &        &           &$\MSS(1,+,6,2098)$   &$\MSS(1,+,6,2182)$\\
              &        &                 &        &           &$\MSS(1,+,7,2138)$   &$\MSS(1,+,7,2278)$\\                     
\hline
$\Sigma(1840)$&$P_{13}$&${\frac{3}{2}^+}$& *      &1690-2125  &$\MSS(3,+,2,1896)$   &$\MSS(3,+,2,2014)$\\[1mm]
              &        &                 &        &           &$\MSS(3,+,3,1961)$   &$\MSS(3,+,3,2100)$\\
              &        &                 &        &           &$\MSS(3,+,4,2011)$   &$\MSS(3,+,4,2115)$\\
              &        &                 &        &           &$\MSS(3,+,5,2044)$   &$\MSS(3,+,5,2138)$\\
$\Sigma(2080)$&$P_{13}$&${\frac{3}{2}^+}$& **     &2040-2180  &$\MSS(3,+,6,2062)$   &$\MSS(3,+,6,2149)$\\
              &        &                 &        &           &$\MSS(3,+,7,2103)$   &$\MSS(3,+,7,2165)$\\
              &        &                 &        &           &$\MSS(3,+,8,2112)$   &$\MSS(3,+,8,2188)$\\
              &        &                 &        &           &$\MSS(3,+,9,2149)$   &$\MSS(3,+,9,2227)$\\
\hline
$\Sigma(1915)$&$F_{15}$&${\frac{5}{2}^+}$& ****   &1900-1935  &$\MSS(5,+,1,1956)$   &$\MSS(5,+,1,1994)$\\[1mm]
              &        &                 &        &           &$\MSS(5,+,2,2027)$   &$\MSS(5,+,2,2107)$\\
$\Sigma(2070)$&$F_{15}$&${\frac{5}{2}^+}$& *      &2026-2080  &$\MSS(5,+,3,2071)$   &$\MSS(5,+,3,2129)$\\
              &        &                 &        &           &$\MSS(5,+,4,2091)$   &$\MSS(5,+,4,2155)$\\
              &        &                 &        &           &$\MSS(5,+,5,2138)$   &$\MSS(5,+,5,2171)$\\
\hline
$\Sigma(2030)$&$F_{17}$&${\frac{7}{2}^+}$& ****   &2025-2040  &$\MSS(7,+,1,2070)$   &$\MSS(7,+,1,2103)$\\[1mm]
              &        &                 &        &           &$\MSS(7,+,2,2161)$   &$\MSS(7,+,2,2165)$\\
\hline
\end{tabular}
\caption{Calculated positions of $\Sigma$ states assigned to the positive parity $2\hbar\omega$ shell
  in comparison to the corresponding experimental mass values taken from
  \cite{PDG00}. Notation as in table \ref{tab:L2hwband}.}
\label{tab:S2hwband}
\end{table}
\begin{table}[!h]
\footnotesize
\begin{center}
\begin{tabular}[h]{|c||c|c|cccc|cc|}
\hline
$J^\pi$&
Model state & 
pos. & 
$\!\!{}^2 8[56]\!\!$&
$\!\!{}^2 8[70]\!\!$&
$\!\!{}^4 8[70]\!\!$&
$\!\!{}^2 8[20]\!\!$&
$\!\!{}^4 10[56]\!\!$&
$\!\!{}^2 10[70]\!\!$\\[1mm]
&
in model $\mathcal{A}$&
neg. & 
$\!\!{}^2 8[56]\!\!$&
$\!\!{}^2 8[70]\!\!$&
$\!\!{}^4 8[70]\!\!$&
$\!\!{}^2 8[20]\!\!$&
$\!\!{}^4 10[56]\!\!$&
$\!\!{}^2 10[70]\!\!$\\
\hline
\hline
$\frac{1}{2}^+$      &$\MSS(1,+,1,1190)$ &   98.7 & {\bf \underline{    94.6}} &      3.9 &      0.0 &      0.0 &      0.0 &      0.1 \\
     & ground-state $\Sigma$     &    1.3 &      0.4 &      0.6 &      0.2 &      0.0 &      0.0 &      0.0\\
\hline
\hline
     &$\MSS(1,+,2,1760)$ &   98.8 & {\bf \underline{    96.1}} &      2.3 &      0.0 &      0.1 &      0.0 &      0.2 \\
     &      &    1.2 &      0.3 &      0.6 &      0.3 &      0.0 &      0.0 &      0.0 \\
\cline{2-9}
     &$\MSS(1,+,3,1947)$ &   99.1 &      6.9 & {\bf \underline{    88.4}} &      0.9 &      0.3 &      0.0 &      2.5 \\
     &      &    0.9 &      0.2 &      0.3 &      0.3 &      0.1 &      0.0 &      0.0 \\
\cline{2-9}
$\frac{1}{2}^+$      &$\MSS(1,+,4,2009)$ &   99.0 &      0.0 &      0.2 &      8.4 &      0.1 & {\bf \underline{    89.9}} &      0.4 \\
     &      &    1.0 &      0.0 &      0.0 &      0.1 &      0.0 &      0.5 &      0.3 \\
\cline{2-9}
     &$\MSS(1,+,5,2052)$ &   99.2 &      0.8 &      1.8 &      1.2 &      1.9 &      0.2 & {\bf \underline{    93.2}} \\
     &      &    0.8 &      0.0 &      0.0 &      0.0 &      0.0 &      0.2 &      0.5 \\
\cline{2-9}
     &$\MSS(1,+,6,2098)$ &   99.2 &      0.2 &      1.1 & {\bf \underline{    88.3}} &      0.4 &      8.5 &      0.6 \\
     &      &    0.8 &      0.0 &      0.0 &      0.7 &      0.0 &      0.0 &      0.0 \\
\cline{2-9}
     &$\MSS(1,+,7,2138)$ &   98.9 &      0.3 &      0.8 &      0.8 & {\bf \underline{    95.2}} &      0.0 &      1.9 \\
     &      &    1.1 &      0.0 &      0.4 &      0.5 &      0.1 &      0.0 &      0.0 \\
\hline
\hline
$\frac{3}{2}^+$     &$\MSS(3,+,1,1412)$ &   99.4 &      0.0 &      0.0 &      0.5 &      0.0 & {\bf \underline{    98.9}} &      0.0 \\
     & ground-state $\Sigma^*$     &    0.6 &      0.0 &      0.0 &      0.0 &      0.0 &      0.3 &      0.2 \\
\hline
\hline
     &$\MSS(3,+,2,1896)$ &   98.8 & {\bf \underline{    73.9}} &     22.2 &      0.6 &      0.1 &      0.0 &      2.0 \\
     &      &    1.2 &      0.4 &      0.3 &      0.5 &      0.0 &      0.0 &      0.0 \\
\cline{2-9}
     &$\MSS(3,+,3,1961)$ &   99.1 &      0.0 &      0.0 &      5.1 &      0.1 & {\bf \underline{    93.9}} &      0.1 \\
     &      &    0.9 &      0.0 &      0.0 &      0.1 &      0.0 &      0.5 &      0.3 \\
\cline{2-9}
     &$\MSS(3,+,4,2011)$ &   99.0 &      1.5 &      1.5 &     17.3 &      1.4 & {\bf \underline{    73.4}} &      4.0 \\
     &      &    1.0 &      0.0 &      0.1 &      0.2 &      0.0 &      0.5 &      0.2 \\
\cline{2-9}
$\frac{3}{2}^+$     &$\MSS(3,+,5,2044)$ &   99.0 &      2.4 &     18.9 &     28.0 &     10.2 &      7.1 & {\bf \underline{    32.5}} \\
     &      &    1.0 &      0.0 &      0.3 &      0.3 &      0.0 &      0.0 &      0.4 \\
\cline{2-9}
     &$\MSS(3,+,6,2062)$ &   99.1 &      4.6 &     19.5 & {\bf \underline{    61.6}} &      2.4 &      6.0 &      5.0 \\
     &      &    0.9 &      0.1 &      0.1 &      0.5 &      0.0 &      0.0 &      0.1 \\
\cline{2-9}
     &$\MSS(3,+,7,2103)$ &   99.1 &     10.2 &     37.0 & {\bf \underline{    40.8}} &      1.4 &      0.8 &      8.8 \\
     &      &    0.9 &      0.0 &      0.3 &      0.4 &      0.0 &      0.0 &      0.1 \\
\cline{2-9}
     &$\MSS(3,+,8,2112)$ &   99.1 &      4.0 &      1.7 &     25.0 &      5.9 &     15.9 & {\bf \underline{    46.6}} \\
     &      &    0.9 &      0.0 &      0.1 &      0.2 &      0.0 &      0.1 &      0.5 \\
\cline{2-9}
     &$\MSS(3,+,9,2149)$ &   99.0 &      0.3 &      0.5 &     19.8 & {\bf \underline{    77.1}} &      1.2 &      0.2 \\
     &      &    1.0 &      0.0 &      0.4 &      0.5 &      0.1 &      0.0 &      0.0 \\
\hline
\hline
     &$\MSS(5,+,1,1956)$ &   98.7 & {\bf \underline{    77.8}} &     18.2 &      0.2 &      0.0 &      0.0 &      2.5 \\
     &      &    1.3 &      0.4 &      0.4 &      0.4 &      0.1 &      0.0 &      0.0 \\
\cline{2-9}
     &$\MSS(5,+,2,2027)$ &   99.0 &      2.9 &      7.8 &     16.3 &      0.0 & {\bf \underline{    65.9}} &      6.0 \\
     &      &    1.0 &      0.0 &      0.1 &      0.1 &      0.1 &      0.5 &      0.2 \\
\cline{2-9}
$\frac{5}{2}^+$     &$\MSS(5,+,3,2071)$ &   98.9 &     14.0 & {\bf \underline{    72.0}} &      7.6 &      0.0 &      4.9 &      0.4 \\
     &      &    1.1 &      0.5 &      0.3 &      0.1 &      0.1 &      0.0 &      0.0 \\
\cline{2-9}
     &$\MSS(5,+,4,2091)$ &   99.1 &      0.1 &      2.4 & {\bf \underline{    53.9}} &      0.0 &      5.2 &     37.5 \\
     &      &    0.9 &      0.0 &      0.2 &      0.2 &      0.2 &      0.3 &      0.1 \\
\cline{2-9}
     &$\MSS(5,+,5,2138)$ &   99.1 &      2.0 &      0.3 &     21.4 &      0.0 &     22.8 & {\bf \underline{    52.7}} \\
     &      &    0.9 &      0.0 &      0.1 &      0.1 &      0.0 &      0.4 &      0.3 \\
\hline
\hline
$\frac{7}{2}^+$      &$\MSS(7,+,1,2070)$ &   99.0 &      0.0 &      0.0 &     29.4 &      0.0 & {\bf \underline{    69.6}} &      0.0 \\
     &      &    1.0 &      0.0 &      0.0 &      0.2 &      0.0 &      0.4 &      0.2 \\
\cline{2-9}
     &$\MSS(7,+,2,2161)$ &   99.2 &      0.0 &      0.0 & {\bf \underline{    70.0}} &      0.0 &     29.2 &      0.0 \\
     &      &    0.8 &      0.1 &      0.1 &      0.4 &      0.1 &      0.1 &      0.1 \\
\hline
\end{tabular}
\end{center}
\caption{Configuration mixing of positive-parity excited $\Sigma$ states in model
  $\mathcal{A}$ assigned to the $2\hbar\omega$ band. For comparison also
  the ground-states $\Sigma\frac{1}{2}^+$ and $\Sigma^*\frac{3}{2}^+$ are listed.}
\label{tab:conf_mix_Sig_2hw}
\end{table}

\begin{figure}[!h]
  \begin{center}
    \epsfig{file={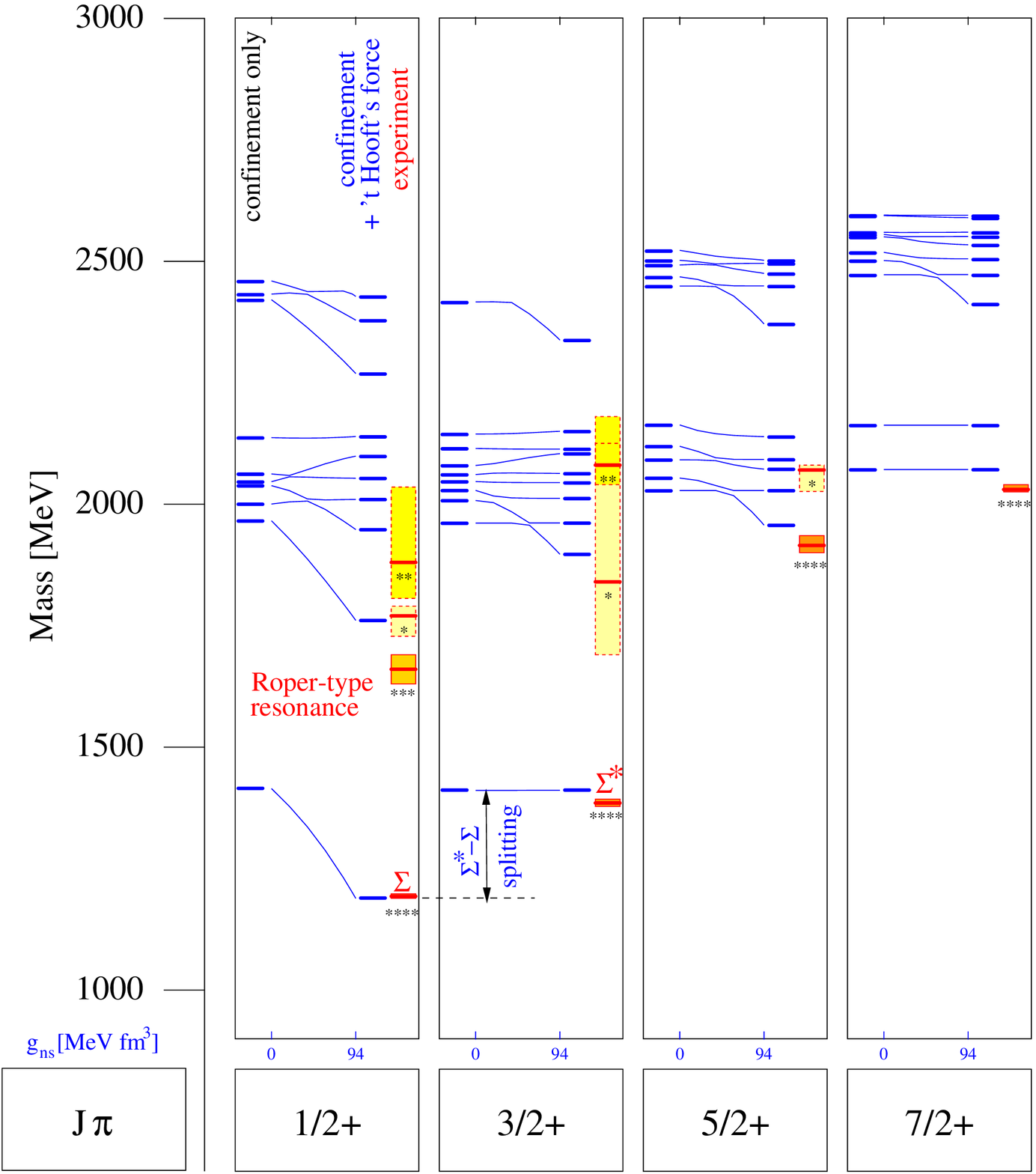},width=110mm}
  \end{center}
  \caption{Instanton-induced hyperfine splittings of the positive-parity
    $\Sigma$ states in model $\mathcal{A}$. Left in each column the spectrum
    from confinement alone is shown.  The curves show the change of the
    spectrum as function of the 't~Hooft coupling $g_{ns}$ which finally is
    fixed (right spectrum) from the hyperon splittings
    $\Sigma^*-\Sigma-\Lambda$ and $\Xi^*-\Xi$.  The rightmost spectrum shows for
    comparison the experimental data with their uncertainties.}
  \label{fig:Model2S+gvar}
\end{figure}

Unfortunately, the assignment here is far less clear than for the
corresponding $\Delta$, $N$ and $\Lambda$ states due to the lack
of well established experimental data in this flavor sector. Although
even more states are expected, less states have been seen by
experiments up to now and the quality of data is even worse. Actually, only
three states are established. These are the two well-established
four-star resonances $\Sigma\frac{5}{2}^+(1915,\mbox{****})$ and
$\Sigma\frac{7}{2}^+(2030,\mbox{****})$ and the three-star Roper-type 
resonance $\Sigma\frac{1}{2}^+(1660,\mbox{***})$. The few
remaining resonances have only one- and two-star ratings.
Consequently, apart from the Roper-type resonance, the
experimental situation concerning the intra-band splittings of the
$2\hbar\omega$ band is rather unclear and less apparent than in the nucleon-
and $\Lambda$-sector. But, in fact this is what we would anticipate
from the much weaker effects of the instanton induced force in this
flavor sector, where 't~Hooft's force acts exclusively in the scalar
non-strange-strange diquark channel with the weaker coupling $g_{ns}$.
In the $\Lambda$-sector, where both types of diquarks occur, we
could nicely demonstrate the fact that owing to flavor-$SU(3)$
symmetry breaking effects the influence of non-strange-strange diquark
correlations on the energy levels is significantly weaker than that
arising from correlations in the scalar non-strange diquark sector.
Consequently, those particular $\Sigma$ states that are lowered by 
't~Hooft's force do not become as clearly isolated
from the majority of unaffected, strongly clustered states as the
corresponding states in the nucleon- and $\Lambda$-spectra. Moreover,
the spectrum of unaffected energy levels is even richer according to
the additional flavor-decuplet states in the $\Sigma$-sector.
Therefore, we expect the hyperfine structures of the $\Sigma$ spectrum
much more difficult to resolve experimentally and from this point
of view the inferior quality of experimental data seems not surprising. 
The significantly weaker effect of 't~Hooft's force on the
positive-parity excited $\Sigma$ states in model $\mathcal{A}$ is
convincingly demonstrated in fig. \ref{fig:Model2S+gvar}.
 
The figure shows for each total spin $J$ the behavior of 
$\Sigma$ energy levels as a function of the 't~Hooft coupling
$g_{ns}$. The leftmost spectrum in each column is that obtained with the
confinement force of model $\mathcal{A}$ alone.  Then the 't~Hooft coupling $g_{ns}$ is
gradually increased up to its value $g_{ns}=94$ MeV fm$^3$ adjusted to
reproduce the hyperon splittings $\Sigma^*-\Sigma-\Lambda$ and $\Xi^*-\Xi$.
The right part of each column then depicts the resulting spectrum obtained
with the full dynamics in comparison to the experimental data.  Indeed, we
again observe the same systematics as found already in the nucleon- and
$\Lambda$-sectors (here compare to ref. \cite{Loe01b} and
fig. \ref{fig:Model2L+gvar}), namely the downward mass shift of
exactly four dominantly $^2 8[56]$ or $^2 8[70]$ states of the $2\hbar\omega$
shell in the sectors $J^\pi=\frac{1}{2}^+$, $\frac{3}{2}^+$ and
$\frac{5}{2}^+$. In analogy to the $N$- and $\Lambda$-sectors, the largest
effect is found for the Roper-like state in the
$J^\pi=\frac{1}{2}^+$ sector, whereas the almost equally large downward mass shift
of the three other states is comparatively weak.  On the one hand, the
weaker instanton induced hyperfine splittings are sufficiently large to
explain quantitatively some of the observed structures along with the well
reproduced $\Sigma\frac{3}{2}^+(1385,\mbox{****})$--$\Sigma\frac{1}{2}^+(1193,\mbox{****})$
ground-state splitting. This is nicely confirmed by the correctly described
mass splitting between the two well-established resonances
$\Sigma\frac{5}{2}^-(1915,\mbox{****})$ and
$\Sigma\frac{7}{2}^-(2030,\mbox{****})$. But on the other hand, the separation
of the lowered states relative to the bulk of unaffected states 
in fact is far less clear than in the $N$- and
$\Lambda$-sectors.  While in the $2\hbar\omega$ band of the $N$- and
$\Lambda$-spectrum the mass gap between the two split shell structures
amounts to roughly 200 MeV, it is here mostly not even 100 MeV.  Indeed, this
might explain the experimentally badly resolved structures especially in the $\frac{3}{2}^+$
sector. In the following discussion we shall investigate the situation for each
spin sector $J^\pi=\frac{1}{2}^+$, $\frac{3}{2}^+$, $\frac{5}{2}^+$ and
$\frac{7}{2}^+$ separately.  We restrict this detailed discussion to the more
realistic model $\mathcal{A}$. The contributions of the different spin-flavor
$SU(6)$-configurations to each $2\hbar\omega$ state in model $\mathcal{A}$ are
additionally tabulated in table \ref{tab:conf_mix_Sig_2hw}. This information
will be useful to identify hyperfine structures of $\Sigma$ states with
corresponding structures in the $\Delta$-, $N$- and $\Lambda$-sectors.

For the ${\bf \Sigma\frac{7}{2}^+}$ {\bf sector} with maximal total spin
$J=\frac{7}{2}$ in the $2\hbar\omega$ band our model predicts two states: the
lowest at 2070 MeV and another, roughly 90 MeV higher, at 2161 MeV.
Both states are totally unaffected by 't~Hooft's force and hence are
determined by the confinement kernel alone, as illustrated in fig.
\ref{fig:Model2S+gvar}. Although being predicted slightly too heavy by roughly
30-40 MeV, the lower-mass resonance can rather clearly be associated with the
well-established four-star resonance $\Sigma\frac{7}{2}^+(2030,\mbox{****})$.
It is dominantly a $^4 10[56]$ state ($\sim 70\%$) with a $30\%$ admixture of
a $^4 8[70]$ configuration and thus $\Sigma\frac{7}{2}^+(2030,\mbox{****})$
may be viewed as the flavor decuplet counterpart to the
$\Delta\frac{7}{2}^+(1950,\mbox{****})$ in the corresponding
$\Delta\frac{7}{2}^+$ sector. The second excited state turns out to be
dominantly $^4 8[70]$ ($\sim 70\%$) with a $30\%$ admixture of $^4 10[56]$.
This state, which corresponds to the $\Lambda\frac{7}{2}^+(2020,\mbox{*})$ and
the $N\frac{7}{2}^+(1990,\mbox{**})$ in the $\Lambda$- and nucleon spectrum,
respectively, is still missing in the experimental $\Sigma$-spectrum.\\
In the ${\bf \Sigma\frac{5}{2}^+}$ {\bf sector} we predict the existence of
five $2\hbar\omega$ states. Similar to the lowest states predicted in the
$N\frac{5}{2}^+$ and $\Lambda\frac{5}{2}^+$ sectors the lowest
$\Sigma\frac{5}{2}^+$ state exhibits a dominant $^2 8[56]$ contribution ($\sim
79\%$) with additional admixture ($19\%$) of $^2 8[70]$. As shown in fig.
\ref{fig:Model2S+gvar}, this state similarly reveals a downward mass shift by
't~Hooft's force of roughly 100 MeV and thus becomes the lowest state
predicted at 1956 MeV which may be clearly associated with the four-star
resonance $\Sigma\frac{5}{2}^+(1915,\mbox{****})$.  Thus, we identify this
resonance as the counterpart to the Regge states
$N\frac{5}{2}^+(1680,\mbox{****})$ and
$\Lambda\frac{5}{2}^+(1820,\mbox{****})$ in the $N$- and $\Lambda$-spectrum.
The mass difference to the first excitation predicted in $\Sigma\frac{7}{2}^+$
amounts to 114 MeV such that the experimental mass splitting of 115 MeV
between the $\Sigma\frac{5}{2}^+(1915,\mbox{****})$ and the
$\Sigma\frac{7}{2}^+(2030,\mbox{****})$ is quite well explained by 't~Hooft's
force.  The remaining four states in $\Sigma\frac{3}{2}^+$ are virtually
not affected by 't~Hooft's force and lie in a mass range between 2000 and 2140
MeV, {\it i.e.} around the one-star resonance
$\Sigma\frac{5}{2}^+(2070,\mbox{*})$ observed as second structure
in the $F_{15}$ partial wave.\\
The ${\bf \Sigma\frac{3}{2}^+}$ {\bf sector} is rather poorly explored. Only
two poorly established structures have been resolved in the $P_{13}$ partial
wave, one in a region around 1840 MeV, which is the one-star
$\Sigma\frac{5}{2}^+(1840,\mbox{*})$, and another one, the slightly better
established two-star resonance $\Sigma\frac{5}{2}^+(2080,\mbox{**})$.  The
observed situation roughly corresponds to that predicted for the eight states
expected in our model $\mathcal{A}$.  Two of the eight states are rather
low-lying at 1896 and 1961 MeV and possibly correspond to the
$\Sigma\frac{5}{2}^+(1840,\mbox{*})$. In analogy to the $\Sigma\frac{5}{2}^+$
sector, the lowest lying state at 1896 MeV is predicted to be dominantly $^2
8[56]$ ($\sim 75\%$)and hence its low position arises from a downward mass
shift by 't~Hooft's force as depicted in fig. \ref{fig:Model2S+gvar}. This
state is the expected flavor octet partner of the first excitations in the
corresponding $N\frac{3}{2}^+$ and $\Lambda\frac{3}{2}^+$ sectors, {\it i.e.}
the $N\frac{3}{2}^+(1720,\mbox{****})$ and the
$\Lambda\frac{3}{2}^+(1890,\mbox{****})$. The second state predicted at 1961
MeV, is dominantly $^4 10[56]$ ($\sim 95\%$) and hence remains totally
unaffected by 't~Hooft's force. But, similar to its $\Delta\frac{3}{2}^+$
counterpart, its comparatively low position arises due to rather strong
relativistic effects induced from the Dirac structure of the confinement
kernel $\mathcal{A}$ (here we refer to the discussion of the
$\Delta$-spectrum in ref. \cite{Loe01b}).  The remaining six
states cluster in a rather narrow mass range between 2010 and 2150 MeV
roughly corresponding to the range
of possible values quoted for the second resonance $\Sigma\frac{5}{2}^+(2080,\mbox{**})$.\\
Finally, let us focus to ${\bf \Sigma\frac{1}{2}^+}$ {\bf sector} with the
scalar/isoscalar excitations of the $\Sigma\frac{1}{2}^+$ ground-state.
Similar to the $N\frac{1}{2}^+$ sector two states are selectively lowered from
the other members of the $2\hbar\omega$ band, as demonstrated in fig.
\ref{fig:Model2S+gvar}. In fact, the systematics is exactly the same as in the
corresponding nucleon sector. Again, the almost pure ($96\%$) $^2 8[56]$
Roper-like state, which is already the lowest state in the pure confinement
spectrum, shows the strongest downward mass shift in fig.
\ref{fig:Model2S+gvar}.  Similar to the corresponding $N$- and $\Lambda$-Roper 
states this mass shift of roughly 200 MeV is as large as that of
the $\Sigma\frac{1}{2}^+$ ground-state. Hence this clearly isolated state
becomes the lowest radial excitation at 1760 MeV, and, although being
predicted roughly 100 MeV too high, this state should be associated with the
$\Sigma\frac{1}{2}^+(1660,\mbox{***})$, the counterpart to Roper
resonance $N\frac{1}{2}^+(1440,\mbox{****})$.  The second radial excitation,
which is dominantly a $^2 8[70]$ state shows a moderate mass shift of roughly
100 MeV and finally is predicted at 1947 MeV within the error range of
the two-star resonance $\Sigma\frac{1}{2}^+(1880,\mbox{**})$. Hence this
resonance is the expected counterpart to the low-lying nucleon resonance
$N\frac{1}{2}^+(1710,\mbox{***})$. There is a further comparatively low-lying
structure observed experimentally in the $P_{11}$ partial wave. This is the
only poorly established one-star resonance
$\Sigma\frac{1}{2}^+(1770,\mbox{*})$ which, however, does absolutely not fit
in the systematics observed so far in the $N$- and $\Lambda$-sector.  Note,
however, that this poorly established structure rests solely on only one
partial wave analysis which is in disagreement with most other analyses (see
ref. \cite{PDG00}). Therefore, the existence of this third low-lying structure is
highly questionable and will presumably be disproved by future experiments.
The remaining four $2\hbar\omega$ states expected in this sector form a
pattern of more or less equidistant states in a region between 2000 and 2140
where so far no structure could be experimentally resolved.

\subsubsection{Negative-parity excited $\Sigma$ states}
Let us now turn to the discussion of negative-parity states, where most of the
resonances observed so far should belong to the $1\hbar\omega$ band with spins
$J^\pi=\frac{1}{2}^-$, $\frac{3}{2}^-$ and $\frac{5}{2}^-$. There is only one
resonance with higher spin $J^\pi=\frac{7}{2}^-$, the
$\Lambda\frac{7}{2}^-(2100,\mbox{*})$ which definitely cannot be a member of
the $1\hbar\omega$ shell. However, its evidence is only poor. Table
\ref{tab:S1hwband} shows the predicted masses in models $\mathcal{A}$ and
$\mathcal{B}$ for the negative-parity states expected in the $1\hbar\omega$
band.  Where possible, the assignment of model states to resonances observed
is made according to a comparison of calculated and measured resonance
positions.
\begin{table}[!h]
\center
\begin{tabular}{ccccccc}
\hline
Exp. state   &PW&${J^\pi}$        & Rating     & Mass range [MeV]& Model state &  Model state \\
\cite{PDG00} &  &                      &            & \cite{PDG00}    & in model $\mathcal{A}$& in model $\mathcal{B}$  \\
\hline
$\Sigma(1620)$&$S_{11}$&${\frac{1}{2}^-}$& **     &1594-1643  &$\MSS(1,-,1,1628)$   &$\MSS(1,-,1,1704)$\\[1mm]
$\Sigma(1750)$&$S_{11}$&${\frac{1}{2}^-}$& ***    &1730-1800  &$\begin{array}{c}
                                                                \MSS(1,-,2,1771)\\
                                                                \MSS(1,-,3,1798)   
                                                                \end{array}$     &$\MSS(1,-,2,1814)$\\
$\Sigma(2000)$&$S_{11}$&${\frac{1}{2}^-}$& *      &1755-2044  &                  &$\MSS(1,-,3,1897)$\\
\hline
$\Sigma(1580)$&$D_{13}$&${\frac{3}{2}^-}$& **     &1578-1587  &                  &\\
$\Sigma(1670)$&$D_{13}$&${\frac{3}{2}^-}$& ****   &1665-1685  &$\MSS(3,-,1,1669)$&$\MSS(3,-,1,1706)$\\[1mm]
              &        &                 &        &           &$\MSS(3,-,2,1728)$&$\MSS(3,-,2,1798)$\\
              &        &                 &        &           &$\MSS(3,-,3,1781)$&$\MSS(3,-,3,1820)$\\
$\Sigma(1940)$&$D_{13}$&${\frac{3}{2}^-}$& ***    &1900-1950  &                  &\\
\hline
$\Sigma(1775)$&$D_{15}$&${\frac{5}{2}^-}$& ****   &1770-1780  &$\MSS(5,-,1, 1770)$   &$\MSS(5,-,1,1792)$ \\
\hline
\end{tabular}
\caption{Calculated positions of $\Sigma$ states assigned to the negative parity $1\hbar\omega$ shell
  in comparison to the corresponding experimental mass values taken from
  \cite{PDG00}. Notation as in table \ref{tab:L2hwband}.}
\label{tab:S1hwband}
\end{table}
\begin{table}[!h]
\begin{center}
\begin{tabular}[h]{|c||c|c|cccc|cc|}
\hline
$J^\pi$&
Model state &
pos. & 
$\!\!{}^2 8[56]\!\!$&
$\!\!{}^2 8[70]\!\!$&
$\!\!{}^4 8[70]\!\!$&
$\!\!{}^2 8[20]\!\!$&
$\!\!{}^4 10[56]\!\!$&
$\!\!{}^2 10[70]\!\!$\\[1mm]
&
in model $\mathcal{A}$&
neg. & 
$\!\!{}^2 8[56]\!\!$&
$\!\!{}^2 8[70]\!\!$&
$\!\!{}^4 8[70]\!\!$&
$\!\!{}^2 8[20]\!\!$&
$\!\!{}^4 10[56]\!\!$&
$\!\!{}^2 10[70]\!\!$\\
\hline
\hline
     &$\MSS(1,-,1,1628)$ &   98.6 &      5.4 & {\bf \underline{    87.4}} &      2.3 &      0.1 &      0.0 &      3.4 \\
     &      &    1.4 &      0.7 &      0.3 &      0.1 &      0.3 &      0.0 &      0.0 \\
\cline{2-9}
$\frac{1}{2}^-$     &$\MSS(1,-,2,1771)$ &   99.3 &      0.2 &      2.9 & {\bf \underline{    94.6}} &      0.2 &      0.3 &      1.1 \\
     &      &    0.7 &      0.0 &      0.0 &      0.5 &      0.1 &      0.0 &      0.0 \\
\cline{2-9}
     &$\MSS(1,-,3,1798)$ &   99.2 &      0.1 &      2.8 &      1.7 &      0.3 &      0.0 & {\bf \underline{    94.4}} \\
     &      &    0.8 &      0.0 &      0.0 &      0.0 &      0.0 &      0.4 &      0.3 \\
\hline
\hline
     &$\MSS(3,-,1,1669)$ &   98.9 &      5.1 & {\bf \underline{    89.0}} &      1.2 &      0.1 &      0.0 &      3.4 \\
     &      &    1.1 &      0.2 &      0.3 &      0.4 &      0.2 &      0.0 &      0.0 \\
\cline{2-9}
$\frac{3}{2}^-$     &$\MSS(3,-,2,1728)$ &   99.2 &      0.1 &      0.1 & {\bf \underline{    82.7}} &      0.1 &      0.2 &     16.0 \\
     &      &    0.8 &      0.1 &      0.2 &      0.3 &      0.1 &      0.0 &      0.1 \\
\cline{2-9}
     &$\MSS(3,-,3,1781)$ &   99.2 &      0.2 &      4.4 &     15.0 &      0.2 &      0.0 & {\bf \underline{    79.3}} \\
     &      &    0.8 &      0.0 &      0.1 &      0.1 &      0.0 &      0.1 &      0.5 \\
\hline
\hline
$\frac{5}{2}^-$     &$\MSS(5,-,1,1770)$ &   99.2 &      0.0 &      0.0 & {\bf \underline{    99.0}} &      0.0 &      0.2 &      0.0 \\
     &      &    0.8 &      0.0 &      0.2 &      0.5 &      0.1 &      0.0 &      0.0 \\
\hline
\end{tabular}
\end{center}
\caption{Configuration mixing of negative-parity $\Sigma$ states in model $\mathcal{A}$
assigned to the  $1\hbar\omega$ band.}
\label{tab:conf_mix_Sig_1hw}
\end{table}

Again let us restrict our detailed discussion to the better
results of model $\mathcal{A}$ depicted in fig. \ref{fig:SigM2}.  Table
\ref{tab:conf_mix_Sig_1hw} explicitly shows to what extent the different
spin-flavor $SU(6)$ configurations are contributing in model $\mathcal{A}$
to each state of the $1\hbar\omega$ band.

\begin{figure}[!h]
  \begin{center}
    \epsfig{file={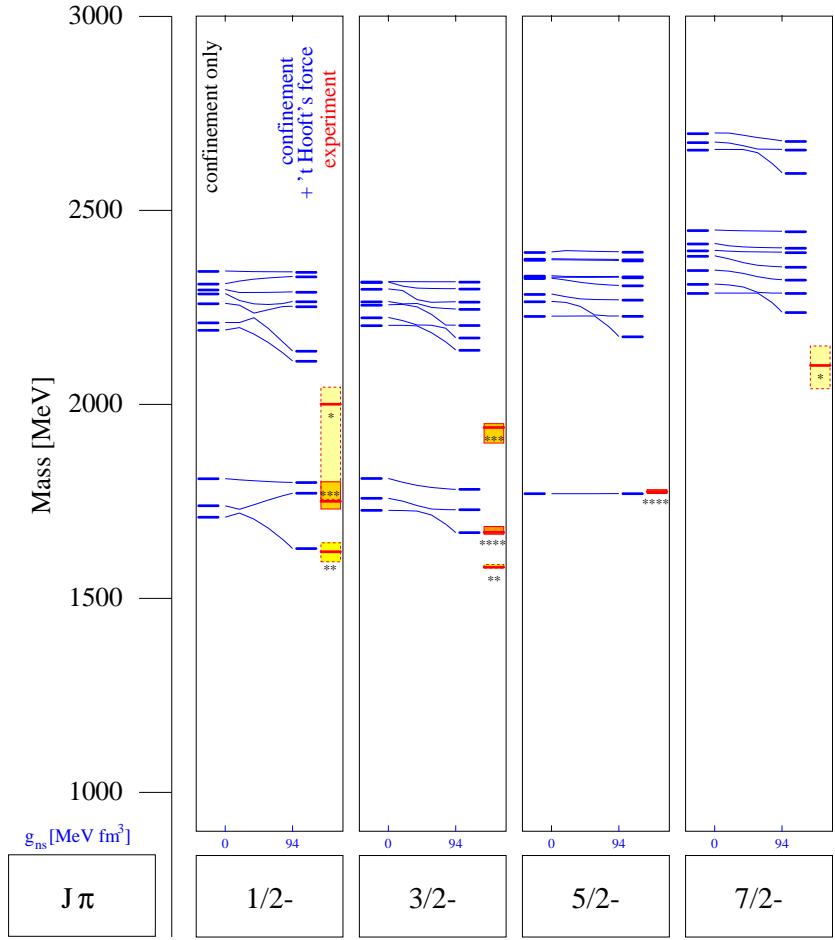},width=110mm}
  \end{center}
  \caption{Instanton-induced hyperfine splittings of the positive-parity $\Sigma$-states
    in model $\mathcal{A}$. See caption of fig. \ref{fig:Model2S+gvar} for further explanations.}
  \label{fig:Model2S-gvar}
\end{figure}

Unlike the situation for the corresponding negative-parity nucleon and
$\Lambda$-resonances, where a unique one-to-one correspondence between our
predictions and the observed states was readily apparent, the experimental
situation here is still quite inconclusive.  Apart from the two
four-star states $\Sigma\frac{3}{2}^-(1670,\mbox{****})$ and
$\Sigma\frac{5}{2}^-(1775,\mbox{****})$ most of the PDG quoted states
\cite{PDG00} are not established and quark models for baryons
\cite{IsKa78,CaIs86,BIL00,BBHMP90} 
in general have difficulties to account for some of the
less well established structures, {\it e.g.} the poorly determined two-star
resonance $\Sigma\frac{3}{2}^-(1580,\mbox{**})$.  In fact, as can be seen in
fig. \ref{fig:SigM2} and in table \ref{tab:S1hwband}, our model as well can
only partially explain the experimental features of the negative-parity
$\Sigma$-sector. Note, however, that the best established four-star states
$\Sigma\frac{3}{2}^-(1670,\mbox{****})$ and
$\Sigma\frac{5}{2}^-(1775,\mbox{****})$ are quite well described.  Figure
\ref{fig:Model2S-gvar} illustrates the effect of the instanton induced
interaction on the energy levels of several negative-parity $\Sigma$ states in
model $\mathcal{A}$.
As before, the figure demonstrates the change of the spectrum with increasing
't~Hooft coupling $g_{ns}$ which finally is fixed to account for the flavor
octet ground-state hyperons.  Note that the effects of 't~Hooft's force are
again significantly smaller than in the corresponding nucleon and $\Lambda$
sectors. But they are sufficiently large to provide a quantitatively correct
explanation for at least some of the observed hyperfine splittings in the
$1\hbar\omega$ shell, as {\it e.g.} that of the two best-established states
$\Sigma\frac{3}{2}^-(1670,\mbox{****})$ and
$\Sigma\frac{5}{2}^-(1775,\mbox{****})$.  In the following discussion we will
investigate in detail the situation for
each spin sector $\frac{1}{2}^-$, $\frac{3}{2}^-$ and $\frac{5}{2}^-$ in turn:\\

\begin{figure}[!h]
\begin{center}
\epsfig{file={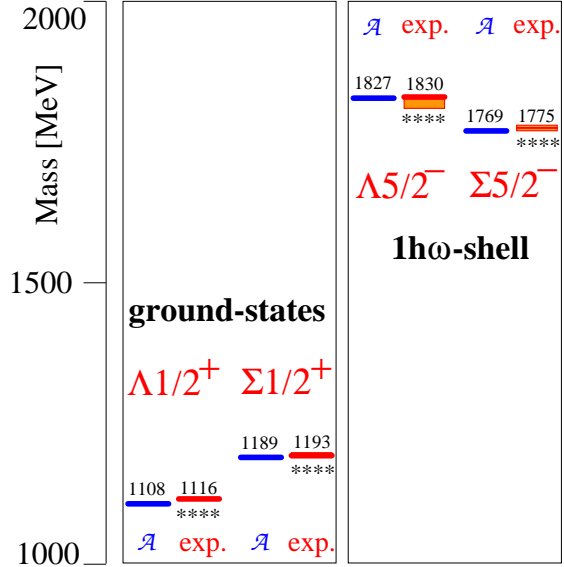},width=74mm}
\end{center}
\caption{Mass splittings between $\Lambda$ and $\Sigma$ states. See text for
  further explanation.}
\label{fig:LamSigcompM2}
\end{figure}

In the ${\bf \Sigma\frac{5}{2}^-}$ {\bf sector} our model $\mathcal{A}$
predicts a single $1\hbar\omega$ state at 1770 MeV which accurately accounts for the
single four-star resonance $\Sigma\frac{5}{2}^-(1775,\mbox{****})$ observed in
the $D_{15}$ partial wave.  Due to its maximally possible spin $J=\frac{5}{2}$
this state, which is predicted to be purely $^4 8[70]$, remains totally
unaffected by 't~Hooft's force as illustrated in fig.
\ref{fig:Model2S-gvar}. This state is the flavor octet partner of the
$\Lambda\frac{5}{2}^-(1830,\mbox{****})$ and the
$N\frac{5}{2}^-(1675,\mbox{****})$ in the $\Lambda$- and nucleon spectrum,
respectively. It is quite interesting to compare this result with that in 
the corresponding $\Lambda\frac{5}{2}^-$ sector where our model likewise
predicted a purely $^4 8[70]$ state in the $1\hbar\omega$ shell at 1828 MeV
that exactly matches the measured position of the
$\Lambda\frac{5}{2}^-(1830,\mbox{****})$.  But note the significant mass
difference of $\sim 55$ MeV between the $\Lambda\frac{5}{2}^-(1830,\mbox{****})$ and
the $\Sigma\frac{5}{2}^-(1775,\mbox{****})$ although both states have exactly
the same flavor content of two non-strange and one strange quark.  This mass
splitting is roughly of the same order of magnitude as the $\Sigma-\Lambda$
ground-state splitting which amounts to $\sim 75$ MeV, but it shows the reversed
order; see fig. \ref{fig:LamSigcompM2} where the mass splittings between
these $\Sigma$ and $\Lambda$ states (model $\mathcal{A}$ and experiment) is
graphically illustrated. 
Due to the accurate predictions for both $J^\pi=\frac{5}{2}^-$ resonances this
mass splitting is remarkably well reproduced with 58 MeV in our model
$\mathcal{A}$. But, unlike the $\Sigma-\Lambda$ ground-state splitting which
originates from 't~Hooft's interaction due to the different 't~Hooft couplings
$g_{nn}>g_{ns}$, the reversed $\Lambda\frac{5}{2}^-$--$\Sigma\frac{5}{2}^-$
splitting is not an instanton induced effect, since both $J^\pi=\frac{5}{2}^-$
states are entirely determined by the confinement potential alone.  It is a
spin-independent effect arising due to the flavor $SU(3)$ breaking from the
non-strange and strange quark mass difference. This effect is qualitatively
understandable considering once again the naive non-relativistic oscillator
model, where the coordinates are chosen such that the two non-strange quarks
move in the $\rho$-oscillator and the strange quark moves relative to the
non-strange quark pair in the $\lambda$-oscillator (see
\cite{IsKa78,CaRo00}). Then, due to the heavier strange quark mass, the
$\lambda$-frequency $\omega_\lambda$ is smaller than the $\rho$-frequency
$\omega_\rho$, namely $\omega_\lambda = \omega_\rho\;(\frac{2 m_n+m_s}{3
  m_s})^\frac{1}{2}< \omega_\rho$.  Owing to Pauli's principle the total $L=1$
wave functions must be symmetric under exchange of the two non-strange quarks.
Both states, $\Lambda\frac{5}{2}^-$ and $\Sigma\frac{5}{2}^-$, contain a
spin-quartet wave function which is symmetric under interchange of any two
quarks. But the crucial difference between the two states is their distinct
isospin: the flavor wave function of $\Lambda\frac{5}{2}^-\;(T=0)$ is
antisymmetric while that of $\Sigma\frac{5}{2}^-\;(T=1)$ is symmetric under
exchange of the two non-strange quarks. To achieve total symmetry the
$\Lambda\frac{5}{2}^-$ must be orbitally excited in the odd $\rho$-coordinate,
whereas $\Sigma\frac{5}{2}^-$ has to be orbitally excited in the even
$\lambda$-coordinate.  Since the $\rho$-oscillator has the higher frequency,
the $\Lambda\frac{5}{2}^-$ thus is heavier than the $\Sigma\frac{5}{2}^-$.
The accurate prediction of this mass difference in our model shows that the
quark mass difference $m_s-m_s$ fixed from the ground-state decuplet in fact
is correctly affecting the energy levels of the first excited
$\Lambda\frac{5}{2}^-$ and $\Sigma\frac{5}{2}^-$ states.  Moreover, it
provides good support for our confinement force which is chosen
flavor-independent.  Finally, it should be noted here that, in nice agreement
with our predictions, the lowest $\Sigma\frac{5}{2}^+$ state does {\it not}
form an approximate parity doublet structure together with the lowest
excitation in $\Sigma\frac{5}{2}^-$.  In this respect the $\Sigma$-spectrum
differs significantly from the nucleon- and $\Lambda$-spectrum which both
reveal a well-established parity doublet structure formed by the lowest
excitations in $N\frac{5}{2}^\pm$ and $\Lambda\frac{5}{2}^\pm$, respectively.
In the $N\frac{5}{2}^+$ and $\Lambda\frac{5}{2}^+$ sectors the effect of a
strong scalar {\it non-strange} diquark correlation was sufficiently large to
lower the first excitations deeply enough to become nearly degenerate with the
lowest excitations $N\frac{5}{2}^-$ and $\Lambda\frac{5}{2}^-$, respectively.
In consistency with the experimental findings
($\Sigma\frac{5}{2}^+(1915,\mbox{****})\leftrightarrow\Sigma\frac{5}{2}^-(1775,\mbox{****})$),
the effect of the scalar {\it non-strange-strange} diquark correlation in the
corresponding lowest $\Sigma\frac{5}{2}^+$ state is, however, too weak to generate
likewise a degeneracy for the lowest $\Sigma\frac{5}{2}^\pm$ states.

In the ${\bf \Sigma\frac{3}{2}^-}$ {\bf sector} the currently available data
of altogether three observed resonances significantly disagrees with our
predictions for the three states expected in the $1\hbar\omega$ shell.
However, the best established of these states, the four-star resonance
$\Sigma\frac{3}{2}^-(1670,\mbox{****})$, is exactly reproduced by the lowest
state predicted in this sector at 1669 MeV. This state turns out to be
dominantly $^2 8[70]$ ($\sim 90\%$) and, as shown in fig.
\ref{fig:Model2S-gvar}, it is slightly lowered by 't~Hooft's interaction.
Hence, it is the counterpart of $\Lambda\frac{3}{2}^-(1690,\mbox{****})$ and
$N\frac{3}{2}^-(1520,\mbox{****})$ in the $\Lambda$- and nucleon spectrum,
respectively. In addition to the $\Sigma\frac{3}{2}^-(1670,\mbox{****})$ there
are two further $D_{13}$ resonances extracted from experiment, one poorly
established two-star resonance $\Sigma\frac{3}{2}^-(1580,\mbox{**})$ lying
even below the $\Sigma\frac{3}{2}^-(1670,\mbox{****})$, and another with
three-star rating, the $\Sigma\frac{3}{2}^-(1940,\mbox{***})$, which lies
exceptionally high at almost 2 GeV.  However, the two remaining
$\Sigma\frac{3}{2}^-$ states predicted in the $1\hbar\omega$ band by no means
can account for these two states observed.  Our result is quite similar to
that of other constituent quark models \cite{IsKa78,CaIs86}.  The calculation
with model $\mathcal{A}$ yields two rather close states at 1728 and 1781 MeV.
The first one is dominantly ($83\%$) a $^4 8[70]$ state and is the expected
flavor octet partner of the $N\frac{3}{2}^-(1700,\mbox{***})$. The second,
dominantly $^2 10[70]$ state ($\sim 80\%$) is the expected flavor decuplet
counterpart of the $\Delta(1700,\mbox{****})$. The rather puzzling
experimental situation of the $1\hbar\omega$ states in the
$\Sigma\frac{3}{2}^-$ sector does absolutely not fit in the systematics
observed in the $N$- and $\Delta$-sector and in general is completely
incomprehensible in a potential model framework.  Concerning the
$\Sigma\frac{3}{2}^-(1940,\mbox{***})$ one could speculate whether it is a
member of the $3\hbar\omega$ band (see table \ref{tab:S3hwband}).  But even
then the measured position remains puzzling. Although 't~Hooft's force
selectively lowers two states of $3\hbar\omega$ band (see fig.
\ref{fig:Model2S-gvar}), the shift is much too weak to account for this state.
The lowest $\Sigma\frac{3}{2}^-$ state of the $3\hbar\omega$ shell is
predicted at 2139 MeV but the $\Sigma\frac{3}{2}^-(1940,\mbox{***})$ lies
exactly in between this lowest predicted $3\hbar\omega$ state and the highest
predicted $1\hbar\omega$ state.  It is worthwhile to mention here that not all
analyses of $\bar K N$ scattering data require this state (see \cite{PDG00}
and references therein). The evidence for the poorly determined
$\Sigma\frac{3}{2}^-(1580,\mbox{**})$ in any case is not compulsory.  Thus,
except for the well-established $\Sigma\frac{3}{2}^-(1670,\mbox{****})$, the
experimental situation in $\Sigma\frac{3}{2}^-$ is quite inconclusive and a
verification of these rather old data by new experiments would be highly
desirable.

The situation appears slightly better for the ${\bf \Sigma\frac{1}{2}^-}$ {\bf
  sector}. The lowest state predicted at 1628 MeV nicely coincides with the
$\Sigma\frac{1}{2}^-(1620,\mbox{**})$. This state may be viewed as the flavor
octet partner of the $\Lambda$- and nucleon states
$\Lambda\frac{1}{2}^-(1670,\mbox{****})$ and
$N\frac{1}{2}^-(1535,\mbox{****})$ since it likewise exhibits a dominant $^2
8[70]$ contribution ($\sim 88\%$). Similar to its $N$- and
$\Lambda$-counterparts this state is lowered by 't~Hooft's force (see fig.
\ref{fig:Model2S-gvar}), here by roughly 80 MeV which is just the right size
to correctly reproduce the position of the
$\Sigma\frac{1}{2}^-(1620,\mbox{**})$ within its experimental uncertainties.
We can reproduce as well the more established three-star state
$\Sigma\frac{1}{2}^-(1750,\mbox{***})$, but our model predicts within its
range of possible values two close states at 1771 and 1798 MeV although so far
only one state is experimentally resolved in this mass region of the $S_{11}$
partial wave. But there is another very poorly determined structure seen in
the $S_{11}$ partial wave, the one-star $\Sigma\frac{1}{2}^-(2000,\mbox{*})$,
whose large range of possible values even overlaps with that of the
$\Sigma\frac{1}{2}^-(1750,\mbox{***})$. The first of the two close states
predicted turns out to be a dominantly $^4 8[70]$ state ($\sim 95\%$).  It
corresponds to the $N\frac{1}{2}^-(1650,\mbox{****})$ and the
$\Lambda\frac{1}{2}^-(1650,\mbox{****})$ in the $N$- and $\Lambda$-spectrum
and similarly shows an upwards mass shift due to the repulsive action of
't~Hooft's force in the pseudo-scalar diquark channel. In this way the
instanton force provides a good explanation for the mass splitting between the
$\Sigma\frac{1}{2}^-(1620,\mbox{**})$ and the
$\Sigma\frac{1}{2}^-(1750,\mbox{***})$, as shown in fig.
\ref{fig:Model2S-gvar}. The second state predicted close to the
$\Sigma\frac{1}{2}^-(1750,\mbox{***})$ is hardly influenced by 't~Hooft's
force, since it consists of an almost pure $^2 10[70]$ configuration. This
state is the expected flavor decuplet partner of the $\Delta$-resonance
$\Delta\frac{1}{2}^-(1620,\mbox{****})$.\\

Finally it remains to comment on the negative-parity states of the
$3\hbar\omega$ band. Our predictions for the lightest few states in this shell
are summarized in table \ref{tab:S3hwband}. 

\begin{table}[!h]
\center
\begin{tabular}{ccccccc}
\hline
Exp. state   &PW&${J^\pi}$        & Rating     & Mass range [MeV]& Model state &  Model state \\
\cite{PDG00} &  &                      &            & \cite{PDG00}    & in model $\mathcal{A}$& in model $\mathcal{B}$  \\
\hline
$\Sigma(2000)$&$S_{11}$&${\frac{1}{2}^-}$& *      &1755-2044  &                  &              \\
              &        &                 &        &           &$\MSS(1,-,4,2111)$    &$\MSS(1,-,4,2249)$\\
              &        &                 &        &           &$\MSS(1,-,5,2136)$    &$\MSS(1,-,5,2326)$\\
              &        &                 &        &           &$\MSS(1,-,6,2251)$    &$\MSS(1,-,6,2377)$\\
              &        &                 &        &           &$\MSS(1,-,7,2264)$    &$\MSS(1,-,7,2382)$\\
              &        &                 &        &           &$\MSS(1,-,8,2288)$    &$\MSS(1,-,8,2413)$\\
\hline
$\Sigma(1940)$&$D_{13}$&${\frac{3}{2}^-}$& ***    &1900-1950  &                  &              \\
              &        &                 &        &           &$\MSS(3,-,4,2139)$    &$\MSS(3,-,4,2229)$\\
              &        &                 &        &           &$\MSS(3,-,5,2171)$    &$\MSS(3,-,5,2325)$\\
              &        &                 &        &           &$\MSS(3,-,6,2203)$    &$\MSS(3,-,6,2368)$\\
              &        &                 &        &           &$\MSS(3,-,7,2244)$    &$\MSS(3,-,7,2372)$\\
              &        &                 &        &           &$\MSS(3,-,8,2263)$    &$\MSS(3,-,8,2383)$\\
\hline
              &$D_{15}$&${\frac{5}{2}^-}$&        &           &$\MSS(5,-,2,2174)$   &$\MSS(5,-,2,2296)$ \\
              &        &                 &        &           &$\MSS(5,-,3,2226)$   &$\MSS(5,-,3,2364)$ \\
              &        &                 &        &           &$\MSS(5,-,4,2268)$   &$\MSS(5,-,4,2377)$ \\
              &        &                 &        &           &$\MSS(5,-,5,2305)$   &$\MSS(5,-,5,2382)$ \\
\hline
$\Sigma(2100)$&$G_{17}$&${\frac{7}{2}^-}$& *      &2040-2150  &                 &              \\
              &        &                 &        &           &$\MSS(7,-,1,2236)$   &$\MSS(7,-,1,2293)$\\
              &        &                 &        &           &$\MSS(7,-,2,2285)$   &$\MSS(7,-,2,2373)$\\
              &        &                 &        &           &$\MSS(7,-,3,2320)$   &$\MSS(7,-,3,2387)$\\
              &        &                 &        &           &$\MSS(7,-,4,2353)$   &$\MSS(7,-,4,2420)$\\
\hline
              &$G_{19}$&${\frac{9}{2}^-}$&        &           &$\MSS(9,-,1,2325)$   &$\MSS(9,-,1,2367)$\\
              &        &                 &        &           &$\MSS(9,-,2,2418)$   &$\MSS(9,-,2,2435)$\\
\hline
\end{tabular}
\caption{Calculated positions of the lightest few $\Sigma$ states assigned to the negative parity $3\hbar\omega$ shell
  in comparison to the corresponding experimental mass values taken from
  \cite{PDG00}. Notation as in table \ref{tab:L2hwband}.}
\label{tab:S3hwband}
\end{table}
All states predicted in this band-structure lie beyond 2100 MeV.
Concerning the instanton-induced hyperfine structures (see
fig. \ref{fig:Model2S-gvar}) we should note that 't~Hooft's force
selectively lowers a group of six (dominantly $^2 8[56]$ and $^2
8[70]$) states in the sectors with spins $J$ from $\frac{1}{2}$ to
$\frac{7}{2}$.  In fact, we find the same systematics observed for the
corresponding group of six states in the nucleon $3\hbar\omega$ band
\cite{Loe01b}. But once again the mass shift here is rather small and
only the two lowered states in the $J^\pi=\frac{1}{2}^-$ sector become
significantly isolated from the other $3\hbar\omega$ states. In the
$\Sigma\frac{1}{2}^-$ and $\Sigma\frac{3}{2}^-$ the downward mass
shift cannot account the positions of the rather poorly established
resonance $\Sigma\frac{1}{2}^-(2000,\mbox{*})$ and
the $\Sigma\frac{3}{2}^-(1940,\mbox{***})$ which already could not be
associated with states predicted in the $1\hbar\omega$ shell.  The
one-star state $\Sigma\frac{7}{2}^-(2100,\mbox{*})$ is the only effect
seen in the $G_{17}$ partial wave.  According to its total spin
$J=\frac{7}{2}$ this state definitely cannot belong to the
$1\hbar\omega$ shell, but must be, if we take it seriously, a member
of the $3\hbar\omega$ band. Our model has the same problems to explain
this structure as all other constituent quark models.  The downward
mass shift of the dominantly $^2 8 [70]$ state in the
$\Sigma\frac{7}{2}^-$ sector is much too small to explain the puzzling
low position of the $\Sigma\frac{7}{2}^-(2100,\mbox{*})$.  Its
position predicted at 2236 MeV is far above the PDG quoted position
for the $\Sigma\frac{7}{2}^-(2100,\mbox{*})$.  We should note in this
respect that the corresponding lowest excitation in
$\Lambda\frac{7}{2}^-$ is observed at the same resonance position, but
remember that the well-established
$\Lambda\frac{7}{2}^-(2100,\mbox{****})$ could be explained due to the
much stronger effect of a scalar {\it non-strange} diquark
correlation. We thus do not worry about this poorly established one-star
resonance whose existence is highly questionable anyway.

\subsubsection{Comment on the results of model $\mathcal{B}$}
\begin{figure}[!h]
  \begin{center}
    \epsfig{file={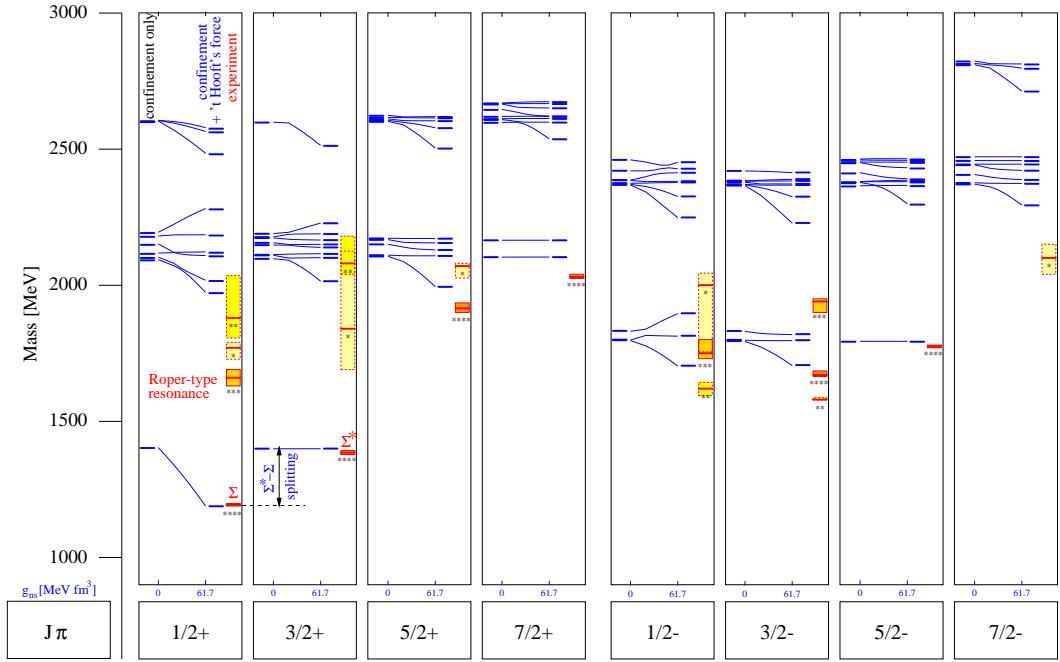},width=140mm}
  \end{center}
  \caption{Influence of the instanton-induced interaction on the energy levels
  of the positive-parity (left) and negative-parity (right) $\Sigma$-states
  in model $\mathcal{B}$. The curves illustrate the variation of energy levels
  with increasing 't~Hooft coupling $g_{ns}$ which is finally
  fixed to $g_{nn}=89.6$ MeV fm$^3$.}
  \label{fig:Model1aSgvar}
\end{figure}

So far we discussed entirely the predictions of model $\mathcal{A}$ which in
the course of our previous investigations of the $N$- and $\Lambda$- spectra
turned out to achieve significantly better agreement with the phenomenology
than model $\mathcal{B}$.  For the sake of completeness we will conclude the
discussion of the $\Sigma$ sector  briefly commenting on the results of model
$\mathcal{B}$.  Comparing the predicted $\Sigma$ spectra of models
$\mathcal{A}$ and $\mathcal{B}$ in figs. \ref{fig:SigM2} and
\ref{fig:SigM1}, respectively, the superior predictive power of model
$\mathcal{A}$ to that of model $\mathcal{B}$ is once again confirmed. 
The $\Sigma$ spectrum predicted in model $\mathcal{B}$ 
reveals essentially the same shortcomings 
that could be already exposed  in discussion of the nucleon- and
$\Lambda$-spectra.  Similar to the $\Lambda$ spectrum the centroids of the
$1\hbar\omega$ and $2\hbar\omega$ band structures again are predicted too high
compared to that of the observed states. Figure \ref{fig:Model1aSgvar}
displays for model $\mathcal{B}$ the effect of 't~Hooft's force on the energy
levels of the positive- and negative parity $\Sigma$ states.
In comparison to the corresponding figs. \ref{fig:Model2S+gvar} and
\ref{fig:Model2S-gvar} of model $\mathcal{A}$, we find once again that the
interplay of 't~Hooft's residual interaction with the relativistic
effects of the confinement kernel $\mathcal{B}$ induces hyperfine structures
which differ significantly from those induced in model $\mathcal{A}$. Similar
to the $N\frac{1}{2}^+$- and $\Lambda\frac{1}{2}^+$-sector the largest
discrepancies between both models show up in the $\Sigma\frac{1}{2}^+$ sector,
where the instanton force in model $\mathcal{B}$ cannot account for the low
position of the Roper-type resonance
$\Sigma\frac{1}{2}^+(1660,\mbox{***})$.

\subsection{Summary for the $\Sigma$-spectrum}
To summarize the discussion of this section, we presented the predicted
$\Sigma$-spectra of both model versions in comparison with the rather scarce currently
available experimental data. Model $\mathcal{A}$ provides a good description
for all well established four-star resonances in the positive- and
negative-parity sectors. Studying the influence of 't~Hooft's residual force,
which here exclusively acts in the scalar non-strange-strange diquark channel, we
found significantly smaller effects than in the $\Lambda$ and $N$- sectors,
where the much stronger effects of the non-strange diquark correlation
emerged. This is consistent with the in fact smaller hyperfine splittings
observed in this flavor sector and in particular it nicely explains the
absence of approximate parity doublet structures, for which there is (in
contrast to the $N$-and $\Lambda$-sectors) hardly any evidence in the
experimental $\Sigma$-spectrum.  Along with the well reproduced
$\Sigma\frac{3}{2}^+(1385,\mbox{****})$--$\Sigma\frac{1}{2}^+(1193,\mbox{****})$
ground-state splitting, 't~Hooft's force in general provides again a rather
good description of the observed intra-band structures as far as they are
experimentally resolved by the partly inconclusive data. In particular, we
found a strong lowering of the lowest $\Sigma\frac{1}{2}^+$ excitation
corresponding to the low-lying $\Sigma\frac{1}{2}^+(1660,\mbox{***})$, the
counterpart of the Roper resonance. 
Finally, we should mention that once again our fully relativistic
approach (model $\mathcal{A}$) shows significant improvements relative to the corresponding
non-relativistic model of Blask {\it et al.} \cite{BBHMP90,Met92}.
\section{The $\Xi$-resonance spectrum}
\label{sec:Xi}
In this section we present our predictions for the excited $\Xi$-baryons with
strangeness $S^*=-2$ and isospin $T=\frac{1}{2}$. Our results in this flavor
sector are almost entirely predictive, due to the lack of experimental data.
Not much is known about $\Xi$ resonances since no direct formation is possible
and the $\Xi$ baryons can only be produced as a part of a final state which in general is
topologically complicated and difficult to study. Moreover the production
cross sections are rather small \cite{PDG00}. Apart from the well established
ground-states $\Xi\frac{1}{2}^+(1318,\mbox{****})$ and
$\Xi\frac{3}{2}^+(1530,\mbox{****})$ there is only one single negative-parity
excited state whose quantum numbers are established, {\it i.e.} the three-star
resonance $\Xi\frac{3}{2}^-(1820,\mbox{***})$. There are further evidences for
$\Xi$-resonances (even with a three-star rating), but in all cases spin and
parity of these states are completely undetermined \cite{PDG00}.\\

\begin{figure}[!h]
  \begin{center}
    \epsfig{file={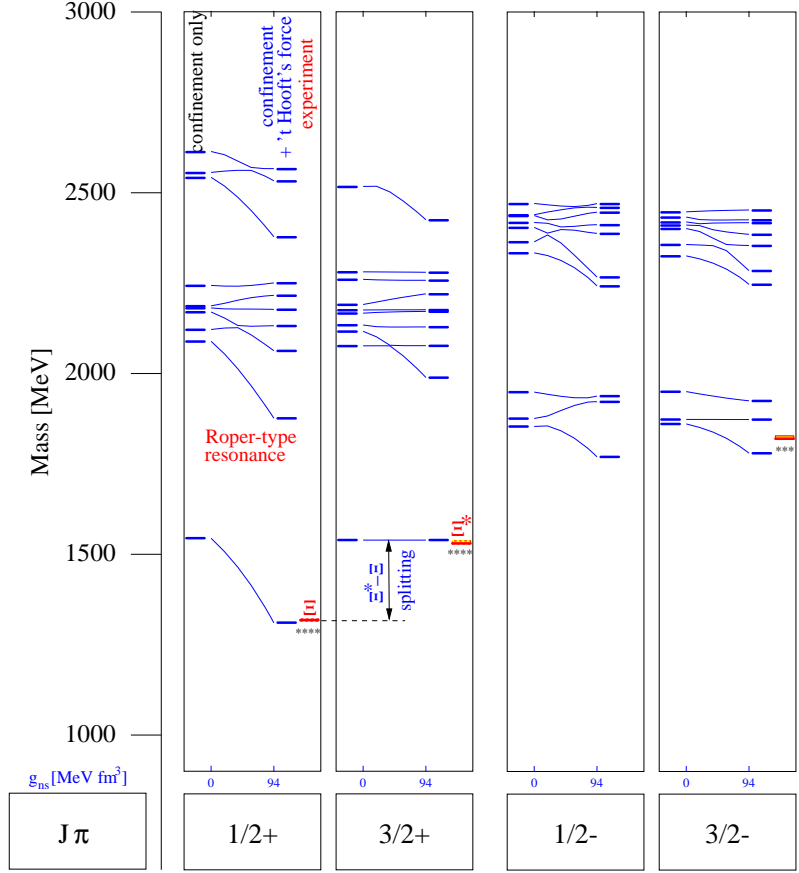},width=105mm}
  \end{center}
  \caption{Instanton-induced hyperfine splittings of the positive- (left) and negative-parity (right)
    $\Xi$ states in model $\mathcal{A}$. Left in each column the spectrum from
    confinement alone is shown.  The curves show the change of the spectrum as
    function of the 't~Hooft coupling $g_{ns}$ which finally is fixed (right
    spectrum) from the hyperon splittings $\Sigma^*-\Sigma-\Lambda$ and
    $\Xi^*-\Xi$.  The rightmost spectrum shows for comparison the experimental
    data.}
  \label{fig:Model2Xgvar}
\end{figure}

\begin{figure}[!h]
  \begin{center}
    \epsfig{file={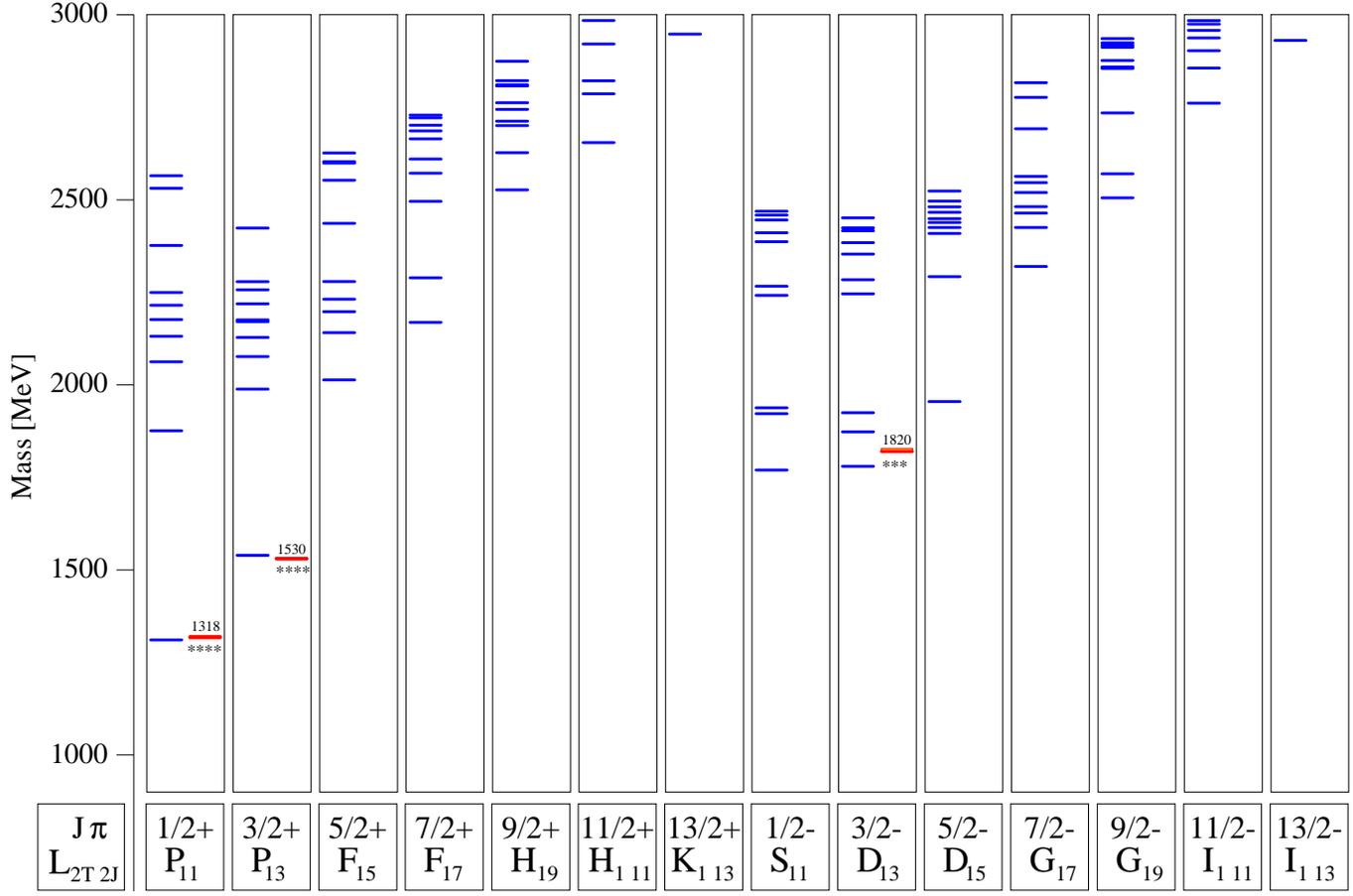},width=180mm}
  \end{center}
\caption{The predicted positive- and negative-parity
\textbf{$\Xi$-resonance spectrum} with isospin $T=\frac{1}{2}$ and
strangeness $S^* = -2$ in \textbf{model $\mathcal{A}$} (left part of each column)
in comparison to the experimental spectrum taken from Particle Data Group \cite{PDG00} (right part of each
column).  The resonances are classified by the total spin
$J$ and parity $\pi$. The experimental resonance position is indicated
by a bar, the corresponding uncertainty by the shaded box; the status of each
resonance is indicated by stars. At most ten radial excitations are shown
in each column.}
\label{fig:XiM2}
\end{figure}

\begin{figure}[!h]
  \begin{center}
    \epsfig{file={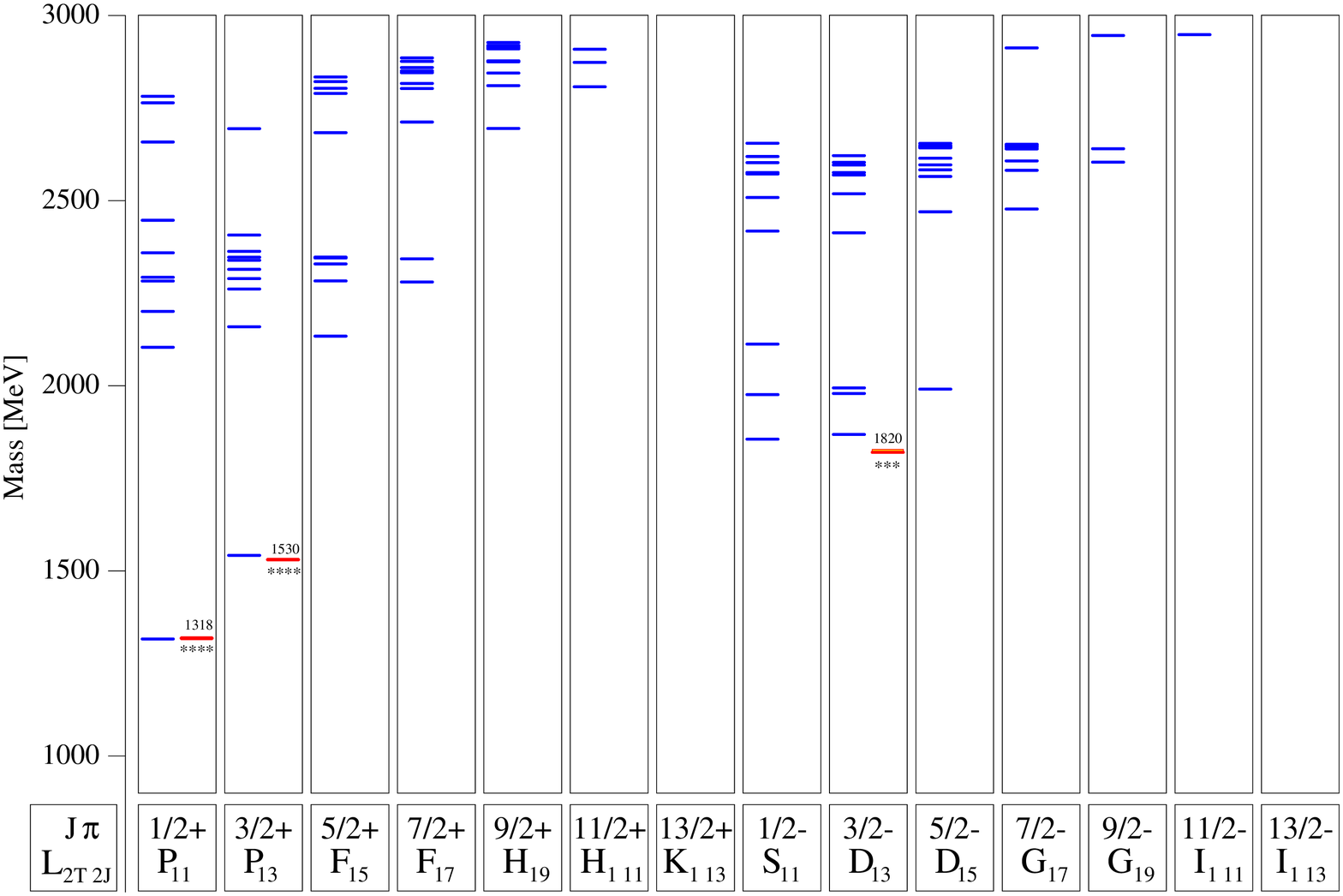},width=180mm}
  \end{center}
\caption{The predicted positive- and negative-parity
\textbf{$\Xi$-resonance spectrum} with isospin $T=\frac{1}{2}$ and
strangeness $S^* = -2$ in \textbf{model $\mathcal{B}$} (left part of each column)
in comparison to the experimental spectrum taken from Particle Data Group \cite{PDG00} (right part of each
column). The resonances are classified by the total spin
$J$ and parity $\pi$. See also caption to fig. \ref{fig:XiM2}.}
\label{fig:XiM1}
\end{figure}

There is not much to say about the structures expected in this flavor sector:
There are exactly the same degrees of freedom as in $\Sigma$ sector discussed in the 
preceding section. The octet and decuplet flavor wave function
corresponding to that in the $\Sigma$ sector only differ by the interchange $\ket{n}\leftrightarrow\ket{s}$ of
the non-strange and strange quark flavors and the Salpeter amplitudes $\Phi_{J^\pi}^\Xi$ of excited
$\Xi$-states with spin and parity $J^\pi$ are obtained by the embedding
map (see ref. \cite{Loe01a})
\begin{equation}
\label{embedSalp_Xi}
\Phi_{J^\pi}^\Xi \;=\; T^{+++} {\varphi_{J^\pi}^\Xi} \;+\; T^{---} {\varphi_{J^{-\pi}}^\Xi}
\end{equation}
of Pauli spinors $\varphi^\Xi_{J^{\pi}}$ and
$\varphi^\Xi_{J^{-\pi}}$ which, analogous to the $\Sigma$-states, 
decompose into the spin-flavor $SU(6)$-configurations
\begin{eqnarray}
\label{Xi_decompPauli}
\ket{\varphi^\Xi_{J^{\pm}}} 
&=&
\phantom{+\;}
\ket{\Xi\;J^\pm,\; ^2 8[56]}
\;+\;
\ket{\Xi\;J^\pm,\;^2 8[70]}
\;+\;
\ket{\Xi\;J^\pm,\;^4 8[70]}
\;+\;
\ket{\Xi\;J^\pm,\;^2 8[20]}\nn
& &
+\;
\ket{\Xi\;J^\pm,\;^4 10[56]}
\;+\;
\ket{\Xi\;J^\pm,\;^2 10[70]},
\end{eqnarray}
with the flavor-octet and decuplet contributions
\begin{eqnarray}
\label{Xi8_configPauli}
\ket{\Xi\;J^\pm,\; ^2 8[56]} &:=& 
\sum\limits_{L} \Bigg[ 
\ket{\psi^{L\;\pm}_\mathcal{S}}\tens
\frac{1}{\sqrt 2}\bigg(
\ket {\chi^\frac{1}{2}_{\mathcal{M}_\mathcal{A}}}
\tens
\ket {\phi^\Xi_{\mathcal{M}_\mathcal{A}}}
+
\ket {\chi^\frac{1}{2}_{\mathcal{M}_\mathcal{S}}}
\tens
\ket {\phi^\Xi_{\mathcal{M}_\mathcal{S}}}
\bigg)\Bigg]^J,\nn[-2mm]
\ket{\Xi\;J^\pm,\;^2 8[70]} &:=& 
\sum\limits_{L} \Bigg[\phantom{+\;}\frac{1}{2}\; 
\ket{\psi^{L\;\pm}_{\mathcal{M}_\mathcal{A}}}\tens 
\bigg(
\ket {\chi^\frac{1}{2}_{\mathcal{M}_\mathcal{A}}}
\tens
\ket {\phi^\Xi_{\mathcal{M}_\mathcal{S}}}
+
\ket {\chi^\frac{1}{2}_{\mathcal{M}_\mathcal{S}}}
\tens
\ket {\phi^\Xi_{\mathcal{M}_\mathcal{A}}}
\bigg)\nn[-2mm]
&& 
\phantom{\sum\limits_{L} \Bigg[}+
\frac{1}{2}\;
\ket{\psi^{L\;\pm}_{\mathcal{M}_\mathcal{S}}}\tens
\bigg(
\ket {\chi^\frac{1}{2}_{\mathcal{M}_\mathcal{A}}}
\tens
\ket {\phi^\Xi_{\mathcal{M}_\mathcal{A}}}
-
\ket {\chi^\frac{1}{2}_{\mathcal{M}_\mathcal{S}}}
\tens
\ket {\phi^\Xi_{\mathcal{M}_\mathcal{S}}}\bigg)\Bigg]^J,\nn[-2mm]
\ket{\Xi\;J^\pm,\;^4 8[70]} &:=& 
\sum\limits_{L} \Bigg[\frac{1}{\sqrt 2} 
\bigg(
\ket{\psi^{L\;\pm}_{\mathcal{M}_\mathcal{A}}}\tens 
\ket{\chi^\frac{3}{2}_\mathcal{S}}\tens  
\ket{\phi^\Xi_{\mathcal{M}_\mathcal{A}}}
-
\ket{\psi^{L\;\pm}_{\mathcal{M}_\mathcal{S}}}\tens 
\ket{\chi^\frac{3}{2}_\mathcal{S}}\tens  
\ket{\phi^\Xi_{\mathcal{M}_\mathcal{S}}}
\bigg)\Bigg]^J,\nn[-2mm]
\ket{\Xi\;J^\pm,\;^2 8[20]} &:=&  
\sum\limits_{L} \Bigg[\ket{\psi^{L\;\pm}_\mathcal{A}}\tens
\frac{1}{\sqrt 2}\bigg(
\ket {\chi^\frac{1}{2}_{\mathcal{M}_\mathcal{A}}}
\tens
\ket {\phi^\Xi_{\mathcal{M}_\mathcal{S}}}
-
\ket {\chi^\frac{1}{2}_{\mathcal{M}_\mathcal{S}}}
\tens
\ket {\phi^\Xi_{\mathcal{M}_\mathcal{A}}}
\bigg)\Bigg]^J,\nn[-2mm]
\ket{\Xi\;J^\pm,\;^4 10[56]} &:=&  
\sum\limits_{L} \Bigg[
\ket{\psi^{L\;\pm}_\mathcal{S}}
\tens
\ket {\chi^\frac{3}{2}_\mathcal{S}}\Bigg]^J
\tens
\ket {\phi^\Xi_\mathcal{S}},\nn[-2mm]
\ket{\Xi\;J^\pm,\;^2 10[70]} &:=&  
\sum\limits_{L} \Bigg[
\frac{1}{\sqrt 2}\bigg(
\ket{\psi^{L\;\pm}_{\mathcal{M}_\mathcal{S}}}
\tens
\ket {\chi^\frac{1}{2}_{\mathcal{M}_\mathcal{S}}}
+
\ket{\psi^{L\;\pm}_{\mathcal{M}_\mathcal{A}}}
\tens
\ket {\chi^\frac{1}{2}_{\mathcal{M}_\mathcal{A}}}
\bigg)
\Bigg]^J
\tens
\ket {\phi^\Xi_\mathcal{S}}.
\end{eqnarray}
Again 't~Hooft's force acts here for non-strange-strange quark pairs which are
antisymmetric in flavor thus affecting the states in the same manner as in the
$\Sigma$-sector. Apart from slightly different spin-orbit effects and the
overall higher mass positions of states due to bigger strangeness content, the
predicted structures of the $\Xi$-spectrum as well as the configuration mixing
of states is thus very similar to the $\Sigma$-sector.  In particular
't~Hooft's force generates quite the same hyperfine structures, as shown in
fig.  \ref{fig:Model2Xgvar} for model $\mathcal{A}$ (compare to the
corresponding figs. \ref{fig:Model2S+gvar} and \ref{fig:Model2S-gvar} for the
$\Sigma$-sector).
In this respect it is worth to note that along with the downward mass shift of
the $\Xi\frac{1}{2}^+$ ground-state, which nicely explains the
$\Xi\frac{3}{2}^+(1530,\mbox{****})$--$\Xi\frac{1}{2}^+(1318,\mbox{****})$
ground-state splitting, we observe again an equally large (roughly 200 MeV)
selective lowering of the lowest positive-parity excited state with the same
quantum numbers $J^\pi=\frac{1}{2}^+$. This state, which we predict to occur at
1876 MeV in model $\mathcal{A}$, can be viewed as the Roper-type analog
of the $\Xi$-resonances. Also in the negative-parity sector the instanton
induced effects are quite the same as in the $\Sigma$-spectrum. Thus,
all our remarks concerning the structure of the positive- and
negative-parity $\Sigma$-spectrum basically hold
also for the excited states predicted in the  $\Xi$-sector.\\
Our predictions for the $\Xi$-resonances with spin and parity $J^\pi$ up to
$\frac{13}{2}^\pm$ are graphically displayed in figs. \ref{fig:XiM2} and
\ref{fig:XiM1}. In addition, the positions of $1\hbar\omega$ and
$2\hbar\omega$ states and the lightest few $3\hbar\omega$ states are
explicitly tabulated in tables \ref{tab:X1hwband}, \ref{tab:X2hwband} and
\ref{tab:X3hwband}. According to our previous results in the other flavor
sectors, we expect the predictions of model $\mathcal{A}$ to be most
reliable, but for the sake of completeness we also present those of model $\mathcal{B}$.

As mentioned before, there is little information on the excited states
available at the moment. The $\Xi\frac{3}{2}^-(1820,\mbox{***})$ is the only
excited resonance with established spin and parity.  We associate this
resonance with the lowest $\Xi\frac{3}{2}^-$ state predicted close to the
$\Xi\frac{3}{2}^-(1820,\mbox{***})$ at 1780 MeV (in model $\mathcal{A}$, see
table \ref{tab:X1hwband}). Concerning the other states quoted by the PDG \cite{PDG00},
whose spins and parities are unknown so far, we should note that a speculative
assignment to model states in general is ambiguous and
inconclusive. New experimental efforts to shed light into this poorly explored
flavor sector would be highly desirable.
\begin{table}[!h]
\center
\begin{tabular}{ccccccc}
\hline
Exp. state   &PW&${J^\pi}$        & Rating     & Mass range [MeV]& Model state &  Model state \\
\cite{PDG00} &  &                      &            & \cite{PDG00}    & in model $\mathcal{A}$& in model $\mathcal{B}$  \\
\hline
           &$S_{11}$&${\frac{1}{2}^-}$&        &           &$\MSX(1,-,1,1770)$   &$\MSX(1,-,1,1855)$\\
           &        &                 &        &           &$\MSX(1,-,2,1922)$   &$\MSX(1,-,2,1976)$\\
           &        &                 &        &           &$\MSX(1,-,3,1938)$   &$\MSX(1,-,3,2112)$\\
\hline
$\Xi(1820)$&$D_{13}$&${\frac{3}{2}^-}$& ***    &1818-1828  &$\MSX(3,-,1,1780)$   &$\MSX(3,-,1,1868)$\\
           &        &                 &        &           &$\MSX(3,-,2,1873)$   &$\MSX(3,-,2,1979)$\\
           &        &                 &        &           &$\MSX(3,-,3,1924)$   &$\MSX(3,-,3,1994)$\\
\hline
           &$D_{15}$&${\frac{5}{2}^-}$&        &           &$\MSX(5,-,1,1955)$   &$\MSX(5,-,1,1991)$\\
\hline
\end{tabular}
\caption{Calculated positions of $\Xi$ states assigned to the negative parity
  $1\hbar\omega$ shell. Notation as in table \ref{tab:L2hwband}.}
\label{tab:X1hwband}
\end{table}
\begin{table}[!h]
\center
\begin{tabular}{ccccccc}
\hline
Exp. state   &PW&${J^\pi}$        & Rating     & Mass range [MeV]& Model state &  Model state \\
\cite{PDG00} &  &                      &       & \cite{PDG00}    & in model $\mathcal{A}$& in model $\mathcal{B}$  \\
\hline
              &$P_{11}$&${\frac{1}{2}^+}$&     &  &$\MSX(1,+,2,1876)$   &$\MSX(1,+,2,2104)$\\                     
              &        &                 &     &  &$\MSX(1,+,3,2062)$   &$\MSX(1,+,3,2201)$\\
              &        &                 &     &  &$\MSX(1,+,4,2131)$   &$\MSX(1,+,4,2283)$\\
              &        &                 &     &  &$\MSX(1,+,5,2176)$   &$\MSX(1,+,5,2293)$\\
              &        &                 &     &  &$\MSX(1,+,6,2215)$   &$\MSX(1,+,6,2359)$\\
              &        &                 &     &  &$\MSX(1,+,7,2249)$   &$\MSX(1,+,7,2446)$\\
\hline
              &$P_{13}$&${\frac{3}{2}^+}$&     &  &$\MSX(3,+,2,1988)$   &$\MSX(3,+,2,2159)$\\
              &        &                 &     &  &$\MSX(3,+,3,2076)$   &$\MSX(3,+,3,2261)$\\
              &        &                 &     &  &$\MSX(3,+,4,2128)$   &$\MSX(3,+,4,2289)$\\
              &        &                 &     &  &$\MSX(3,+,5,2170)$   &$\MSX(3,+,5,2314)$\\
              &        &                 &     &  &$\MSX(3,+,6,2175)$   &$\MSX(3,+,6,2338)$\\
              &        &                 &     &  &$\MSX(3,+,7,2219)$   &$\MSX(3,+,7,2347)$\\
              &        &                 &     &  &$\MSX(3,+,8,2257)$   &$\MSX(3,+,8,2363)$\\
              &        &                 &     &  &$\MSX(3,+,9,2279)$   &$\MSX(3,+,9,2407)$\\
\hline
              &$F_{15}$&${\frac{5}{2}^+}$&     &  &$\MSX(5,+,1,2013)$   &$\MSX(5,+,1,2134)$\\
              &        &                 &     &  &$\MSX(5,+,2,2141)$   &$\MSX(5,+,2,2283)$\\
              &        &                 &     &  &$\MSX(5,+,3,2197)$   &$\MSX(5,+,3,2329)$\\
              &        &                 &     &  &$\MSX(5,+,4,2231)$   &$\MSX(5,+,4,2345)$\\
              &        &                 &     &  &$\MSX(5,+,5,2279)$   &$\MSX(5,+,5,2347)$\\
\hline
              &$F_{17}$&${\frac{7}{2}^+}$&     &  &$\MSX(7,+,1,2169)$   &$\MSX(7,+,1,2280)$\\
              &        &                 &     &  &$\MSX(7,+,2,2289)$   &$\MSX(7,+,2,2342)$\\
\hline
\end{tabular}
\caption{Calculated positions of $\Xi$ states assigned to the positive parity $2\hbar\omega$ shell. 
Notation as in table \ref{tab:L2hwband}.}
\label{tab:X2hwband}
\end{table}
\begin{table}[!h]
\center
\begin{tabular}{ccccccc}
\hline
Exp. state   &PW&${J^\pi}$        & Rating     & Mass range [MeV]& Model state &  Model state \\
\cite{PDG00} &  &                      &            & \cite{PDG00}    & in model $\mathcal{A}$& in model $\mathcal{B}$  \\
\hline
              &$S_{11}$&${\frac{1}{2}^-}$&       &  &$\MSX(1,-,4,2241)$  &$\MSX(1,-,4,2417)$\\
              &        &                 &       &  &$\MSX(1,-,5,2266)$  &$\MSX(1,-,5,2508)$\\
              &        &                 &       &  &$\MSX(1,-,6,2387)$  &$\MSX(1,-,6,2571)$\\
              &        &                 &       &  &$\MSX(1,-,7,2411)$  &$\MSX(1,-,7,2575)$\\
              &        &                 &       &  &$\MSX(1,-,8,2445)$  &$\MSX(1,-,8,2602)$\\
\hline
              &$D_{13}$&${\frac{3}{2}^-}$&       &  &$\MSX(3,-,4,2246)$  &$\MSX(3,-,4,2413)$\\
              &        &                 &       &  &$\MSX(3,-,5,2284)$  &$\MSX(3,-,5,2518)$\\
              &        &                 &       &  &$\MSX(3,-,6,2353)$  &$\MSX(3,-,6,2569)$\\
              &        &                 &       &  &$\MSX(3,-,7,2384)$  &$\MSX(3,-,7,2567)$\\
              &        &                 &       &  &$\MSX(3,-,8,2416)$  &$\MSX(3,-,8,2596)$\\
\hline
              &$D_{15}$&${\frac{5}{2}^-}$&       &  &$\MSX(5,-,2,2292)$  &$\MSX(5,-,2,2469)$\\
              &        &                 &       &  &$\MSX(5,-,3,2409)$  &$\MSX(5,-,3,2565)$\\
              &        &                 &       &  &$\MSX(5,-,4,2425)$  &$\MSX(5,-,4,2583)$\\
              &        &                 &       &  &$\MSX(5,-,5,2438)$  &$\MSX(5,-,5,2596)$\\

\hline
              &$G_{17}$&${\frac{7}{2}^-}$&       &  &$\MSX(7,-,1,2320)$  &$\MSX(7,-,1,2477)$\\
              &        &                 &       &  &$\MSX(7,-,2,2425)$  &$\MSX(7,-,2,2581)$\\
              &        &                 &       &  &$\MSX(7,-,3,2464)$  &$\MSX(7,-,3,2607)$\\
              &        &                 &       &  &$\MSX(7,-,4,2581)$  &$\MSX(7,-,4,2639)$\\
\hline
              &$G_{19}$&${\frac{9}{2}^-}$&       &  &$\MSX(9,-,1,2505)$  &$\MSX(9,-,1,2604)$\\
              &        &                 &       &  &$\MSX(9,-,2,2570)$  &$\MSX(9,-,2,2640)$\\
\hline
\end{tabular}
\caption{Calculated positions of the lightest few $\Xi$ states assigned to the negative parity $3\hbar\omega$ shell. 
Notation as in table \ref{tab:L2hwband}.}
\label{tab:X3hwband}
\end{table}
\clearpage
\section{The $\Omega$-spectrum}
\label{sec:Om}
\begin{figure}[!h]
  \begin{center}
    \epsfig{file={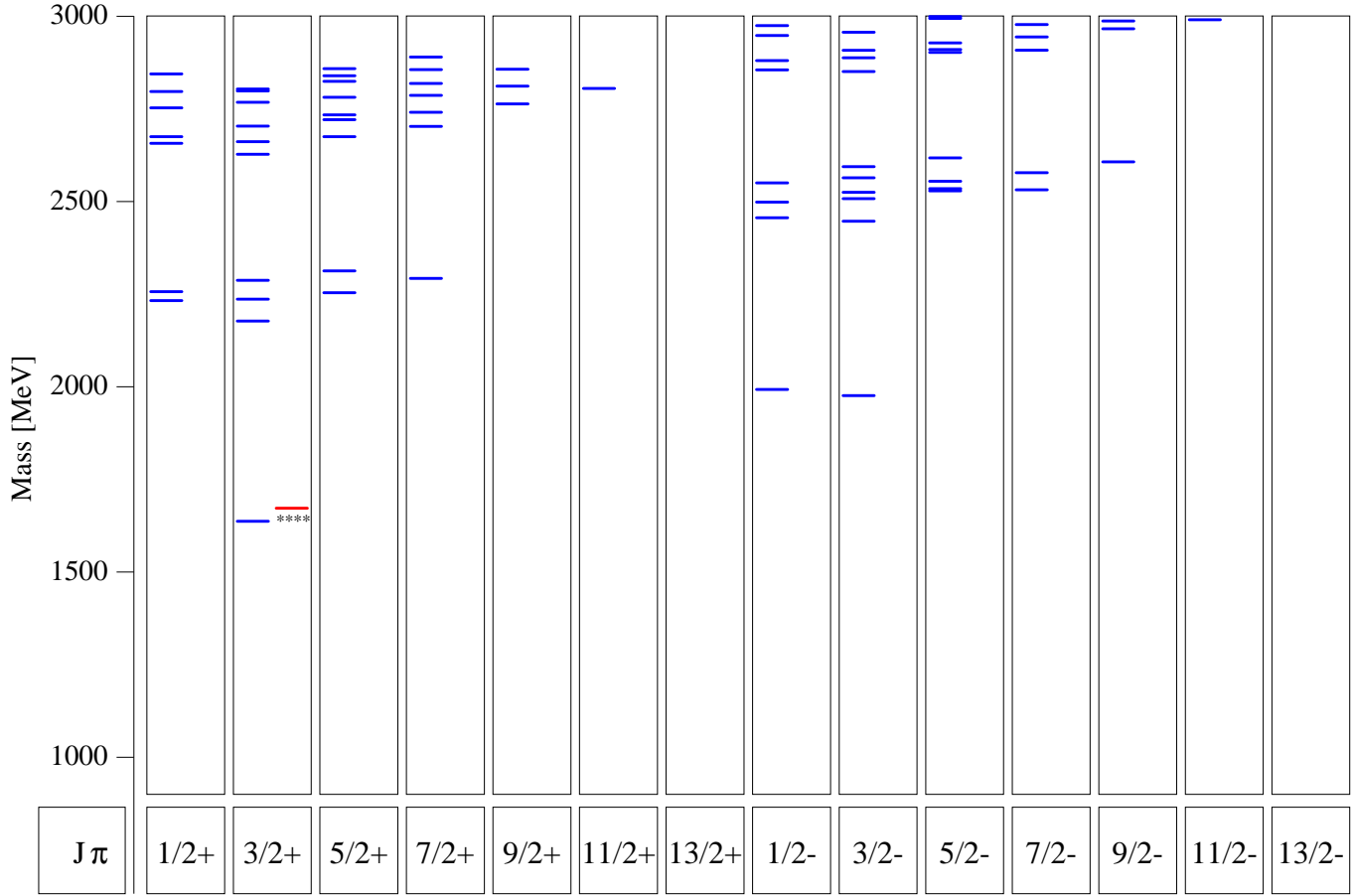},width=180mm}
  \end{center}
\caption{The predicted positive- and negative-parity
\textbf{$\Omega$-baryon spectrum} with isospin $T=0$ and
strangeness $S^* = -3$ in \textbf{model $\mathcal{A}$} (left part of each column)
in comparison to experimental data \cite{PDG00} (right part of the
column). The resonances are classified by the total spin $J$ and parity $\pi$. At most ten radial excitations are shown
in each column.}
\label{fig:OmegaM2}
\end{figure}

\begin{figure}[!h]
  \begin{center}
    \epsfig{file={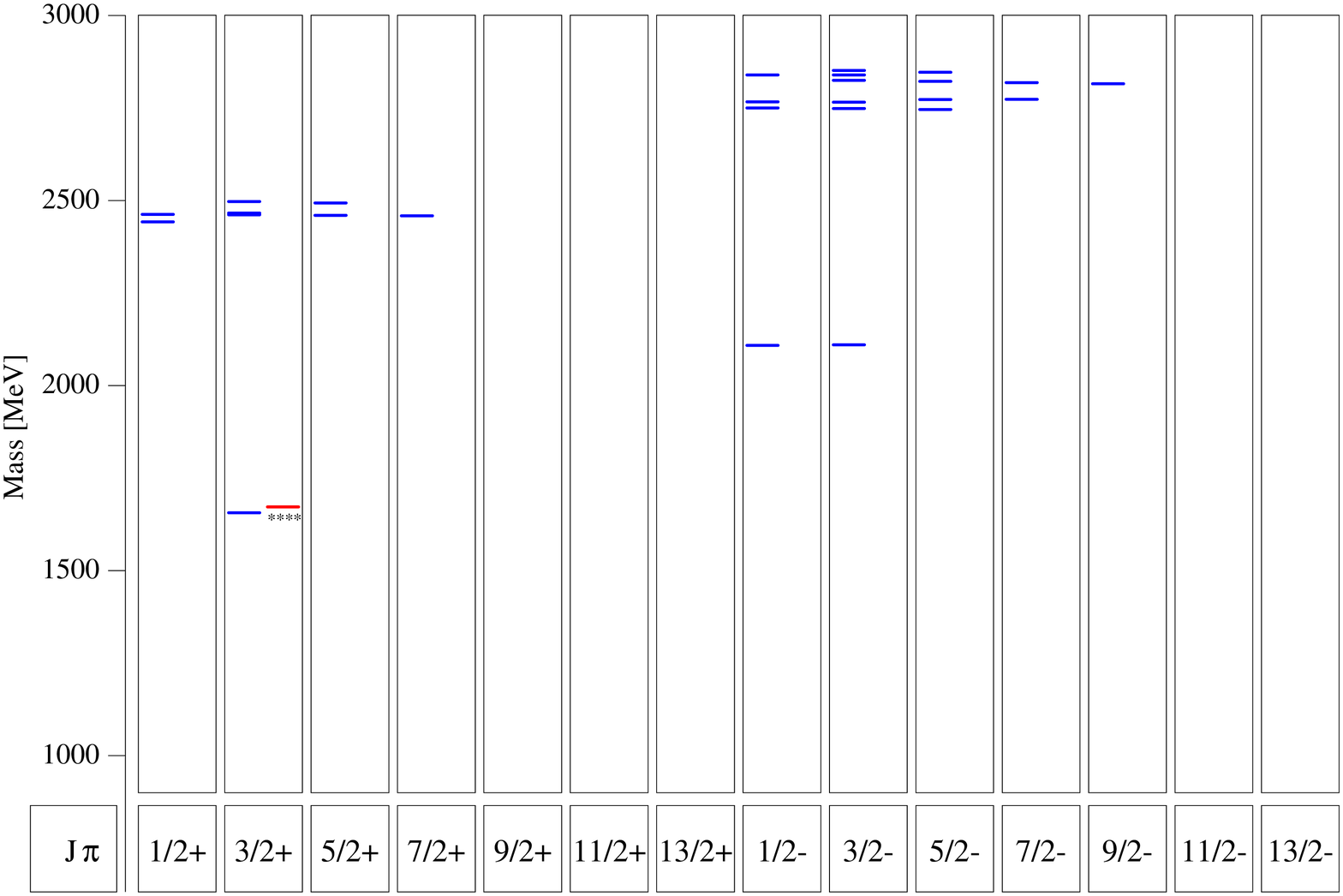},width=180mm}
  \end{center}
\caption{The predicted positive- and negative-parity
  \textbf{$\Omega$-baryon spectrum} with isospin $T=0$ and strangeness $S^* =
  -3$ in \textbf{model $\mathcal{B}$} (left part of each column) in comparison
  to experimental data \cite{PDG00} (right part of the column). The
  resonances are classified by the total spin $J$ and parity $\pi$. At most
  ten radial excitations are shown in each column.}
\label{fig:OmegaM1}
\end{figure}

Finally, let us conclude our discussion of the light baryon spectrum
by presenting our predictions for the $\Omega$ baryons with
strangeness $S^*=-3$ and isospin $T=0$. Almost nothing is known
experimentally about the excited $\Omega$ spectrum even 35 years after
the unambiguous discovery of the $\Omega^-$ ground-state in
1964. Apart from the ground-state only three excitations with $S^*=-3$
have been found \cite{PDG00}, but all of them without established spin
and parity: $\Omega?^?(2250,\mbox{***})$, $\Omega?^?(2380,\mbox{***})$
and $\Omega?^?(2470,\mbox{***})$. But note that even the quantum
numbers of the $\Omega\frac{3}{2}^+(1672,\mbox{****})$ have not
actually been measured but follow from the assignment to the
ground-state decuplet.\\ Similar to the $\Delta$-sector, in our
approach also the $\Omega$ states are determined by the three-body
confinement force alone, since 't~Hooft's force does not act on their
common totally symmetric flavor-decuplet wave-function, which differs
to that of the $\Delta$ states only by the replacement
$\ket{n}\rightarrow\ket{s}$ of all non-strange flavors by the heavier
strange quark flavors.  Thus, apart from the overall higher positions
and slightly smaller spin-orbit effects the predicted structures are
essentially the same as in the $\Delta$-spectrum. Our predictions for
the $\Omega$ baryons are depicted in figs. \ref{fig:OmegaM2} and
\ref{fig:OmegaM1} for model $\mathcal{A}$ and $\mathcal{B}$,
respectively, where again those of version $\mathcal{A}$ should be
most reliable. The calculated masses for the $1\hbar\omega$,
$2\hbar\omega$ and $3\hbar\omega$ states are explicitly summarized in
tables \ref{tab:O1hwband}, \ref{tab:O2hwband} and \ref{tab:O3hwband},
respectively. Note that the three excited states observed roughly fit
to the structures predicted, but a possible spin assignment would be
rather ambiguous.

\begin{table}[!h]
\center
\begin{tabular}{ccccccc}
\hline
Exp. state   &PW&${J^\pi}$        & Rating     & Mass range [MeV]& Model state &  Model state \\
\cite{PDG00} &  &                      &            & \cite{PDG00}    & in model $\mathcal{A}$& in model $\mathcal{B}$  \\
\hline
             &$S_{31}$&${\frac{1}{2}^-}$&     &  & $\MSO(1,-,1,1992)$  & $\MSO(1,-,1,2108)$\\
\hline
             &$D_{33}$&${\frac{3}{2}^-}$&    &  & $\MSO(3,-,1,1976)$  & $\MSO(3,-,1,2110)$ \\
\hline
\end{tabular}
\caption{Calculated positions of $\Omega$-states assigned to the negative parity $1\hbar\omega$ shell.
Notation as in table \ref{tab:L2hwband}.}
\label{tab:O1hwband}
\end{table}
\begin{table}[!h]
\center
\begin{tabular}{ccccccc}
\hline
Exp. state   &PW&${J^\pi}$        & Rating     & Mass range [MeV]& Model state &  Model state \\
\cite{PDG00} &  &                      &            & \cite{PDG00}    & in model $\mathcal{A}$& in model $\mathcal{B}$  \\
\hline
&$P_{31}$& ${\frac{1}{2}^+}$&    &  & $\MSO(1,+,1,2232)$ & $\MSO(1,+,1,2442)$\\
&        &                  &    &  & $\MSO(1,+,1,2256)$ & $\MSO(1,+,1,2462)$\\
\hline
&$P_{33}$& ${\frac{3}{2}^+}$&    &  & $\MSO(3,+,2,2177)$  & $\MSO(3,+,2,2461)$\\
&        &                  &    &  & $\MSO(3,+,3,2236)$  & $\MSO(3,+,3,2466)$\\
&        &                  &    &  & $\MSO(3,+,4,2287)$  & $\MSO(3,+,4,2497)$\\
\hline
&$F_{35}$& ${\frac{5}{2}^+}$&    &  & $\MSO(5,+,1,2253)$  & $\MSO(5,+,1,2460)$\\
&        &                  &    &  & $\MSO(5,+,2,2312)$  & $\MSO(5,+,2,2493)$\\
\hline
&$F_{37}$& ${\frac{7}{2}^+}$&    &  & $\MSO(7,+,1,2292)$  & $\MSO(7,+,1,2458)$\\
\hline
\end{tabular}
\caption{Calculated positions of $\Omega$-states assigned to the positive parity $2\hbar\omega$ shell. 
Notation as in table \ref{tab:L2hwband}.}
\label{tab:O2hwband}
\end{table}
\clearpage
\begin{table}[!h]
\center
\begin{tabular}{ccccccc}
\hline
Exp. state   &PW&${J^\pi}$        & Rating     & Mass range [MeV]& Model state &  Model state \\
\cite{PDG00} &  &                      &            & \cite{PDG00}    & in model $\mathcal{A}$& in model $\mathcal{B}$  \\
\hline
      &$S_{31}$&${\frac{1}{2}^-}$&     &  & $\MSO(1,-,2,2456)$&$\MSO(1,-,2,2750)$\\
      &        &                 &     &  & $\MSO(1,-,3,2498)$&$\MSO(1,-,3,2766)$\\
      &        &                 &     &  & $\MSO(1,-,4,2550)$&$\MSO(1,-,4,2839)$\\
\hline
      &$D_{33}$&${\frac{3}{2}^-}$&     &  & $\MSO(3,-,2,2446)$&$\MSO(3,-,2,2748)$\\
      &        &                 &     &  & $\MSO(3,-,3,2507)$&$\MSO(3,-,3,2765)$\\
      &        &                 &     &  & $\MSO(3,-,4,2524)$&$\MSO(3,-,4,2824)$\\
      &        &                 &     &  & $\MSO(3,-,5,2564)$&$\MSO(3,-,5,2839)$\\
      &        &                 &     &  & $\MSO(3,-,6,2594)$&$\MSO(3,-,6,2851)$\\
\hline
      &$D_{35}$&${\frac{5}{2}^-}$&     &  &$\MSO(5,-,1,2528)$ &$\MSO(5,-,1,2746)$\\
      &        &                 &     &  &$\MSO(5,-,2,2534)$ &$\MSO(5,-,2,2772)$\\
      &        &                 &     &  &$\MSO(5,-,3,2554)$ &$\MSO(5,-,3,2822)$\\
      &        &                 &     &  &$\MSO(5,-,4,2617)$ &$\MSO(5,-,4,2846)$\\
\hline
      &$G_{37}$&${\frac{7}{2}^-}$&     &  &$\MSO(7,-,1,2531)$ &$\MSO(7,-,1,2773)$\\
      &        &                 &     &  &$\MSO(7,-,2,2577)$ &$\MSO(7,-,2,2818)$\\
\hline
      &$G_{39}$&${\frac{9}{2}^-}$&     &  &$\MSO(9,-,1,2606)$ &$\MSO(9,-,1,2815)$\\
\hline
\end{tabular}
\caption{Calculated positions of negative-parity
$\Omega$ states in the $3\hbar\omega$ shell. Notation as in
table \ref{tab:L2hwband}.}
\label{tab:O3hwband}
\end{table}

\section{Summary and conclusion}
\label{sec:summary}
Extending our previous work \cite{Loe01b} on non-strange baryons we
have presented in this paper a calculation of the strange baryon
spectrum together with a detailed comparison with experiment.  Within
a relativistic quark model with instantaneous forces we are able to
describe the known Regge trajectories and the hyperfine structure in
detail.  't~Hooft's instanton-induced interaction played a central
role, but it was also crucial to establish a suitable confinement
interaction (model $\mathcal{A}$) and, in particular, the Dirac
structure of this force. In comparison with nonrelativistic or
''relativized'' quark models, the phenomenological success of the
present relativistic model is remarkable; our earlier paper
\cite{Loe01b} and the present work demonstrate that the complete known
light baryon spectrum with roughly 100 resonance masses \cite{PDG00}
can be uniformly described with the help of seven model parameters.

Our model potentials are purely phenomenological as far as confinement
is concerned.  The residual interaction has a QCD background in
't~Hooft's instanton-induced quark force; our paper, however, only
tests the operator structure of this force, in particular, the flavor
dependence. For purpose, we only determine the strengths by a fit to
the baryon spectrum. In the appendix we present a poor mans
consistency check: by fixing a common cut-off for instanton sizes we
could roughly reproduce the quark model masses and couplings under the
assumptions of spontaneous chiral symmetry breaking. This is certainly
only a first step towards a more complete incorporation of instanton
effects \cite{SS98,DP84,DP85,DP86,DP92,NVZ89,NRZ96}.

Based on the Bethe-Salpeter amplitudes electro-weak decays and formfactors
were already computed \cite{Kr01} in the Mandelstam formalism \cite{Ma55}. The
results will soon be published as well as some calculations of heavy flavored
(charm and bottom) baryons, where as a new feature, the importance of
one-gluon-exchange cannot be ruled out. Work on the perturbative calculation
of strong baryon decays \cite{Ma55} is more difficult but in progress.

\begin{acknowledgement}
  \textbf{Acknowledgments:} We have profited very much from scientific
  discussions with V.~V.~Anisovich, G.~E.~ Brown, E.~Klempt, K.~Kretzschmar,
  A.~Sarantsev and E.~V.~Shuryak to whom we want to express our gratitude. We
  also thank the Deutsche Forschungsgemeinschaft (DFG) for financial support.
\end{acknowledgement}

\appendix
\section{Appendix: Checking the consistency with QCD-relations}
\label{sec:QCDcheck}
So far, the effective constituent quark masses $m_n$, $m_s$ and the effective
't~Hooft coupling strengths $g_{nn}$ and $g_{ns}$ together with the effective
range $\lambda$ of the regularized instanton-induced four-fermion interaction
have been treated as free parameters of our two models $\mathcal{A}$ and
$\mathcal{B}$. They have been adjusted to fit the experimental spectra.  To be
more precise, the parameters $g_{nn}$, $g_{ns}$ and $\lambda$ of the
't~Hooft interaction have been fixed to the values given in ref. \cite{Loe01b} in
order to reproduce the correct hyperfine structure of the octet and decuplet
ground-state baryons, {\it i.e.} the mass splittings $N-\Delta$,
$\Sigma^*-\Sigma$, $\Xi^*-\Xi$ and $\Sigma-\Lambda$ and as a very nice feature
of model $\mathcal{A}$ it turned out, that at the same time the effect of
't~Hooft's force with these values fixed could even account also for
substantial structures of the excited baryon spectra such as {\it e.g.} the low
position of the Roper resonance and its strange partners or the appearance of
approximately degenerate parity doublets in the experimental
nucleon- and $\Lambda$-spectrum.  Thus we could impressively
demonstrate the possibility that besides a proper confinement mechanism (as
given by model $\mathcal{A}$), instanton-induced interactions indeed may play
the essential role in the determination of the spectra of light-flavored
baryons. An additional nice feature of the instanton force is, that it
intrinsically provides a constituent quark mass generation via the Nambu
mechanism and hence it appears even natural to work directly with constituent
quark masses $m_n$ and $m_s$ as done in the present framework.  In order to
confirm this picture of light baryons, it is important to investigate whether our
phenomenologically adjusted parameters $g_{nn}$, $g_{ns}$, $\lambda$, $m_n$
and $m_s$ are really consistent with the theory of instantons and its relation
to QCD as discussed in detail in ref \cite{Loe01b}. There we have shown that due to
the process of chiral symmetry breaking 't~Hooft's instanton-induced
interaction leads to relations for the effective constituent quark masses
$m_n$, $m_s$ and the effective 't~Hooft coupling constants $g_{nn}$, $g_{ns}$.
The QCD-relations have been derived by normal ordering of the original
instanton-induced six-quark vertex \cite{SVZ80} with respect to the true physical
QCD-vacuum which exhibits the non-vanishing quark condensates $\langle
\overline\Psi_n \Psi_n\rangle$ and $\langle \overline\Psi_s \Psi_s\rangle$ for
non-strange and strange quark fields, respectively. They are functions of the
critical maximum instanton size $\rho_c$, whose value we expect to be in rough
qualitative\footnote{Note that the regularization procedure in the 't~Hooft
  kernel is rather arbitrary. The meaning of the effective range strongly
  depends on the regularizing function chosen. Thus only a rough
  correspondence between the effective range of the 't~Hooft interaction and
  the effective instanton size may be expected.}  agreement with the effective
range $\lambda=0.4$ fm of the regularized instanton force in our model. Let us
briefly recall these $\rho_c$-depended relations of ref. \cite{Loe01b}:
\begin{itemize}
\item
The Wick contraction of two fermion lines gives the \textbf{effective
  constituent quark masses} 
\begin{eqnarray}
\label{GapEqu_rep}
m_n(\rho_c) &:=&  m_n^0 + \int_0^{\rho_c} \textrm{d}\rho\; \frac{d_0(\rho)}{\rho^5}\;\;
\frac{4}{3}\pi^2 \rho^3 \;
\Big(m^0_n\rho - \frac{2}{3} \pi^2 \rho^3\; \Expect{\overline\Psi_n \Psi_n}\Big)\;
\Big(m^0_s\rho - \frac{2}{3} \pi^2 \rho^3\; \Expect{\overline\Psi_s \Psi_s} \Big),\nn
m_s(\rho_c) &:=& m_s^0 + \int_0^{\rho_c} \textrm{d}\rho\; \frac{d_0(\rho)}{\rho^5}\;\;
\frac{4}{3}\pi^2 \rho^3 \;
\Big(m^0_n\rho - \frac{2}{3} \pi^2 \rho^3\; \Expect{\overline\Psi_n \Psi_n} \Big)^2.
\end{eqnarray}
\item
The Wick contraction of a single fermion line gives the
\textbf{effective 't~Hooft couplings}
\begin{eqnarray}
\label{geff_rep}
g_{nn}(\rho_c)
&:=&
\frac{3}{8}
\int_0^{\rho_c} \textrm{d}\rho\; \frac{d_0(\rho)}{\rho^5}\;\;
\left( \frac{4}{3}\pi^2 \rho^3 \right)^2 \;
\Big(m^0_s\rho - \frac{2}{3} \pi^2 \rho^3\; \Expect{\overline\Psi_s \Psi_s} \Big),\nn
g_{ns}(\rho_c)
&:=&
\frac{3}{8}
\int_0^{\rho_c} \textrm{d}\rho\; \frac{d_0(\rho)}{\rho^5}\;\;
\left( \frac{4}{3}\pi^2 \rho^3 \right)^2 \;
\Big(m^0_n\rho - \frac{2}{3} \pi^2 \rho^3\; \Expect{\overline\Psi_n \Psi_n}\Big).
\end{eqnarray}
\end{itemize}
Here the instanton density $d_0(\rho)$ for three colors and three flavors reads
\begin{equation}
\label{instdens_rep}
d_0(\rho) = 3.63\;10^{-3}\left(\frac{8\pi^2}{g^2(\rho)}\right)^{6}
\exp\left(-\frac{8\pi^2}{g^2(\rho)}\right), 
\end{equation}
where $g(\rho)$ is the $\rho$-dependent running coupling constant, which in
two-loop accuracy is given by \cite{Shu82}:
\begin{equation}
\label{thooftcoupling_rep}
\frac{8\pi^2}{g^2(\rho)} 
=
9 \ln\left(\frac{1}{\rho\;\Lambda_{\rm QCD}}\right)
+ \frac{32}{9} \ln\left[\ln\left(\frac{1}{\rho\;\Lambda_{\rm QCD}} \right)\right].
\end{equation}
In this way the parameters $m_n$, $m_s$, $g_{nn}$ and $g_{ns}$, which we fixed
from a fit to the phenomenology of the experimental light baryon spectrum, are
related to standard QCD parameters, {\it i.e.}  the current quark masses
$m_n^0$ and $m_s^0$, the quark condensates $\langle \overline\Psi_n
\Psi_n\rangle$ and $\langle \overline\Psi_s \Psi_s\rangle$ and the QCD scale
parameter $\Lambda_{\rm QCD}$.  Typical phenomenological values of these QCD
parameters taken from \cite{RRY85} are listed in table \ref{tab:QCDparameter}.
\begin{table}[!h]
\center
\begin{tabular}{l|lccrl}
\hline
current quark masses& non-strange &$m_n^0$ &&   9 & MeV \\[2mm]
                    & strange    &$m_s^0$ && 150 & MeV \\[2mm]
\hline
quark condensates   & non-strange &$\langle \overline\Psi_n \Psi_n\rangle$ && $-225^3$ & ${\rm MeV^3}$\\[2mm]
                    & strange    &$\langle \overline\Psi_s \Psi_s\rangle$ && $0.8$      & $\langle
\overline\Psi_n \Psi_n\rangle$\\[2mm]
\hline
QCD scale parameter &&$\Lambda_{\rm QCD}$ && $200$ & MeV\\
\hline
\end{tabular}
\caption{Phenomenological standard values of QCD parameters, compare to ref. \cite{RRY85}}
\label{tab:QCDparameter}
\end{table}

\begin{figure}[!h]
  \begin{center}
    \epsfig{file={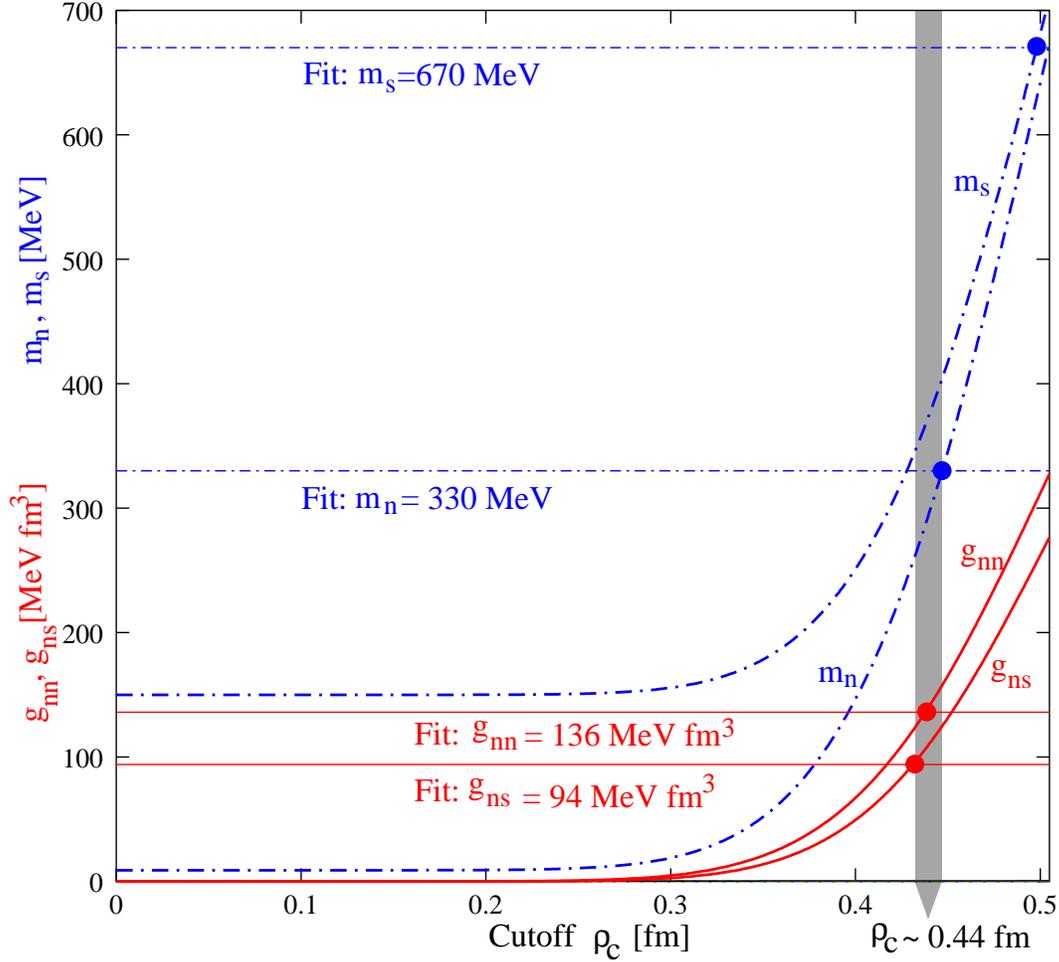},width=140mm}
  \end{center}
\caption{The effective non-strange, strange constituent quark
  masses $m_n$, $m_s$ (dashed dotted lines) and the effective coupling
  constants $g_{nn}$, $g_{ns}$ (dashed lines) as functions of the critical
  maximum instanton size $\rho_c$ due to eqs. (\ref{GapEqu_rep}) and
  (\ref{geff_rep}), respectively. The calculation has been performed in
  two-loop approximation using the standard QCD parameters given in table
  \ref{tab:QCDparameter}. The horizontal lines represent the corresponding
  phenomenologically adjusted values of model $\mathcal{A}$.}
\label{fig:InstantonCheck}
\end{figure}

It is now quite interesting to study to what extent these expressions for the
coupling constants (\ref{geff_rep}) and the constituent quark masses
(\ref{GapEqu_rep}) are consistent with the phenomenologically determined
values in our covariant Salpeter equation-based quark model. In other
words, the question is, whether there is a common uniform instanton
cutoff $\rho_c$ which approximately agrees with the effective range
$\lambda=0.4$ fm of our model, such that the corresponding constituent quark
masses $m_n(\rho_c)$, $m_s(\rho_c)$ and couplings $g_{nn}(\rho_c)$,
$g_{ns}(\rho_c)$ obtained by the gap-equations with the standard QCD values
given in table \ref{tab:QCDparameter} are in fair agreement with our
phenomenologically fixed values ?  We restrict here our discussion to the
confinement model $\mathcal{A}$, which in connection with 't~Hooft's force
yields consistently better results for the excited baryon spectrum than model $\mathcal{B}$
and hence seems to be the more realistic model. The corresponding situation is
presented in fig. \ref{fig:InstantonCheck}.

The plotted curves show the constituent quark masses (dashed dotted curve) and
't~Hooft couplings (dashed curve) as function of the critical instanton size
$\rho_c$ in two-loop approximation. They have been calculated with the QCD
parameters given in table \ref{tab:QCDparameter}. The horizontal lines
represent the corresponding fitted values in model $\mathcal{A}$. Apart from
the strange quark mass $m_s$, we indeed find that the two fitted coupling
strengths $g_{nn}=136$ MeV fm$^3$ and $g_{ns}=94$ MeV fm$^3$ together with the
fitted non-strange quark mass $m_n=330$ MeV can be nicely reproduced by the
gap-equations with an approximately uniform critical instanton size $\rho_c
\simeq 0.44$ fm, which is in a satisfactory qualitative agreement with the
effective range $\lambda=0.4$ fm of the 't~Hooft interaction.  This value
$\rho_c\simeq 0.44$ fm also conforms qualitatively with recent lattice
investigations \cite{ST98,Ne99,RiSch99} on the topological structure of the
QCD vacuum which in fact predict a strong suppression of tunneling events for
larger instanton sizes $\rho>0.45$ fm.  Concerning the 't~Hooft couplings
remember that the phenomenological value and sign of the $\Sigma-\Lambda$
splitting (see ref. \cite{Loe01b}) required the instanton-induced
attraction between a non-strange-strange quark pair to be weaker than between
a non-strange quark pair, {\it i.e.} $g_{ns} < g_{nn}$.  In fact this
requirement is fairly well confirmed by the QCD-relations (\ref{geff_rep}) due
to the different Wick contractions of the original six-point 't~Hooft vertex
with one incoming and outgoing quark of each flavor (u,d,s): $g_{nn}$ involves
the integration over the strange (s) quark loop, while in $g_{ns}$ the lighter
non-strange quark (u and d, respectively) is integrated over.

We should further mention that for a value $\rho=\rho_c = 0.44$ fm we find
only a $ 6\;\%$ difference between the one-loop and two-loop approximation of
the strong $\rho$-dependent running coupling constant $g(\rho)$ given in
eq. \ref{thooftcoupling_rep}, which still is small enough to be within
the scope of the two-loop formula.

Unfortunately, the phenomenologically adjusted strange-quark mass is
definitely too high to be compatible with the gap-equation result at this
critical instanton size $\rho_c \simeq 0.44$.  But note, that the derivation
of the effective 't~Hooft interaction assumes zero-mode dominance in the case
of (almost) massless current quarks.  Compared to the almost vanishing
non-strange current quark mass $m_n^0\simeq 9$ MeV, the strange quark mass
$m^0_s\simeq 150$ MeV, however, is quite big, so that in this case the
zero-mode approximation might become less valid and consequently contributions
to the strange quark mass, that stem from non-zero modes could become
important to explain a part of this missing mass.  Moreover, the instanton
induced interaction is not necessarily the only source for contributions to
the effective constituent quark masses.  In principle there might be other
contributions and hence it is hard to decide whether this discrepancy in the
strange quark mass actually reflects a serious inconsistency.

In table \ref{tab:gap_eq_res} we additionally displayed the explicit absolute
values for the masses and coupling constants obtained by the gap-equations
with a cutoff $\rho_c \simeq 0.44$ fm, in comparison to the corresponding
fitted values of model $\mathcal{A}$.
\begin{table}[!h]
\center
\begin{tabular}{cl|cl}
\hline
gap equations  && empirical fit &  (model $\mathcal{A}$)\\
\hline
$\rho_c = 0.44$ & fm \\
                &    & $\lambda = 0.4$ & fm\\ 
\hline 
$g_{nn} = 141$  & MeV fm$^3$ &$g_{nn} = 136$  & MeV fm$^3$\\
$g_{ns} = 110$  & MeV fm$^3$ &$g_{ns} = 94$  & MeV fm$^3$\\
\hline
$m_n = 297$     & MeV        &$m_n = 330$     & MeV\\
$m_s = 375$     & MeV        &$m_s = 630$     & MeV\\
\hline
\end{tabular}
\caption{Empirically fitted 't~Hooft couplings and constituent
  quark masses of model $\mathcal{A}$ (right column) in comparison with the corresponding gap equation
   results (left column) with $\rho_c \simeq 0.44$ fm (compare to
   fig. \ref{fig:InstantonCheck}). 
   The calculations have been performed in two-loop approximation with the
   standard QCD parameters given in table \ref{tab:QCDparameter}.}
\label{tab:gap_eq_res}
\end{table}

Except for the much too low strange quark mass $m_s$ the agreement with model
$\mathcal{A}$ indeed is satisfying: In view of the sensitive dependence on the
condensate values and the arbitrary regularization procedure, the deviations
of $\leq 15\;\%$ of the calculated values for $g_{nn}$, $g_{ns}$ and $m_n$
from the phenomenological values are rather small. For the ratio
$g_{ns}/g_{nn}$ between the non-strange-strange and the non-strange coupling
constants the QCD-relation yields the explicit value $g_{ns}/g_{nn}= 0.78$
which also is in a satisfactory accordance with the value $g_{ns}/g_{nn}=
0.69$ obtained by the phenomenologically determined couplings.\\

In summary, we thus find a rather good consistency of our model $\mathcal{A}$
with the expectations from the theory of instantons and its relation to QCD,
which strongly supports the spirit of our model that non-perturbative gluon
configurations, {\it i.e.} the instantons, play the dominant role in the
description of spin-spin forces for the light baryons. In comparison with the
analogous couplings fitted in the meson calculations of ref.
\cite{KoRi00,RiKo00} our couplings are, however, too large. This admittedly
indicates that our phenomenological approach needs further theoretical
justifications.

%\begin{figure}
%\resizebox{0.75\textwidth}{!}{%
%  \includegraphics{leer.eps}
%\vspace{5cm}       % Give the correct figure height in cm
%\end{figure}
%\begin{figure*}
%\vspace*{5cm}       % Give the correct figure height in cm
%\end{figure*}
%\begin{table}
%\begin{tabular}{lll}
%\hline\noalign{\smallskip}
%\end{tabular}
% Or use
%\vspace*{5cm}  % with the correct table height
%\end{table}
%
% BibTeX users please use
% \bibliographystyle{}
% \bibliography{}

\end{document}